\documentclass[letter]{aastex701}

\usepackage{graphicx}
\usepackage{graphbox}
\usepackage{hyperref}
\usepackage{amssymb,amsmath}
\usepackage{stmaryrd}
\usepackage{booktabs}
\usepackage{float}

\let\tablenum\relax

\usepackage[print-unity-mantissa=false, separate-uncertainty=true, multi-part-units=single]{siunitx}
\DeclareSIUnit\erg{erg}
\DeclareSIUnit\angstrom{\text {Å}}
\DeclareSIUnit\jansky{\text {Jy}}

\def\lea{\mathrel{<\kern-1.0em\lower0.9ex\hbox{$\sim$}}}
\def\gea{\mathrel{>\kern-1.0em\lower0.9ex\hbox{$\sim$}}}
\newcommand{\lta}{{\>\rlap{\raise2pt\hbox{$<$}}\lower3pt\hbox{$\sim$}\>}}
\newcommand{\gta}{{\>\rlap{\raise2pt\hbox{$>$}}\lower3pt\hbox{$\sim$}\>}}

\begin{document}
\title{Multiwavelength Classification of Chandra Sources in Omega Centauri}

\shorttitle{GC CSC TD}

\author{Steven Chen}
\affiliation{Department of Physics, The George Washington University, 725 21st St. NW, Washington, DC 20052}
\email{schen70@gwmail.gwu.edu}

\author{Oleg Kargaltsev}
\affiliation{Department of Physics, The George Washington University, 725 21st St. NW, Washington, DC 20052}
\email{kargaltsev@email.gwu.edu}

\author{Jeremy Hare}
\affiliation{NASA Goddard Space Flight Center, Greenbelt, MD, 20771}
\affiliation{NASA Postdoctoral Program Fellow}
\email{harej10@gmail.com}

\author{Hui Yang}
\affiliation{IRAP, Université de Toulouse/CNRS/CNES, 9 avenue du colonel Roche, 31028 Toulouse, France}
\email{huiyang@gwmail.gwu.edu}

\author{Oleg Smirnov}
\affiliation{Centre for Radio Astronomy Techniques and Technologies, Department of Physics \& Electronics, Rhodes University, Makhanda, 6140, South Africa}
\affiliation{South African Radio Astronomy Observatory, Cape Town, 7925, South Africa}
\affiliation{Institute for Radioastronomy, National Institute of Astrophysics (INAF IRA), Via Gobetti 101, 40129 Bologna, Italy}
\affiliation{Astrophysics, Department of Physics, The University of Oxford, Keble Road, Oxford, OX1 3RH, UK}
\affiliation{Breakthrough Listen, Astrophysics, Department of Physics, The University of Oxford, Keble Road, Oxford, OX1 3RH, UK}
\email{o.smirnov@ru.ac.za}

\author{Ian Heywood}
\affiliation{SKA Observatory, Jodrell Bank, Lower Withington, Macclesfield SK11 9FT, UK}
\affiliation{Centre for Radio Astronomy Techniques and Technologies, Department of Physics and Electronics, Rhodes University, Makhanda 6140, South Africa}
\affiliation{South African Radio Astronomy Observatory, 2 Fir Street, Observatory 7925, South Africa}
\affiliation{Jodrell Bank Centre for Astrophysics, University of Manchester, Oxford Road, Manchester UK}
\affiliation{Department of Physics, University of Oxford, Denys Wilkinson Building, Keble Road, Oxford OX1 3RH, UK}
\email{ian.heywood@skao.int}

\author{Talon Myburgh}
\affiliation{Centre for Radio Astronomy Techniques and Technologies, Department of Physics \& Electronics, Rhodes University, Makhanda, 6140, South Africa}
\email{myburgh.talon@gmail.com}

\begin{abstract}

We present a comprehensive study of X-ray sources in the globular cluster NGC 5139 to date. Over 200 Chandra sources are cross-matched to multiwavelength catalogs from HST, JWST, and MeerKAT. We employ a machine-learning classification pipeline (MUWCLASS) for the first time in such a crowded stellar environment and introduce the globular-cluster-specific training dataset, which compiles previously classified sources from 25 Galactic globular clusters. Given the challenges presented by the high-density environment, we compare the MUWCLASS results with traditional classification methods. We confirm counterpart associations suggested in previous works and also identify new confident and candidate associations. We discuss the implications of our classification results in light of models of cluster dynamical evolution and binary formation/destruction.

\end{abstract}


\section{Introduction}

Globular clusters are ancient, massive, and dense star clusters, mostly situated outside the plane of galaxies \citep{renaud_origin_2017}. About $150$ GCs are known in the Milky Way \citep{kharchenko_global_2013}. While densely populated with main sequence (MS) stars, GCs are also stellar graveyards, containing thousands of compact objects, including white dwarfs (WDs), neutron stars (NSs), and black holes (BHs) in isolation or in binary systems \citep{pooley_globular_2010, heinke_xray_2010,benacquista_relativistic_2013,berg_x-ray_2019}. Over the past few decades, extensive multiwavelength observations of GCs have revealed large populations of X-ray-emitting sources, including tidally locked coronally active binaries (ABs), cataclysmic variables (CVs), millisecond pulsars (MSPs), and low-mass X-ray binaries in quiescent (qLMXBs) and active states \citep{kremer_compact_2026}. These classifications are traditionally made by analyzing X-ray spectra and light curves and comparing source positions on color-magnitude and hardness-ratio diagrams with those of sources of known classes. However, these methods require manually identifying likely optical counterparts, and are difficult for faint X-ray sources. Therefore, 80\% of $\sim 2000$ CXO sources in all Galactic GCs remain unclassified. Some of these sources are likely to be compact objects, providing insights into stellar and binary evolution, as well as cluster dynamics.

Over the last decade, machine learning (ML) techniques have been incorporated in astronomy for classification and regression tasks, including in X-ray astronomy \citep{tranin_probabilistic_2022}. Such a classification pipeline specifically for Chandra sources has been recently developed and applied to a large fraction of sources from CSCv2.0 \citep[The MUWCLASS pipeline, ][]{yang_classifying_2022}. This approach requires a training dataset (TD) of reliably classified sources. However, it is difficult to classify X-ray sources solely from X-ray properties, especially since most unclassified sources are X-ray faint. Therefore, obtaining accurate multiwavelength properties, both for classified sources in the TD and unclassified sources we are attempting to classify, is crucial. Due to the challenges presented by very crowded environments, \cite{yang_classifying_2022} avoided GCs as well as other crowded environments for constructing their TD and classifying unknown sources. In this work, we use our dataset of reliably classified X-ray sources in Galactic GCs with their multiwavelength properties collected from CSCv2.1 \citep{evans_chandra_2020} and HUGS \citep{nardiello_hubble_2018}. The dataset was first presented in \citep{chen_dataset_2023} \footnote{ available at \url{https://home.gwu.edu/~kargaltsev/XCLASS_GC} in an interactive environment}. 

Omega Centauri (NGC 5139, $\omega$ Cen), the most massive GC known in the Milky Way ($M=4\times10^6 M_\odot$), is located at a distance of 5.49 kpc \citep{haberle_omegacat_2025}. Its estimated age is $>11.5$ Gyr, although, as it contains multiple stellar populations, the era of star formation may span several Gyr \citep{tailo_mosaic_2016, haberle_omegacat_2024}. Its unusual properties are theorized to be due to its origin as the nuclear star cluster of a dwarf galaxy \citep{ibata_identification_2019, massari_origin_2019, haberle_omegacat_2024}. Recent Gaia data have been used to identify the tidal stream of stars left over as remnants of the merger event and to categorize $\sim 35\%$ of Galactic GCs as products of similar mergers \citep{ibata_identification_2019, massari_origin_2019, pfeffer_accreted_2021}.

$\omega$ Cen has been thoroughly studied in X-rays and optical wavelengths, with a combined Chandra observation time of almost 300 ks and deep HST observations in multiple bands. Despite a combined total of 244 X-ray sources detected in $\omega$ Cen in past works \citep{cool_hstacs_2013, henleywillis_deep_2018}, the number of confidently classified X-ray sources remains relatively small and includes one qLMXB, 6 CVs, 5 spider-type MSPs, and 6 non-spider-type MSPs \citep{henleywillis_deep_2018, zhao_chandra_2023}. These MSPs were only recently discovered and timed using modern radio telescopes \citep{dai_discovery_2020, chen_meerkat_2023, dai_timing_2023}, with X-ray counterparts identified by \cite{zhao_chandra_2023}, emphasizing the importance of multiwavelength data in securing classifications. Thus, around 90\% of X-ray sources in $\omega$ Cen remain unclassified. If the class distribution of these fainter X-ray sources can be determined, or even reasonably constrained, it would in turn constrain models of compact binary evolution and GC dynamical evolution.

This paper is structured as follows: In Section \ref{TD}, we present the training dataset of reliably classified X-ray sources. In Section \ref{methods}, we discuss the modifications made to the MUWCLASS pipeline to adapt it to classifying sources in GCs. In Section \ref{analysis}, we begin analysis by evaluating MUWCLASS classification performance on the TD in Subsection \ref{LOO}, before continuing our analysis on both the previously classified and unknown X-ray sources in $\omega$ Cen. We conduct our analysis class by class, and supplement our ML classifications with spectral, timing, and multiwavelength information not used in MUWCLASS. In Section \ref{discussion}, we discuss the current performance and limitations of MUWCLASS, update the population census of X-ray sources in $\omega$ Cen, and examine the implications thereof. Finally, in Section 6, we summarize our classification findings.

\section{Training Dataset}
\label{TD}

\subsection{Crossmatching GCs}

We compiled source positions, positional uncertainties (PUs; at 95\% confidence), and reliable classifications from 50+ publications\footnote{referenced in our dataset} on Chandra sources in 25 GCs. We then sorted the X-ray sources into 6 main classes: AGNs, ABs, CVs, MSPs, Spider MSPs, and qLMXBs. For the 19 clusters where HST counterparts to X-ray sources have been published, we also compiled HST counterpart coordinates, photometry, and instruments/filters, when available. When multiple epochs of published data are available (e.g., for 47 Tuc, Omega Cen, and NGC 6397), we used the latest values. A summary of the classified TD sources in each cluster and the works we used is presented in Table \ref{tab:td_references}.

We found that the reduction and analysis of the CXO and HST data was quite heterogeneous among different publications. For uniformity, we cross-matched the published Chandra coordinates to the CSC, and the published HST coordinates to several catalogs, including the HST UV Globular Cluster Survey \citep[HUGS,][]{nardiello_hubble_2018} for most GCs, and oMEGACat \citep{haberle_omegacat_2024} for Omega Cen. These catalogs are all astrometrically tied to the Gaia catalogs (Gaia DR2 for CSC, DR1 for HUGS, and DR3 for oMEGACat), enabling reliable cross-matching between them. We also use several catalogs on the Chandra Deep Fields, discussed in \ref{TD AGN}, while constructing our sample of AGNs.
 
We first cross-matched the published coordinates of Chandra sources in GCs to CSC sources, using both PUs combined in quadrature. For cases where the published PUs were less than $0.5\arcsec$ or not reported, we used $0.5\arcsec$. We found systematic (i.e., the same for all sources from a particular publication) astrometric offsets between the published and CSC source coordinates to be 0.1--0.3$\arcsec$ in most GCs. The offsets can be attributed to the differences between the relative astrometric corrections performed in the individual publications and the absolute corrections used to tie the CSC frame to the Gaia CRF3\footnote{See \url{https://cxc.cfa.harvard.edu/csc/memos/files/Martinez-Galarza_stack_astrometry_translation_spec_doc.pdf} for details}. We dropped a few sources that had CSC detection significance $<2$. After cross-matching, we adopted the CSC properties, including coordinates, multi-band fluxes, and variability indices. 

Similarly, we cross-matched the published coordinates of confidently identified HST counterparts to HUGS, which provides magnitudes in 5 photometric bands: WFC3/UVIS F275W, F336W, F438W, and ACS/WFC F606W, F814W.\footnote{Since HUGS uses Vega magnitudes, we use Vega magnitudes throughout this work.} Since the density of HST sources is much higher than that of X-ray sources in GCs, and the absolute HST astrometry in some (especially older) publications could be inaccurate, we used an iterative approach to find the correct HUGS counterpart. We first cross-matched published coordinates of HST sources to the HUGS source with the smallest mean difference between their magnitudes (in the closest comparable bands) within a $1.5\arcsec$ radius, and plotted the separations. If a clear systematic offset between the published coordinates and the HUGS coordinates is seen, the median offset was applied as an astrometric correction to their published coordinates. Then, the closest HUGS source is taken if its separation is $<0.1\arcsec$ (or $<0.2\arcsec$) after the correction and the photometry differs by $<1$ mag in at least one (or two, if separation is $<0.2\arcsec$) comparable band(s).

This procedure was applied to 47 Tuc, where a known $1.3\arcsec$ astrometric offset \citep{clement_variable_2001} between published HST coordinates in \cite{albrow_frequency_2001} and HUGS coordinates was confirmed, and to NGC 6397, where a $0.3\arcsec$ offset was found. Other clusters either had good astrometry or too few HST counterparts for determining if there is a systematic shift, and a simple positional cross-match was used if it satisfied the above criteria. If no HST counterpart to an X-ray source was previously identified, then crossmatching was not done, and only X-ray properties were used in the training dataset. 

\subsection{Crossmatching for $\omega$ Cen}
\label{omegacat}

We crossmatch the X-ray source positions from \cite{henleywillis_deep_2018}, who analyzed 222 ks Chandra ACIS observations, and \cite{cool_hstacs_2013}, who analyzed 69 ks ACIS observations, to CSC 2.1. For convenience, we reference CSC sources by their labels given in \cite{henleywillis_deep_2018} and \cite{cool_hstacs_2013}, when available.

Out of 244 sources in these publications, 222 were unambiguously crossmatched to the CSC. Notably, 4 of the 12 MSPs (MSPs A, J, N, R) with X-ray counterparts compiled in \cite{zhao_chandra_2023} were too faint to be detected in the CSC. One other source, 21b (2CXO J132635.3-472758), was classified as a potential counterpart to MSP Q in \cite{zhao_chandra_2023}, and was classified as a faint blue CV with an optical counterpart in the earlier work of \cite{cool_hstacs_2013}, before the MSP was discovered in the radio. We remove this source from the TD, and independently classify the counterpart ourselves.

Omega Centauri is not included in HUGS. An earlier catalog \cite{bellini_new_2010} includes all bands used in HUGS, but this study only covers the central $\sim 5\arcmin$ (in diameter) of the cluster, leaving out many Chandra sources.

In 2024, an updated catalog of HST sources in Omega Cen (oMEGACat) was published by \cite{haberle_omegacat_2024}, using observations in 7 bands. The photometric coverage is nearly complete out to the half-light radius of $\sim \SI{5}{\arcmin}$, and proper motion (PM) measurements are reported for more than 1 million sources. The catalog includes the 5 HUGS bands, except it uses ACS/WFC F435W instead of WFC3/UVIS F438W. However, the F606W observations only cover the central 2.5\arcmin region. Since the nearby F625W band has much wider coverage, we use Gaussian process interpolation to predict the F606W band magnitudes from nearby bands (see Section \ref{GP} for details). oMEGACat also includes the F658N H$\alpha$ band, which is useful for identifying CVs and other accreting sources and for checking the reliability of our counterpart matches. However, since only 9 clusters have coverage in F658N, and 10 clusters in F625W, we opt to not include these bands as features in MUWCLASS. Furthermore, for convenience of comparing to TD sources, we treat the F435W magnitudes as F438W magnitudes. Lastly, oMEGACat includes crossmatches to Gaia DR3 \citep{gaia_collaboration_gaia_2023} and the Gaia Focused Product Release \citep[FPR, ][]{weingrill_gaia_2023}, which include independent PM and parallax measurements for the brighter stars. The limiting magnitude of the observations in all bands is around 25 mag, with large increases in photometric errors above this magnitude, especially in the bluer bands \citep{haberle_omegacat_2024}.

\begin{figure*}
    \centering
    \includegraphics[width=0.5\textwidth]{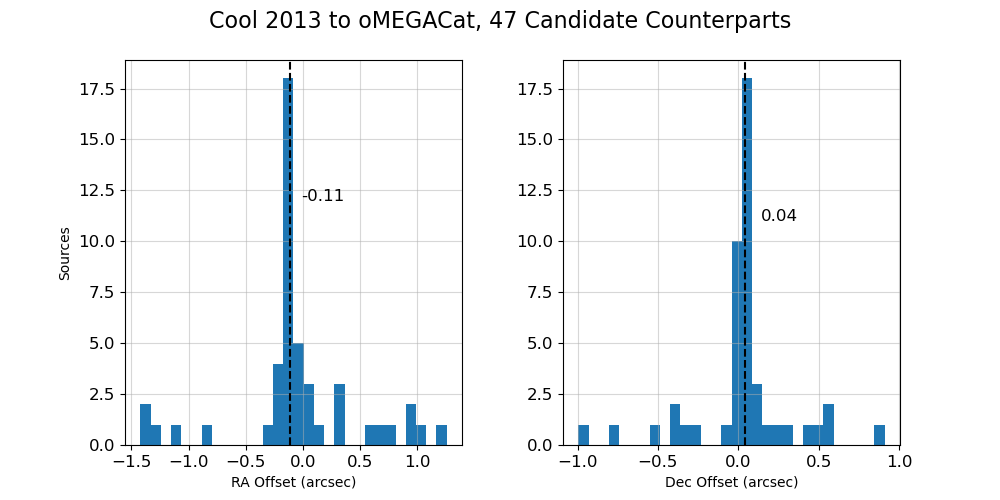}
    \includegraphics[width=0.5\textwidth]{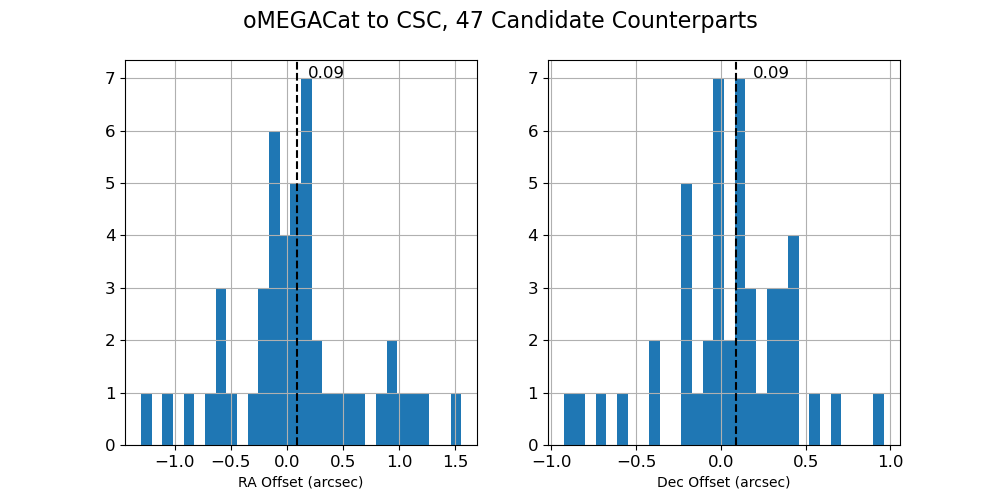}
    \includegraphics[width=0.5\textwidth]{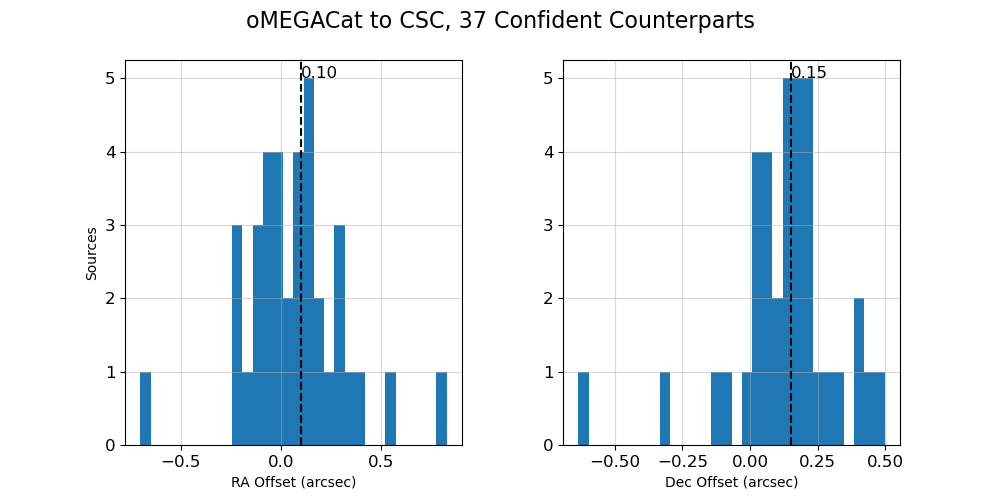}
    \caption{Top: Astrometric offsets between coordinates of 47 counterparts to X-ray sources reported in \cite{cool_hstacs_2013} and oMEGACat. Middle: Preliminary astrometric offsets between CSC and oMEGACat coordinates of counterparts suggested in \cite{cool_hstacs_2013}. Bottom: Final astrometric offsets between CSC and oMEGACat coordinates of 37 confidently classified associations, after correcting for PMs. }
    \label{fig:ngc5139_lit_omegacat_astrometry_offset}
\end{figure*}

We crossmatched CSC positions to known and candidate HST counterparts in oMEGACat using an iterative procedure, similar to the one we use to cross-match to HUGS in other GCs. We first matched the coordinates of HST counterparts in \cite{cool_hstacs_2013} to oMEGACat sources with the smallest magnitude difference in comparable bands within \SI{1}{\arcsec}. While a systematic offset between the two sets of HST coordinates exists (see row 1 of Figure \ref{fig:ngc5139_lit_omegacat_astrometry_offset}), an unambiguous match was found in almost all cases. We then crossmatched the oMEGACat to the CSC, finding a systematic offset of $\Delta_{\alpha}=$0.09\arcsec$, \Delta_{\delta}=0.09\arcsec$, which we then apply to all oMEGACat sources, before performing probabilistic crossmatching to the CSC. However, as many of the HST counterparts in \cite{cool_hstacs_2013} are only candidates, and may not be the true counterpart, this correction is preliminary. A more accurate correction can only be calculated using a set of confident optical counterparts with measured PMs. We thus use the results of our analysis from Section \ref{analysis}, specifically, 37 PM-selected AGN and foreground stars, CV, and MSP counterparts, to refine the astrometric correction. Cluster bulk PMs and individual star PMs in oMEGACat were used to propagate source coordinates to the epoch of CSC 2.1 (2016.0). After these corrections, a final systematic offset of $\Delta_{\alpha}\approx0.$1\arcsec$, \Delta_{\delta}\approx0.15\arcsec$ was found. The preliminary and final offset distributions are shown in Figure \ref{fig:ngc5139_lit_omegacat_astrometry_offset}. The scatter in the final offset distribution can be explained by the X-ray source PSFs. Figure \ref{fig:chandra_omegacat_footprint} shows the approximate footprints of the Chandra and oMEGACat images used for this work. 

\begin{figure*}
    \centering
    \includegraphics[width=0.8\textwidth]{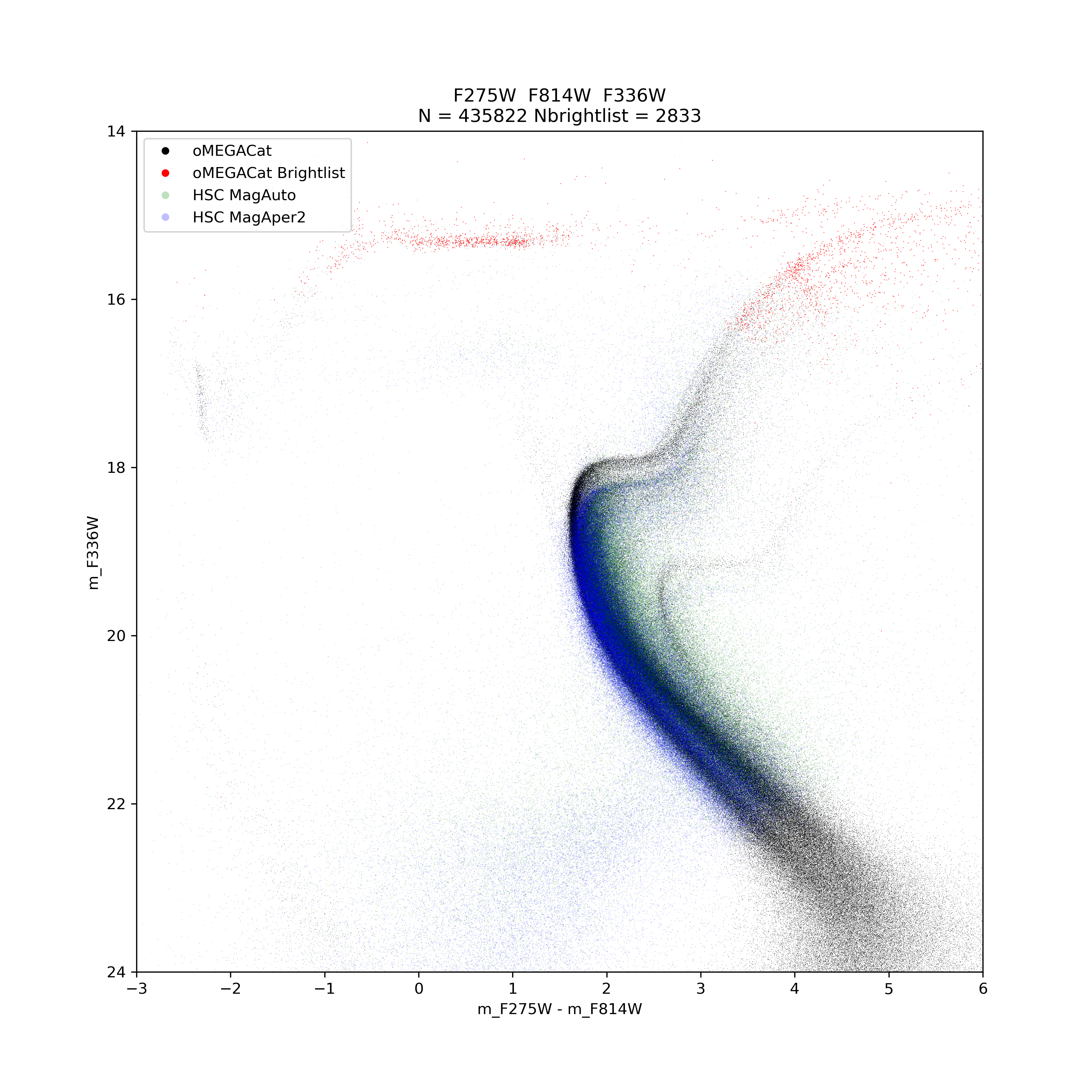}
    \caption{Photometric comparison between oMEGACat and HSC (using both photometry extraction methods). Significant differences can be seen at brighter magnitudes, where there are significantly fewer stars in HSC after the MS turnoff. Some fine structure, including the multiple MSs, is also less resolved in HSC. We therefore use oMEGACat, as it is better suited for our analysis.}
    \label{fig:ngc5139_omegacat_hsc_comparison}
\end{figure*}

\begin{figure*}
    \centering
    \includegraphics[width=0.8\textwidth]{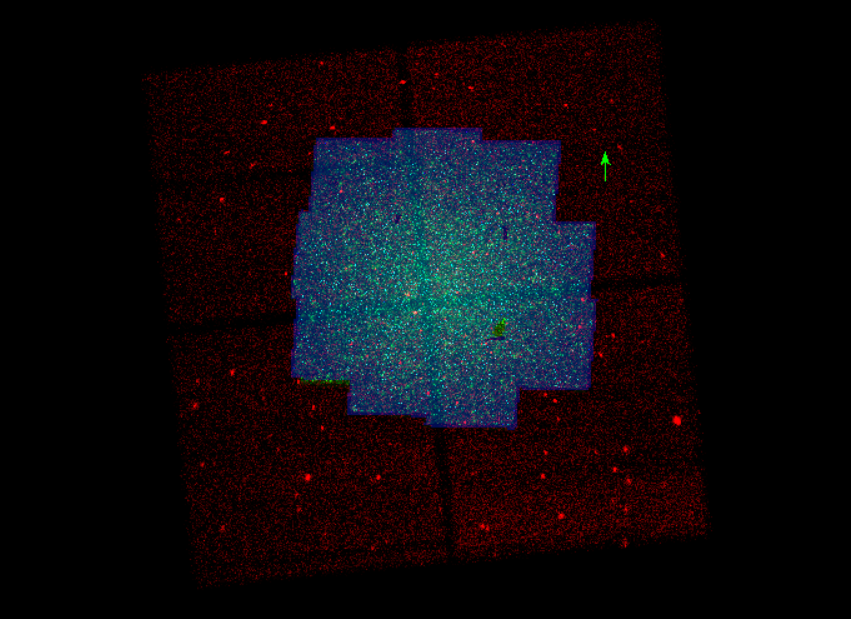}
    \caption{Chandra (red) and oMEGACat F275W (blue) and F814W (green) images of the central regions of $\omega$ Cen, showing the approximate coverage of both catalogs. Green arrow pointed north is $1\arcmin$ in length.}
    \label{fig:chandra_omegacat_footprint}
\end{figure*}

We also experimented with the Hubble Source Catalog v3 (HSC; \cite{lubow_hubble_2017}), but found that the photometry, particularly for the brighter stars, tended to be less accurate than that of oMEGACat. See Figure \ref{fig:ngc5139_omegacat_hsc_comparison} for a comparison CMD of the different catalogs. 

\subsection{AGNs}
\label{TD AGN}

AGNs are expected to make up a significant percent of X-ray sources in GCs, with \cite{henleywillis_deep_2018} estimating $\sim 140$ out of the 233 X-ray sources they detected in Omega Cen ($\sim 60\%$) to be AGNs based on AGN densities in the Chandra Deep Field South (CDFS). Therefore, it is necessary to include a reliable sample of AGNs in our TD. We crossmatch classified AGNs from the Chandra Deep Field South 7-Ms catalog \citep{luo_chandra_2017} and the Chandra Deep Field North 2-Ms catalog \citep{xue_2_2016} to the CSC. These works classified   $>1000$ CXO sources as AGN and galaxies and crossmatched them  to optical counterparts in various HST surveys, including GOODS-S \citep{guo_candels_2013}, CANDELS/3D-HST \citep{skelton_3d-hst_2014, barro_candelsshards_2019}, and HDUV \citep{oesch_hduv_2018}, covering all HUGS bands.

The classified AGN in the CDF fields tend to be significantly harder and dimmer than the AGN in our non-GC specific TD \citep{yang_visualizing_2021}. The difference in flux can be explained by the fact that the deep fields have been observed by Chandra to much greater depth, while the other AGN were classified using shallower observations all over the (non-Galactic plane) sky. The difference in hardness ratio can be explained by AGN at higher redshift being more obscured than nearer ones.
As the AGN in the CDF fields are not subject to these selection effects, they are more representative of the AGN population we expect to classify in the deep CXO exposures of $\omega$ Cen. Therefore, we use these CDF AGNs in our TD.

\subsection{Constructing the TD}

We construct our training dataset (TD) by combining the reliably classified X-ray sources in Omega Cen, 24 other GCs, and the AGNs in the Chandra Deep Fields (CDFs). While the crossmatching procedure is done for all Chandra sources present in the publications, we only use the ones with counterparts in the CSC and confidently classified as AB, AGN, CV, MSP, and qLMXB in the TD. 

We also merge spider-type and non-spider-type MSPs (including binary and isolated systems) in our TD. These MSPs were discovered in the radio, with companion properties (e.g., companion mass and orbital separation) estimated through timing solutions. However, only $\sim 25\%$ of these systems have identified optical counterparts, either due to lack of positive identification in the observations, lack of sufficient optical coverage, or intrinsic faintness (e.g. for isolated systems). Given the small number of sources (and smaller number of identified counterparts), and the relative similarity of MSPs compared to other classes, it is more practical to merge them for our purposes of ML classification.

Additionally, since the Chandra Deep Fields have significantly greater observational depths than $\omega$ Cen, we only use AGN in the TD that have CSC broadband flux (0.5-7.0 keV) $F_b \geq \qty{1e-16}{\erg\per\second\per\cm^2}$, which is comparable with the flux of the faintest sources observed in $\omega$ Cen. This cut reduces the number of AGN from 816 to 621. The sources collected in our TD in each GC (or CDF field), along with the referenced publication, are summarized in Table \ref{tab:td_references}. The number of sources per class in all GCs, and in Omega Cen specifically, along with the percentages of them that have identified HST counterparts, are shown in Table \ref{table:TD_counts}.

\begin{table}
\begin{tabular}{lccccccr}
\toprule
Cluster & AB & AGN & CV & MSP & Spider & qLMXB & References \\
\midrule
CDFN &  & 349 &  &  &  &  & (1),(2),(3) \\
CDFS &  & 452 &  &  &  &  & (4),(5),(2) \\
NGC 104 & 53 &  & 30 & 17 & 4 & 4 & (6),(7),(8),(9),(10) \\
NGC 2808 &  &  & 1 & 1 &  & 1 & (11),(12),(13) \\
NGC 288 & 1 & 1 &  &  &  &  & (14) \\
NGC 362 & 6 &  &  &  &  & 1 & (15) \\
NGC 5139 & 1 & 6 & 7 & 2 & 5 & 1 & (16),(17),(18),(19) \\
NGC 5904 &  &  &  & 2 & 1 &  & (20) \\
NGC 6093 &  &  & 5 &  &  &  & (21) \\
NGC 6121 & 26 & 1 & 2 & 2 &  &  & (22),(23) \\
NGC 6205 &  &  & 1 & 2 & 1 & 1 & (24),(25),(26) \\
NGC 6266 &  &  & 3 &  & 2 & 5 & (27),(19) \\
NGC 6304 &  &  &  &  &  & 1 & (28) \\
NGC 6341 &  & 1 & 4 &  & 1 &  & (29) \\
NGC 6388 &  &  & 4 &  &  &  & (30) \\
NGC 6397 & 21 & 1 & 11 &  & 2 & 1 & (31),(32),(19) \\
NGC 6440 &  &  &  & 1 &  & 9 & (33),(34),(35) \\
NGC 6626 &  &  & 3 & 2 & 4 &  & (36),(19) \\
NGC 6652 &  &  & 1 &  &  & 2 & (33),(37),(38),(39) \\
NGC 6656 &  &  &  &  & 1 &  & (40) \\
NGC 6752 & 2 & 5 & 7 & 5 &  &  & (41),(19) \\
NGC 6809 &  &  & 1 &  &  &  & (42) \\
NGC 6838 & 3 &  & 1 &  & 2 &  & (43),(44),(18) \\
NGC 7078 &  &  & 1 &  &  & 3 & (45),(46),(47) \\
NGC 7099 &  &  & 2 &  &  & 1 & (48) \\
TERZAN 5 &  &  & 5 & 3 & 3 & 7 & (49),(50),(19) \\
TERZAN 6 &  &  &  &  &  & 1 & (51) \\
\bottomrule
\end{tabular}
\caption{Overview of classified sources in each cluster used for the TD, and their references. (1) \cite{xue_2_2016}, (2) \cite{oesch_hduv_2018}, (3) \cite{barro_candelsshards_2019}, (4) \cite{guo_candels_2013}, (5) \cite{luo_chandra_2017}, (6) \cite{heinke_deep_2005}, (7) \cite{rivera-sandoval_discovery_2015}, (8) \cite{bhattacharya_chandra_2017}, (9) \cite{rivera_sandoval_new_2018}, (10) \cite{bao_periodic_2023}, (11) \cite{dieball_far-ultraviolet_2005}, (12) \cite{servillat_combined_2008}, (13) \cite{sanna_discovery_2017}, (14) \cite{kong_chandra_2006}, (15) \cite{kumawat_chandra_2024}, (16) \cite{cool_hstacs_2013}, (17) \cite{henleywillis_deep_2018}, (18) \cite{zhao_chandra_2023}, (19) \cite{bao_chandra_2024}, (20) \cite{zhang_discovery_2023}, (21) \cite{dieball_far-ultraviolet_2010}, (22) \cite{bassa_x-ray_2004}, (23) \cite{lugger_exotica_2023}, (24) \cite{servillat_dwarf_2011}, (25) \cite{shaw_radius_2018}, (26) \cite{zhao_chandra_2021}, (27) \cite{oh_multi-epoch_2020}, (28) \cite{guillot_chandra_2009}, (29) \cite{lu_chandra_2011}, (30) \cite{maxwell_x-ray_2012}, (31) \cite{cohn_identification_2010}, (32) \cite{dieball_far-ultraviolet_2017}, (33) \cite{int_zand_identification_2001}, (34) \cite{pooley_chandra_2002}, (35) \cite{heinke_discovery_2010}, (36) \cite{vurgun_neutron_2022}, (37) \cite{stacey_examination_2012}, (38) \cite{paduano_maveric_2021}, (39) \cite{gautam_upgraded_2022}, (40) \cite{amato_search_2019}, (41) \cite{cohn_deep_2021}, (42) \cite{bassa_x-ray_2008}, (43) \cite{elsner_chandra_2008}, (44) \cite{huang_study_2010}, (45) \cite{white_discovery_2001}, (46) \cite{hannikainen_x-ray_2005}, (47) \cite{heinke_discovery_2009}, (48) \cite{zhao_deep_2020}, (49) \cite{heinke_faint_2006}, (50) \cite{bogdanov_deep_2021}, (51) \cite{vats_quiescent_2018}. }
\label{tab:td_references}
\end{table}

\begin{deluxetable}{lrrrr}
\tabletypesize{\footnotesize}
\tablecaption{Number of sources per class in the TD, along with the percentage of them that have HST counterparts, in all environments and in $\omega$ Cen specifically. \label{table:TD_counts}}
\tablehead{
\colhead{Class} &
\colhead{Number} &
\colhead{Counterpart \%} &
\colhead{Omega Cen} &
\colhead{Counterpart \%}
}
\startdata
MSP      & 37  & 22  & 2 & 0 \\
Spider   & 25  & 28  & 5 & 0 \\
CV       & 88  & 82  & 6 & 100  \\
AB       & 111 & 100 & 1 & 100  \\
qLMXB    & 37  & 30  & 1 & 100  \\
AGN      & 621 & 57  & 6 & 100  \\
\enddata
\end{deluxetable}

\section{Methods}
\label{methods}

\subsection{Probabilistic Crossmatching to HST Counterparts}
\label{nway}

\begin{figure*}
    \centering
    \includegraphics[width=0.5\textwidth]{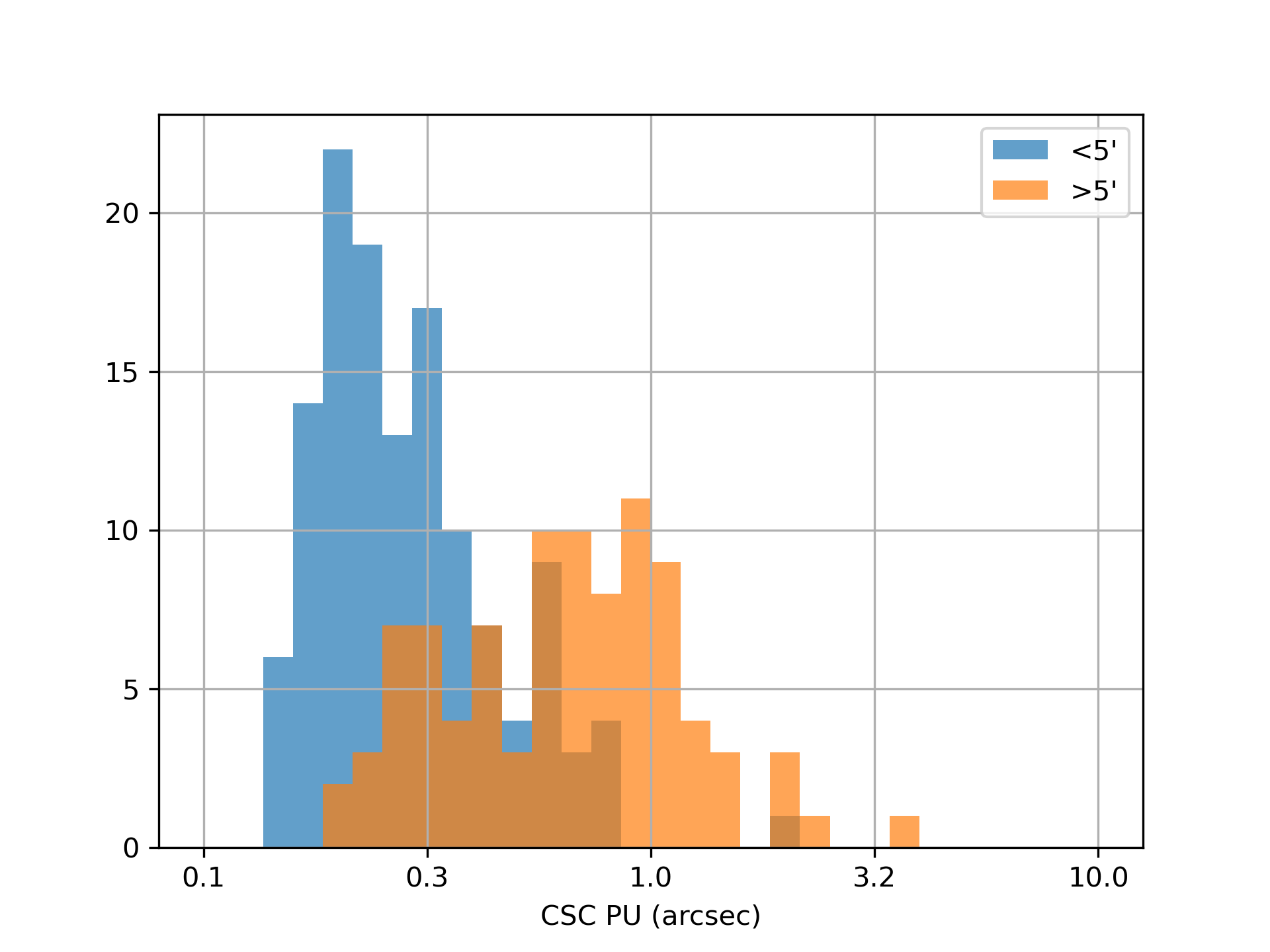}
    \caption{Distribution of CSC $1 \sigma$ source PUs inside and outside the 5\arcmin radius of the $\omega$ Cen cluster center.  The latter is broader due to the larger PSF at larger off-axis angles. The horizontal axis is on a log scale. }
    \label{fig:NGC5139_csc_pu_hist}
\end{figure*}

\begin{figure*}
    \centering
    \includegraphics[width=0.5\textwidth]{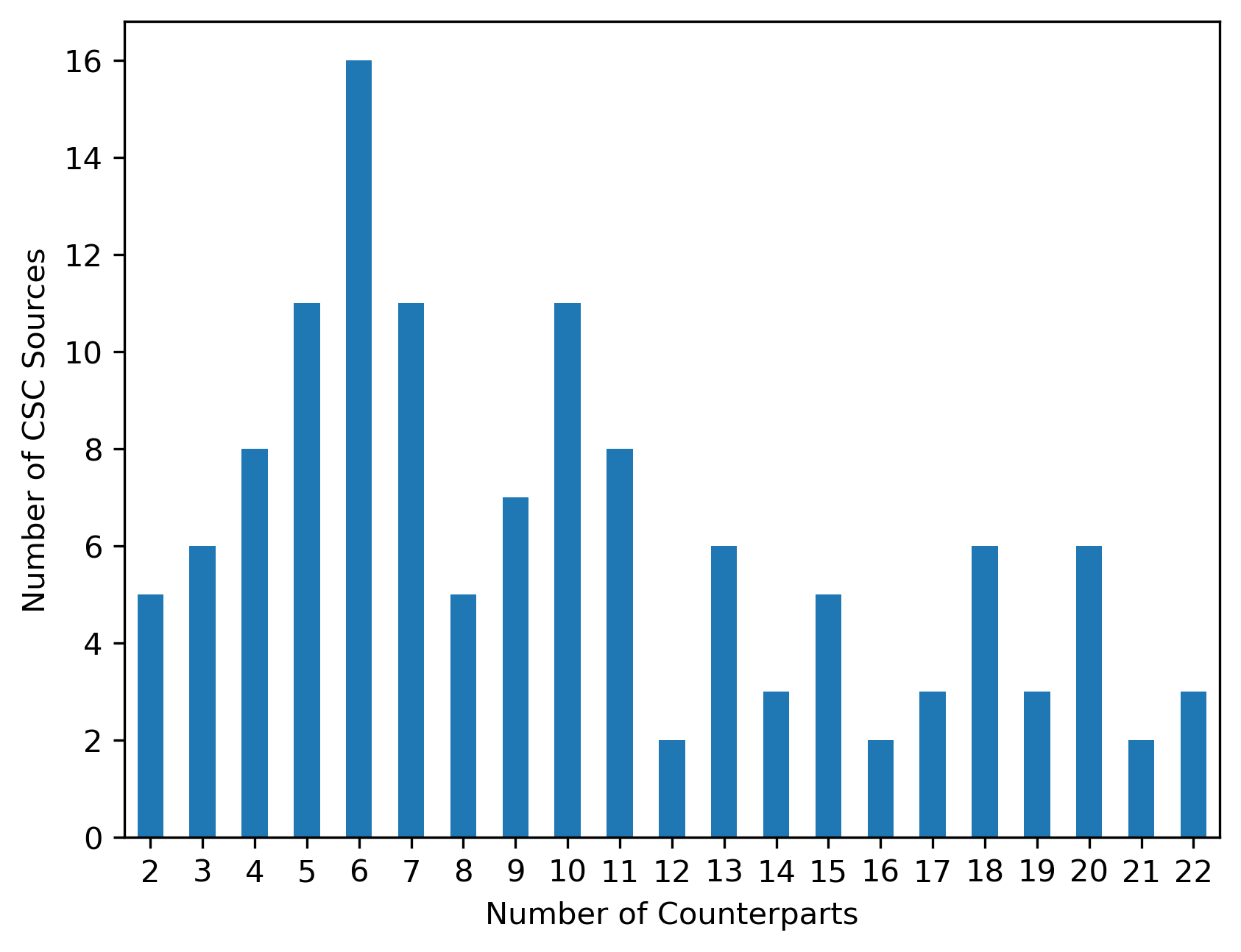}
    \caption{Plot of the number of counterparts each CSC source is matched to, up to $1\arcsec$ or $3\sigma$ CSC PU. Some sources have more than 22 potential counterparts, these are omitted for clarity.}
    \label{fig:csc_omegacat_counterparts_hist}
\end{figure*}

As GCs are very dense environments, even the sub-arcsecond resolution of Chandra gives multiple HST sources within the error circles of unclassified CSC sources. Since manually evaluating each crossmatching association (CSC source - HST source pair) becomes impractical for increasingly large datasets, we adopt a probabilistic approach to counterpart assignment, and the subsequent classification of the resulting associations. For probabilistic crossmatching, we use the Bayesian crossmatching package \textsc{NWAY} \citep{salvato_finding_2018}. For a given primary catalog (here, the CSC) and a secondary catalog (oMEGACat), \textsc{NWAY} finds all secondary catalog counterparts within the maximum radius (chosen to be $1\arcsec$) of a primary catalog source and assigns a probability $P_i$ for each association to be true. \textsc{NWAY} also calculates the probability that at least one of these associations is true, $P_{\text{any}}$, based on source densities in each catalog. Therefore, $1-P_{\text{any}}$ is the probability that all associations are spurious, and $(1-P_{\text{any}}) + \sum_{i=1}^{n} P_i P_{\text{any}} =1$ for $n$ potential counterparts. To reduce the number of unlikely counterparts we need to analyze, we discard all counterparts outside $3\sigma$ positional uncertainty (PU) of the CSC source position\footnote{Note that the CSC reports 95\% ($2\sigma$) PUs by default}. Figure \ref{fig:NGC5139_csc_pu_hist} shows the distribution of the PUs for CSC sources within and outside 5\arcmin of the cluster center. Figure \ref{fig:csc_omegacat_counterparts_hist} shows the distribution of the number of counterparts matched to CSC sources within $1\arcsec$ or $3\sigma$ PU, whichever is less.

\textsc{NWAY} calculates crossmatching probabilities using the astrometric information about sources in both catalogs, including their PUs, separations, and densities. \textsc{NWAY} also supports adding custom priors on other source properties, such as magnitudes and colors, to weigh  association probabilities. These priors were used in, e.g., \cite{salvato_finding_2018} to find the counterparts to X-ray emitting stars and AGN based on their distinct distributions in infrared color-magnitude space. However, we choose not to incorporate these additional priors into the crossmatching procedure, since that would introduce additional biases and uncertainties.

\subsection{JWST data processing and alignment}

JWST observations of Omega Cen were conducted as part of GO-4343 (PI Kargaltsev) in 2024. The observations included 644s exposures with NIRCam F200W and F444W, a 2.1 ks exposure with MIRI F770W, and a 6.9 ks exposure on MIRI F1500W. Two additional calibration observations from ID 1478 (PI Boyer) and ID 6632 (PI Brooks) include 215 s exposures with NIRCAM F212N, F444W, F470N, and 419 s exposures with NIRCAM F115W and F277W. These observations have substantially smaller fields of coverage than oMEGACat.

We downloaded the data from MAST, and extracted photometry with \textsc{DOLPHOT} following the standard prescription \footnote{See \url{https://dolphot-jwst.readthedocs.io/en/latest/index.html}}. 

As the JWST observations are conducted at a different epoch than the reference epoch used to align the oMEGACat observations (2002.5), oMEGACat PMs were used to propagate oMEGACat source coordinates to the epoch of the JWST observations. Both cluster absolute PM and individual source PMs, when available, were used. These propagated HST coordinates were then used to correct the absolute astrometry of the JWST data using \textsc{Tweakreg}. After corrections, JWST sources were crossmatched to oMEGACat sources. Details of the crossmatching procedure can be found in \cite{chen_intermediate-mass_2026}. 

\subsection{Crossmatching to MeerKAT Radio Sources}

\begin{figure*}
    \centering
    \includegraphics[width=0.5\textwidth]{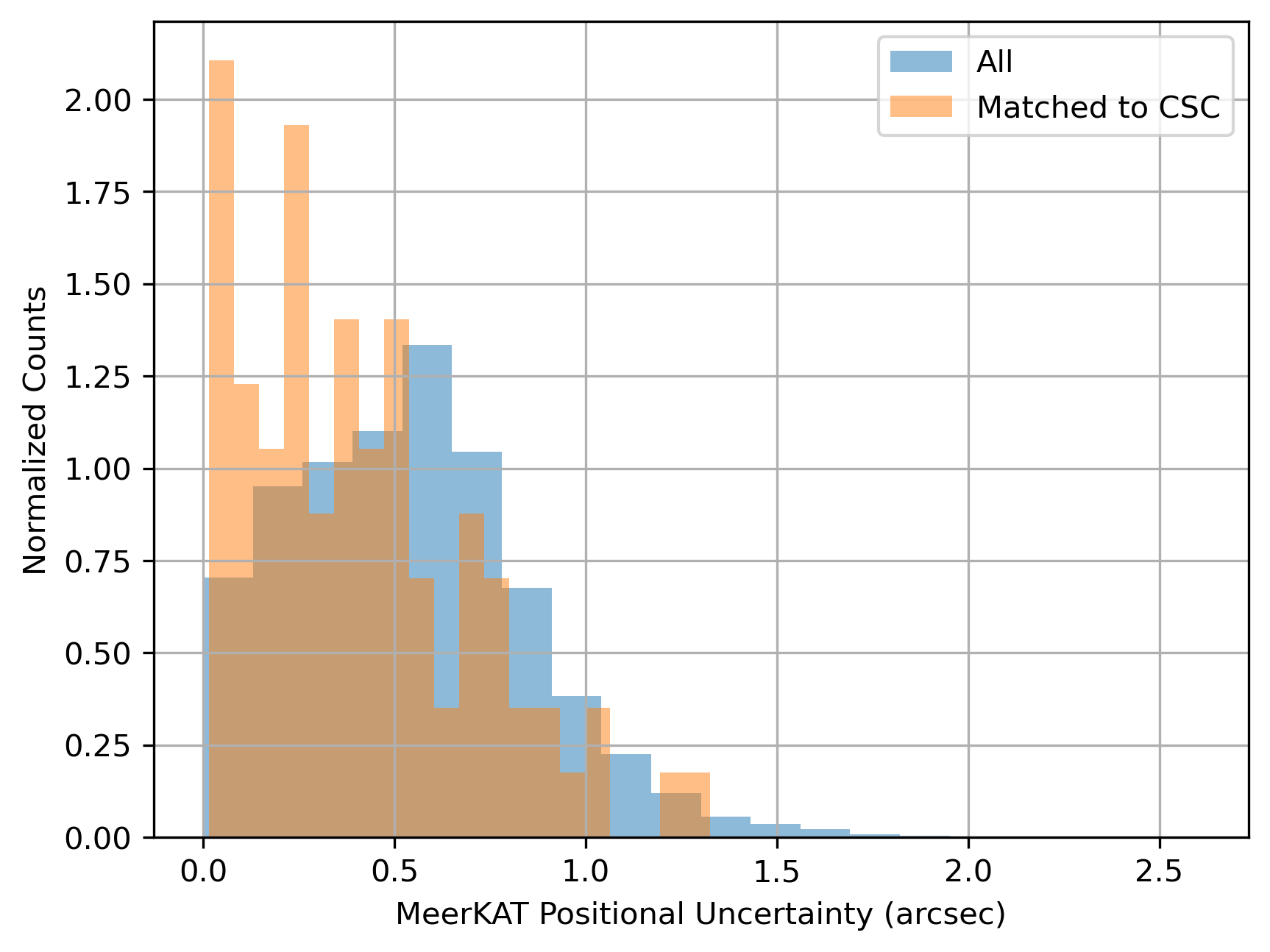}
    \caption{PUs of all MeerKAT point sources produced by \textsc{PyBDSF}, and of those crossmatched to a CSC source within $3\sigma$ combined PU. }
    \label{fig:meerkat_positional_uncertainty_hist}
\end{figure*}

In February 2025, the first results of deep MeerKAT studies, focused on the detection of medium timescale (minutes to hours) transients and variable sources were published by \cite{smirnov_mining_2025} for three GCs, including Omega Cen. Preliminary point source catalogs of the observations were produced using \textsc{PyBDSF} \citep{mohan_pybdsf_2015}, reaching down to $\lesssim \qty{10}{\micro\jansky}$ (corresponding to $\lesssim\qty{1e-19}{\erg\per\second\per\square\cm}$ and $\lesssim\qty{5e26}{\erg\per\second}$ at the cluster distance). The field of view in these observations was $\sim 20\arcmin \times 20\arcmin$, which exceeds the area covered by Chandra observations of Omega Cen. The typical PUs of the radio point sources are under $1\arcsec$, while those matched to CSC sources are under $0.5\arcsec$, see Figure \ref{fig:meerkat_positional_uncertainty_hist}. As the radio source density is relatively low ($\sim 1300$ sources in the area of coverage of the X-ray observations), we perform a simple crossmatching of MeerKAT sources with CSC sources, within $3\sigma$ CSC and MeerKAT PUs combined in quadrature, obtaining 87 matches. The chance coincidence probability for a MeerKAT source to be matched to a CSC source within a $2\arcsec$ radius is $\sim 2\%$. Since most combined PUs are significantly smaller than $2\arcsec$, we expect almost all of these matches to be real. 

\begin{figure*}
    \centering
    \includegraphics[width=0.8\textwidth]{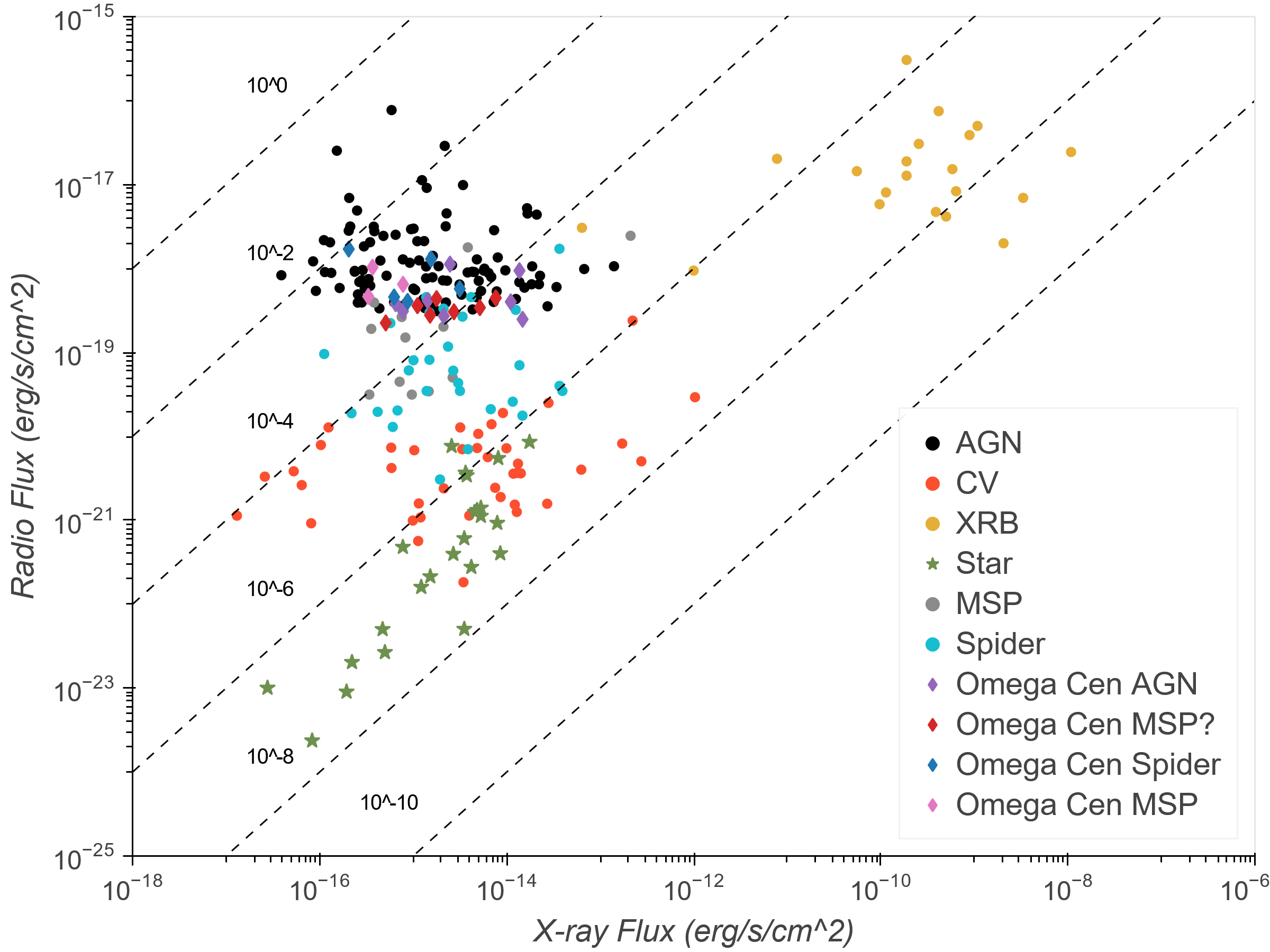}
    \caption{X-ray vs. radio fluxes of several classes of sources, scaled to the distance of $\omega$ Cen. Lines of constant flux ratio are shown. AGN are from CDFN \citep{xue_2_2016, morrison_very_2010}, CVs from \cite{barrett_radio_2020}, XRBs from \cite{dage_radio_2025}, coronally active stars from \cite{vedantham_peculiar_2022}, MSPs and spider MSPs from the ATNF and SpiderCat catalogs \citep{manchester_australia_2005, koljonen_spidercat_2025}}.
    \label{fig:radio_xray_lum}
\end{figure*}

We compiled X-ray and radio sources with known classifications and distances from several publications, and plotted their X-ray vs. radio fluxes, scaled to the distance of $\omega$ Cen in Figure \ref{fig:radio_xray_lum}. As can be seen, while CVs, ABs, and coronally emitting stars can be radio sources, at the cluster distance their fluxes are significantly below the current detection threshold. The presence of radio counterparts to X-ray sources thus provides strong evidence in favor of AGN or MSP classifications. Without radio observations, the X-ray and optical properties of undiscovered AGNs and MSPs may appear similar to CVs or ABs. Note that radio source properties were not used in the MUWCLASS classification pipeline.

\subsection{Luminosities and Extinctions}
\label{luminosity}

As distances to GCs are well-constrained, we can use luminosities instead of fluxes as features in MUWCLASS. Sources in the TD and unclassified sources in Omega Cen have luminosities calculated at the distance of their host GC. TD AGN "luminosities" are calculated for the Omega Cen distance. 
Additionally, as GCs (and the CDFs) are mostly located outside of the Galactic plane, with low, well-constrained absorption, we deredden the TD and unclassified sources using extinction values corresponding to the host GCs. We use dust and gas maps collected in the \textsc{gdpyc} package \citep{ruiz_ruizcagdpyc_2018}, and apply reddening to optical bands using the \textsc{extinction} package \citep{barbary_extinction_2016}. In this work, all X-ray fluxes and optical magnitudes reported for both TD and $\omega$ Cen sources are corrected for reddening and absorption based on their cluster position. This enables a direct comparison between sources from different clusters. For $\omega$ Cen, we use $E(B-V)=0.12$. 

\subsection{Gaussian Process Interpolation of Missing Values}
\label{GP}

In both the TD and field data, many sources are missing magnitudes in one or more HUGS bands used by MUWCLASS. These missing values can create large biases during classification. For example, as only a quarter of MSPs have counterparts in the TD, while all ABs do, a real MSP with its true counterpart may be classified as an AB just because the presence of counterpart makes it more similar to ABs in the multiwavelength feature space.

We take several steps to mitigate this bias. First, the publications from which we constructed the TD, as well as oMEGACat and the CDF catalogs, often contain photometry in bands other than HUGS bands. Often, these bands are adjacent in wavelength to the band(s) with missing measurement(s). We, therefore, construct SEDs (in units of $\SI{}{\erg\per\second\per\square\cm\per\angstrom}$) of the sources in the TD and those to be classified using all available HST photometry, and fill in the missing measurements in the HUGS bands with Gaussian processes (GPs). For more details on the implementation, see Appendix Section \ref{gp_details}.

Since GP predictive accuracy decreases with increasing distance from existing data, and since the GP may rapidly diverge outside the domain of the existing data, 
we only interpolate flux values in bands that have existing fluxes in bands on both sides. Thus, the F275W and F814W bands are never fitted in oMEGACat data, and rarely fitted in the TD due to the scarcity of observations in bands of shorter and longer wavelengths, respectively. In addition, many observations were conducted in narrow bands (N), which may contain absorption or emission features. These features may impact the fits disproportionately, and so, fluxes in narrow bands have their uncertainties increased to the equivalent of 0.5 mags, such that they contribute less to the GP fit. Fluxes in long-pass (LP) and extremely wide (X) bands are 
not used in the fit. We further reject fits that have a GP error greater than 1 mag, leaving the feature empty.

We test this procedure by removing magnitudes from a sample of oMEGACat sources in Omega Cen, then comparing fitted GP magnitudes to the true ones. We observe that the median differences are the largest for the F275W and F814W, at 0.56 and 1.1 mags respectively, while they are $<$0.15 mags for the other bands. This justifies our decision to only interpolate fluxes. Using this procedure, we filled in 2, 13, 57, 63, 7 flux values in the TD (out of 1114 TD sources), and 0, 64, 37, 599, 0 flux values in the oMEGACat data (out of 1521 potential counterparts to CSC sources) at the F275W, F336W, F438W, F606W, F814W bands, respectively. The large number of fill-ins in F606W is due to the lack of coverage in this band outside the core region, as discussed in Section \ref{omegacat}.

\subsection{Class Balance}

Having an uneven number of sources for each class in the TD can lead to biases, where sources are preferentially classified as the most populous classes. There are different ways\footnote{see, for example, \url{https://imbalanced-learn.org/stable/user_guide.html}} of mitigating this issue. MUWCLASS corrects class imbalance by oversampling under-represented classes via the commonly used Synthetic Minority Oversampling Technique (SMOTE, \cite{chawla_smote_2002}), which creates synthetic sources between two similar sources of the same class in the full feature space. However, for heterogeneous classes with few sources, SMOTE can create synthetic sources in the feature space between two sources far apart in feature space, leading to the creation of numerous synthetic sources occupying unrealistic regions. This problem becomes even more acute after dropping sources missing a feature from the TD (see Section \ref{procedure}). Thus, for classes having $<50$ sources in our TD, we first create 4 synthetic sources from each real source by randomly sampling from their feature uncertainties. SMOTE then creates additional synthetic sources of the same class using 4 neighbors (either real, or synthetic sources sampled from uncertainties). 

\subsection{MUWCLASS Classification Procedure}
\label{procedure}

Even after imputation with GPs, there are still sources that lack coverage in one or more HUGS bands, for which GP interpolation is not appropriate. Many X-ray sources in the TD were also not assigned a counterpart in the publications. For the TD, the band-averaged feature completeness per class is 87\%, 40\%, 58\%, 14\%, and 18\% for ABs, AGNs, CVs, qLMXBs, and MSPs, respectively. The differences are due to the varying optical brightness of the sources across different classes. For example, isolated MSPs are simply not detectable by HST at kpc distances, but those with WD companions may be detected as faint sources.

For sources missing values in one or more bands, classification can be biased due to the classifier assigning importance to the missing data\footnote{Note that v.1.3 and later versions of the \textsc{scikit-learn} random forest algorithm allow missing values in the data. However, the implementation produces results similar to imputing all missing values with some arbitrary value, e.g., -100, which does not eliminate the bias.}. To mitigate this bias, we introduce the following procedure: for each source to be classified, if a HUGS feature is present, we drop all sources in the TD without that feature; if that feature is missing, we drop that feature so that it is not used in ML classification. This procedure ensures that classification decisions are only based on the density of sources of each class in feature space, relative to the other classes, and not on the percent of sources of each class which have a particular feature measured.

Due to crowding in GCs, each X-ray source typically has multiple counterparts. We classify each counterpart association separately. We also classify a ``null association,'' where only CSC features are used, representing cases of no HST coverage, as well as the ``upper-limit association,'' representing cases where HST coverage exists but the true counterpart is below the detection limit in all bands with coverage. This ``upper-limit association'' simply uses a value of 30 mag for all bands, as the faintest sources in the HUGS and oMEGACat catalogs are close to 30 mag. The addition of such an upper-limit association ensures that at least one association we classify represents the true case, even if it is a non-detection. Following \textsc{NWAY} convention, the null association is named as the CSC source name appended by ``-0", each counterpart association is named sequentially from ``-1" in order of increasing separation, and the upper limit association is appended with ``\_UL".

Since classifying an association with full (or nearly full) HUGS bands coverage drops many sources for some classes in the TD (only 34 MSPs and Spiders combined, and 12 qLMXBs have full coverage), the small number of remaining TD sources in some classes could lead to biases during classification. On the other hand, classifying the null association does not drop any sources, but only uses X-ray features. Therefore, the classification results of both cases are complementary. For example, if an association with full HUGS band coverage is classified as an MSP, the null association should also have significant $P_{\text{MSP}}$, otherwise it may indicate biases introduced by the optical features. We therefore consider the classification probabilities of counterpart, upper limit, and null associations in our analysis (see Section \ref{analysis}).

Furthermore, in \cite{yang_classifying_2022}, CSC fluxes and their upper and lower limits were converted to symmetric fluxes and uncertainties and sampled from a (symmetric) Gaussian function. We changed the sampling procedure to sample directly from a split Gaussian distribution, defined in Eq. \ref{split_gaussian}:

\begin{align}
    & A\exp \left(-{\frac {(x-\mu )^{2}}{2\sigma _{1}^{2}}}\right)\quad {\text{if }}x<\mu, \\
    & A\exp \left(-{\frac {(x-\mu )^{2}}{2\sigma _{2}^{2}}}\right)\quad {\text{otherwise,}}
    \label{split_gaussian}
\end{align}

where $\quad A={\sqrt {2/\pi }}(\sigma _{1}+\sigma _{2})^{-1}$ is the normalization, $\mu$ is the CSC flux value, and $\sigma_1$ and $\sigma_2$ are the differences between the flux value and the lower and upper limits, respectively. 

Besides these changes, our classification procedure is the same as in \cite{yang_classifying_2022, yang_multiwavelength_2024}. We say that a source association is classified by MUWCLASS as class $\rm C$ with probability $\rm P_C$ if $ \rm P_C$ is the highest among the probabilities for all classes.

\subsection{Spectra and Lightcurves}

We supplement the ML classification results with analysis of Chandra spectra and lightcurves. CSC 2.1 provides the observation-level spectra and lightcurves for each CSC source from obsIDs 13726 and 13727, while the earlier two obsIDs were not used by the CSC. Spectral files from the CSC (.pha3, .rmf3, .arf3 files) were downloaded and fitted using the \textsl{Sherpa} package \citep{fruscione_ciao_2006}. Sources with more than 50 counts were fit with the thermal plasma emission model ({\tt mekal}) and the power-law {\tt xspowerlaw} (PL) models modified by interstellar photoelectric absorption using the {\tt xsphabs} (phabs) model \citep{wilms_absorption_2000}. The \texttt{wstat} statistic was used in all of the fits performed with {\sl Sherpa}.\footnote{see https://cxc.cfa.harvard.edu/sherpa/}.

The light-curve files from the CSC rely on the Bayesian-block approach for binning, and the shapes of the CSC light curves are difficult to interpret in many cases, especially for faint sources. Therefore, we extracted light curves for each observation (obsIDs 653, 1519, 13726, 13727) ourselves using the \texttt{dmextract} function in CIAO tools. We used the stack source region files for each source to define the extraction region. The light curves are binned in 1 ks bins. A summary of variable sources (with Kuiper variability significance $\gtrsim 3\sigma$), along with their classifications, is shown in Table \ref{tab:variable}. The corresponding light curves are shown in Figure \ref{fig:variable}.

\section{Analysis}
\label{analysis}

\subsection{Leave-One-Out Evaluation}
\label{LOO}

\begin{figure*}
    \centering
    \includegraphics[width=0.8\textwidth]{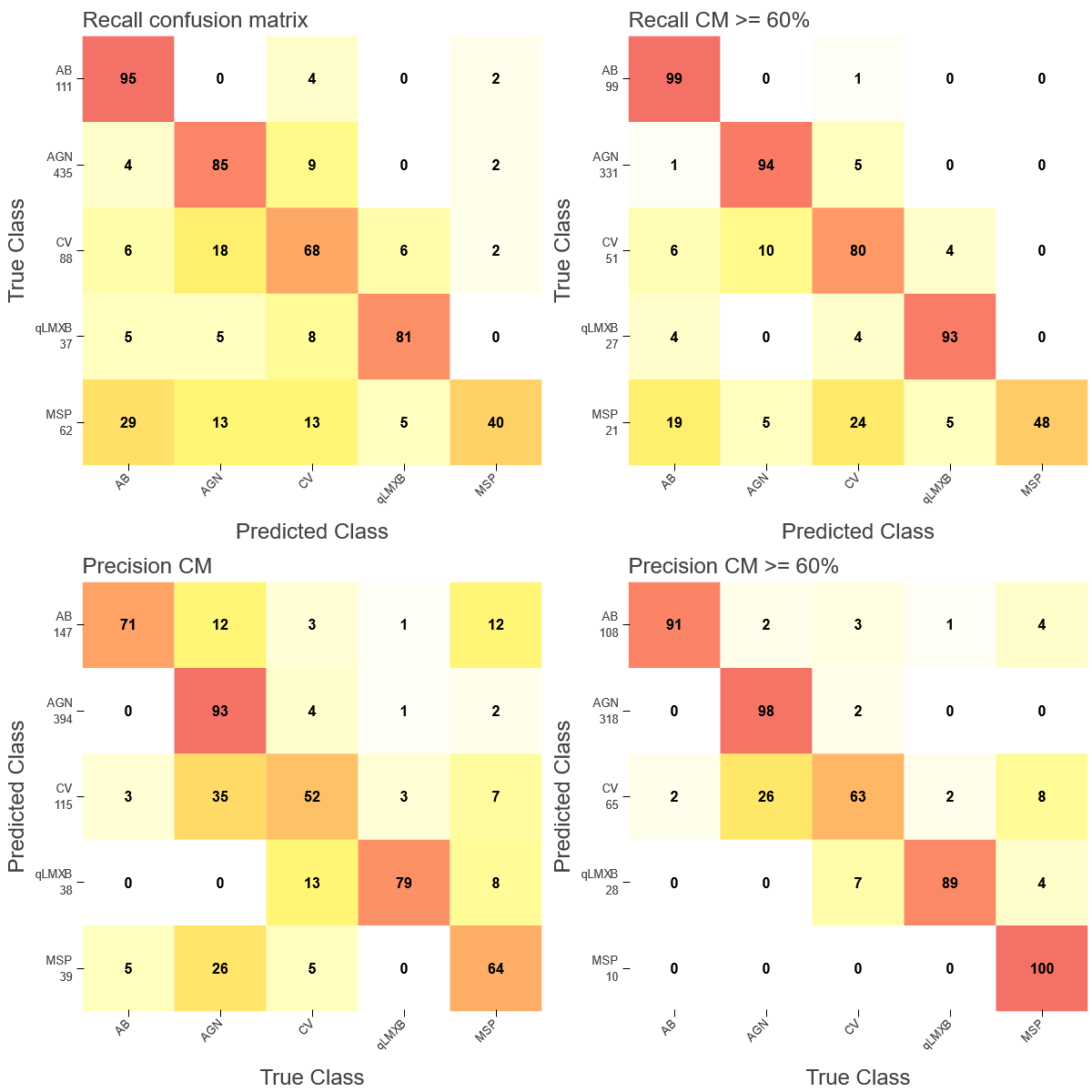}
    \caption{Confusion matrices (CM) showing performance of the LOO evaluation. Top row shows recall CMs, bottom row shows precision CMs. Left side shows all TD sources, right side shows only sources classified with probability $>60\%$.}
    \label{fig:cm_gc}
\end{figure*}

We first evaluate the performance of MUWCLASS using the GC-specific TD via a modified Leave-One-Out (LOO) evaluation procedure. In LOO, each source in the TD is sequentially removed from the TD, and classified using the model trained with all other sources in the TD. In our classification procedure, each source is classified using the subset of the TD that shares the same features as that source, see Section \ref{procedure}. AGNs in the TD have ``luminosities'' calculated at the $\omega$ Cen distance of 5.49 kpc. We show confusion matrices (CM) of the LOO procedure in Figure \ref{fig:cm_gc}. The top two matrices show the \textit{recall} CMs, which show in each cell the percent of sources of the true class, shown on the left, that are classified as each of the classes shown on the bottom. On the other hand, the bottom two matrices show the \textit{precision} CMs, which show in each cell the percent of sources of the predicted class (left) that actually belong to each of the classes shown on the bottom. The CMs on the right side show only sources classified with probability $>60\%$. The numbers below the class labels along the y-axes show the number of sources in each class. 

In \cite{yang_classifying_2022}, the TD was also based on the CSC, but it was focused on non-GC Galactic X-ray sources and AGNs, and crossmatched to optical and infrared all-sky surveys instead of HST. Comparing the performance of the LOO evaluations, we have similar performance in stellar-type classes (AB vs. low-mass stars and YSOs), AGNs, and CVs. We observe much better performance in the qLMXB class vs. the LMXB class in \cite{yang_classifying_2022}, because the well-constrained distances to the GCs makes the higher intrinsic luminosities of qLMXBs more distinct. The GC-specific MSP class (which includes Spider MSPs) also performs similarly to the NS class \citep{yang_classifying_2022} at $\sim 60\%$ precision. This moderate performance can be attributed to the fact that while many MSPs have soft X-ray spectra, some have X-ray features more typical of other classes such as AGN, and most MSPs lack identified optical counterparts. Isolated MSPs, and many spider systems are too faint to be detected in the optical, while binary MSPs with a WD companion are more reliably detected.

\subsection{Identification of Counterparts}
\label{counterparts}

The greatest challenge in securely classifying X-ray sources in GCs is identifying their true optical counterparts, or establishing the lack thereof. PM measurements are crucial for counterpart identification, as they separate cluster members from foreground and extragalactic sources. The PM measurements in oMEGACat \cite{haberle_omegacat_2024} provide a dramatic improvement over the previously available PMs for the $\omega$ Cen field, which used ground based CCD imaging down to $B\sim20$ \citep{bellini_ground-based_2009}. A companion work \citep{nitschai_omegacat_2023} provides radial velocity measurements using spectroscopy from the Multi-Unit Spectroscopic Explorer (MUSE), with MUSE sources cross-matched to oMEGACat sources. However, these measurements are only 50\% complete up to $\sim19$ mag, limiting their utility in evaluating our classifications. The metallicity measurements from the same survey can be used to evaluate a hypothesis of \cite{cool_hstacs_2013} that several giant and subgiant candidate counterparts are from the same metal-rich giant branch, see Section \ref{Giant and Subgiant ABs}. 

Another MUSE study \citep{saracino_studying_2025} focused on investigating the properties of binaries in the center of the cluster, reaches down to 20.5 mag in F625W. Of the eight crossmatched to confident and candidate optical counterparts of CSC sources, only 1 (foreground star 2CXO J132644.0-472856-2 (see Section \ref{foreground}), has significantly variable radial velocity, although that source is nonvariable according to the photometric variability metric.

\cite{henleywillis_deep_2018} estimated the number of X-ray sources that are cluster members, AGN, and foreground stars based on the X-ray source radial distribution around the cluster vs.\ X-ray source flux (see  Figures 4-6 of that paper). They estimated that a significant fraction of their 233 X-ray sources in the $\omega$ Cen field are AGN, specifically, $\sim 60\%$ of the 169 sources with 0.5-2 keV flux above $\SI{5e-16}{\erg\per\square\cm\per\s}$. The same flux cut gives a total of 129 CSC sources, 62 of which have oMEGACat coverage. Assuming the same 60\% fraction of these sources are AGN, we estimate $\sim 77$ AGN in the whole field, and $\sim 37$ AGN in the region with oMEGACat coverage. About 70\% of the TD AGNs in CDFS and CDFN, within 5\arcmin of the field centers (to ensure HST coverage), with the same CSC flux cut, have HST counterparts brighter than 24 mag in F814W ($\sim 95\%$ completeness in oMEGACat). Therefore, we expect to observe $\sim 26$ extragalactic counterparts to 37 AGN in the region with oMEGACat coverage. This number can be treated as an upper limit, because the 60\% AGN fraction is based on the whole Chandra field, while the region of the cluster with oMEGACat coverage is closer to the cluster center, and has a larger percentage of non-AGN sources compared to the whole field. 

\cite{henleywillis_deep_2018} also estimate 30-40 X-ray sources could be associated with foreground stars, based on densities of candidate coronal sources in the XMM-Newton SSC survey of the Galactic plane \citep{Gomez-Moran_XMM-Newton_2013} at the same galactic latitude of $\omega$ Cen, but with a larger range of Galactic longitudes. Subtracting expected foreground X-ray sources and AGN, \cite{henleywillis_deep_2018} leaves $60 \pm 20$ X-ray sources to be cluster members. Of the 45 X-ray sources with confident or tentative optical counterparts identified in \cite{henleywillis_deep_2018}, 29 were reported to be cluster members. 

To identify cluster members and extragalactic sources using the oMEGACat PMs, we use PM cuts of $((\mu_\alpha \cos{\delta})^2 + (\mu_\delta)^2)^\frac{1}{2}<3$ and $((\mu_\alpha \cos{\delta}-3.3)^2 + (\mu_\delta-6.6)^2)^\frac{1}{2}<2$ mas/yr, respectively, based on Figure 7 in \citep{haberle_omegacat_2024}. Note that oMEGACat reports PM measurements after correcting for the motion of the entire cluster (3.3, 6.6 mas/yr). We assume foreground stars have PMs inconsistent with either of these cuts.

Out of $\sim 1.5$ million sources in oMEGACat, 92.6\% pass the PM cuts for cluster membership, $\approx$2.0\% are non-members, and 5.4\% have no PM measurements. Out of the non-members, only 680 (0.05\% of total) show PM consistent with background galaxies, while the rest are foreground stars. Among the 1521 potential counterparts to the 137 CSC sources within $3\sigma$ CSC PU and within $1\arcsec$ separation (11 matches per X-ray sources on average), 36 (2.4\%) are foreground stars, 72 (4.7\%) have no PM measurements, and 17 (1.1\%) have PMs consistent with background galaxies. The remaining 1396 sources (91.8\%) are cluster members. Some faint sources, bright sources, and sources close to bright stars do not have PM measurements, and have bad quality flags in oMEGACat. Upon inspecting the images, some of these sources appear to be spurious, caused by PSF artifacts from nearby stars. Other faint sources are not detected in the catalog, but are nevertheless visible in one or more images. As we discuss later in this section, in some cases, these undetected sources are likely to be the true counterpart. 

\begin{figure*}
    \centering
    \includegraphics[width=\textwidth]{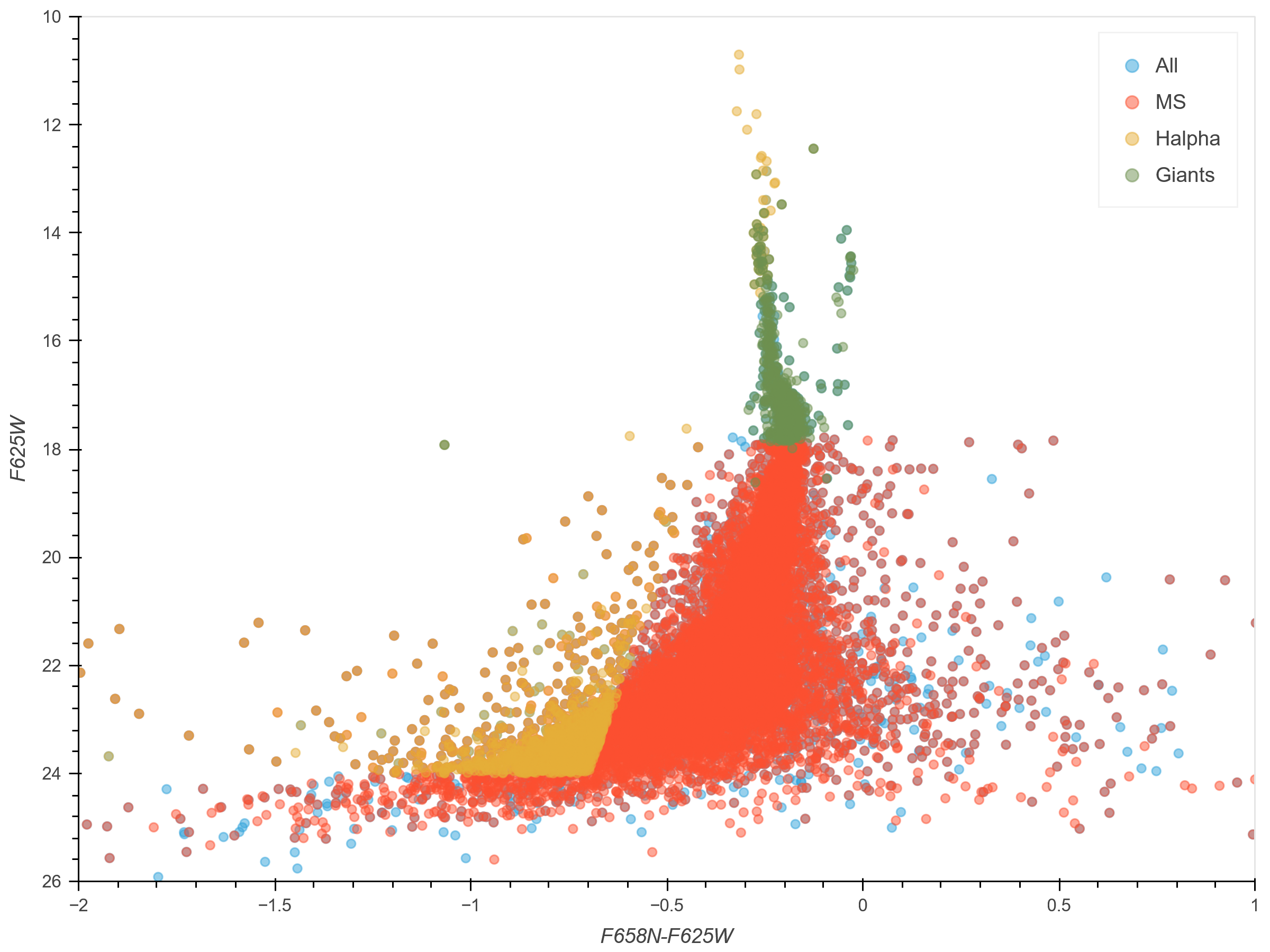}
    \caption{F658N-F625W vs. F625W CMD of 20,000 randomly sampled oMEGACat sources. Colors indicate the rough cuts made to divide these sources into cluster member MS stars, giants, and stars with H$\alpha$ excess over the MS. }
    \label{fig:omegacat_cmd}
\end{figure*}

\begin{figure*}
    \centering
    \includegraphics[width=\textwidth]{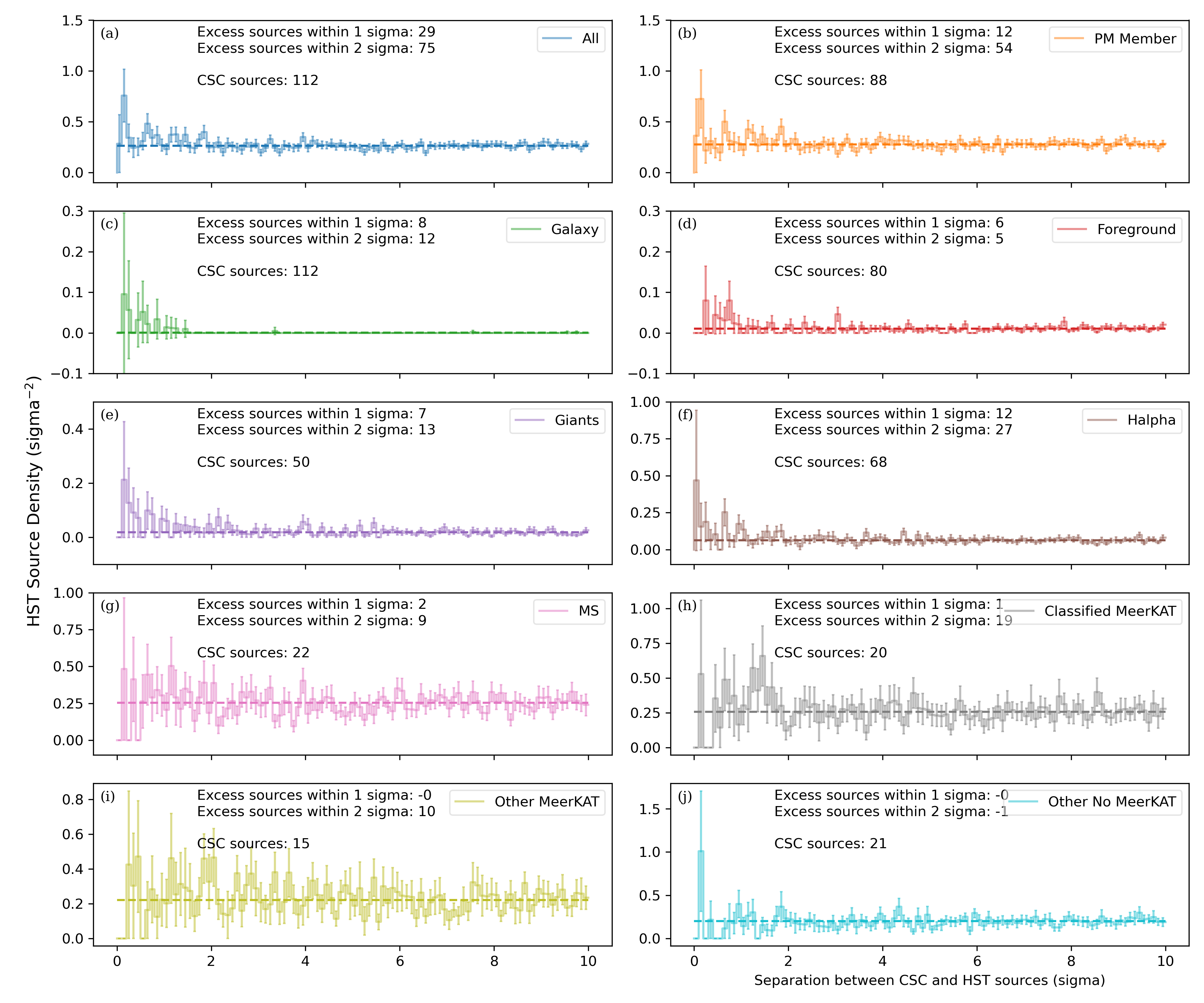}
    \caption{Density of various types of oMEGACat sources, detected in F275W, as a function of separation from CSC sources, averaged per CSC source, see Section \ref{counterparts}. The separation is normalized to the CSC source PU, with each bin representing an annulus of width equal to $0.05\sigma$ CSC PU. Dashed lines show mean source density calculated between $3\arcsec$ to $5\arcsec$ away from each CSC source. Excess number of sources above the mean density within $0.2\arcsec$ and $0.8\arcsec$ are shown.}
    \label{fig:csc_omegacat_separation_density}
\end{figure*}

Figure \ref{fig:csc_omegacat_separation_density} shows the source density of oMEGACat sources per square arcsec, as a function of separation from CSC sources, in units of CSC PU ($\sigma$), averaged per CSC source. Only CSC sources with $1\sigma$ PUs $<0.5\arcsec$ are included. Each panel from (a) to (g) shows the distribution of different types of oMEGACat sources, including all sources, PM cluster members, background galaxies, foreground non-members, member giant stars, member stars with H$\alpha$ or blue excess, and member MS stars. The cluster member stars in panels (e), (f), and (g) were divided based on simple cuts in CMD space, shown in Figure \ref{fig:omegacat_cmd}. 
For each panel, CSC sources associated with other types of optical sources in previous panels are not included, i.e.\ AGN associations are excluded from the cluster member panels. The last three panels shows the distribution of oMEGACat sources of all types around (h) CSC sources with MeerKAT counterparts and confident or candidate oMEGACat associations, (i) other CSC sources with MeerKAT counterparts, and (j) CSC sources without confident or candidate oMEGACat associations or MeerKAT counterparts. The dashed lines show the mean source density of each type between $5\sigma$ and $10\sigma$ separation from CSC sources. 

Additionally, only oMEGACat sources detected in F275W are included in the figures, except panel (d) for foreground stars. We find this restriction greatly increases the apparent excess of source densities at small separations (and presumably, the true counterpart fraction) compared to including all oMEGACat sources. This indicates that for our source classes, the mechanisms responsible for producing X-rays also produce excess UV emission compared to non X-ray sources.

\begin{figure*}
    \centering
    \includegraphics[width=0.8\textwidth]{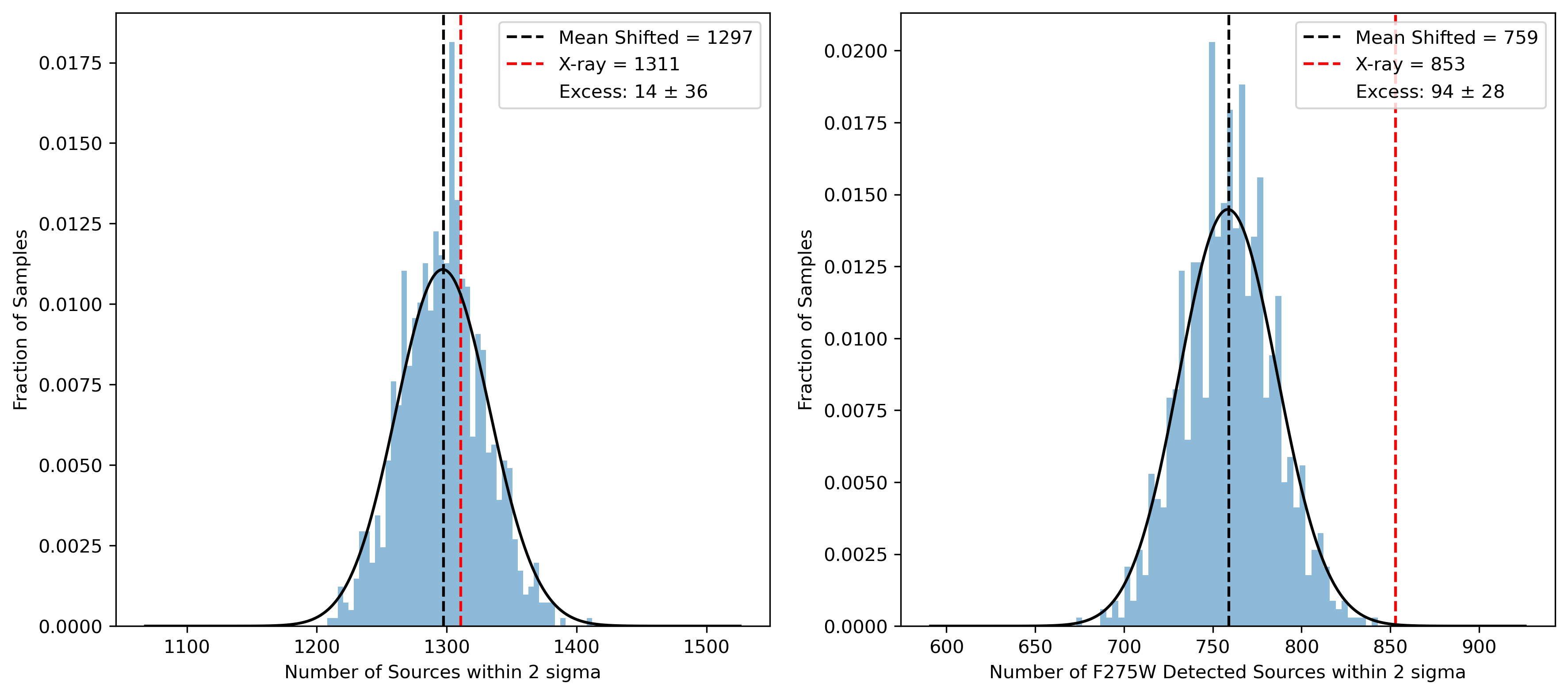}
    \caption{Distribution of the summed number of sources within $2\sigma$ CSC PU separation of random positions close to X-ray sources. Poisson distributions with $\lambda$ set to the mean of the distributions shown in black. The red dashed line shows the summed number of sources within $2\sigma$ PU of actual CSC sources. 
    The left and right panels correspond to all and F275W detections, respectively. }
    \label{fig:csc_omegacat_sigma_excess}
\end{figure*}

An alternative estimate of the number of true counterparts can be seen in Figure \ref{fig:csc_omegacat_sigma_excess}. For each of the 129 CSC sources with oMEGACat coverage, we shift it to a random nearby position (between $2\arcsec$ to $5\arcsec$) and count the number of optical sources within the radius corresponding to the $2\sigma$ PU of the respective CSC source. The number of matched optical sources is then summed over all 129 CSC sources, and the process is repeated 1,000 times. The resulting distribution of counted optical source for the shifted positions is consistent with a Poisson distribution. The mean number of sources over the 1000 trials is compared to the summed number of sources within $2\sigma$ separation of the actual CSC positions. 

The excess is again greatly enhanced if only F275W detected sources are counted, and the number is compatible with the excess for the first panel in Figure \ref{fig:csc_omegacat_separation_density}. Increasing the radius to $3\sigma$ PU does not significantly increase the excess. The square root of the mean of the distributions serves as an estimate of the error on the excess. For sources with F275W detection, the excess is significant at $3.4\sigma$ level.

We thus estimate the number of true counterparts to CSC sources detected in oMEGACat to be $\sim75$. The panels of Figure \ref{fig:csc_omegacat_separation_density} give the expected number of true counterparts of different types, which, when summed, is consistent with the total number from Figure \ref{fig:csc_omegacat_sigma_excess} (right panel). The number of CSC sources that are cluster members with optical counterparts is 33, roughly half of the 60 cluster members estimated from X-ray source distributions in \cite{henleywillis_deep_2018}. 

In the next several subsections, We discuss each of these classes of sources in detail, including previously identified associations, associations which we confidently classify, and other associations of note. When classifying previously identified associations with MUWCLASS, i.e. TD sources belonging to $\omega$ Cen, each source being classified is dropped from the TD to prevent overfitting. Since PM information, along with other information crucial to counterpart identification are not used by MUWCLASS, we also provide our manual classifications of these associations, which should be considered as the more definitive classification in this work. 

We first discuss non-cluster members (i.e. AGN, foreground stars) and cluster member classes (i.e. giant stars, CVs) where we can make confident associations. We then discuss previously known and candidate MSPs and spiders, and search for possible counterparts. The remaining CSC sources are expected to largely consist of fainter MS ABs and AGN without visible galaxy counterparts. We identify possible members of these classes, and analyze their bulk properties. For the associations we discuss, we show a table of astrometric and multiwavelength properties most relevant for classification. We also show a table of variable sources in Table \ref{tab:variable}, and their corresponding lightcurves in Figure \ref{fig:variable}. We refer to counterpart associations by the order of their separation from the CSC source (corresponding to the suffix in the association name). For very bright and faint CSC sources, we also give the X-ray photon counts as a reference. Unless otherwise stated, any discussed counterparts have PMs consistent with cluster membership. 

\subsection{PM-Identified AGNs}
\label{AGNs}

We first discuss previously and newly classified AGN identified using oMEGACat PM measurements. Panel 3 of Figure \ref{fig:csc_omegacat_separation_density} shows a clear excess of PM-identified galaxies over the expected number of chance coincidences within $2\sigma$ PU of CSC sources. Using $\sim 600$ PM-identified galaxies within 5\arcmin of the cluster center in oMEGACat, and a typical $2\sigma$ CSC PU of $0.5\arcsec$, the chance coincidence rate is $\lesssim 0.2\%$. Thus, it is probable that all 18 of the galaxies within $3\sigma$ PUs of the CSC sources are associated with them.

\textbf{CXO sources previously classified as AGN:}
Sources 22f, 31b, 41g, 54d, 54g, 62b were classified as AGN in \cite{henleywillis_deep_2018} based on ground-based PM measurements in \cite{bellini_ground-based_2009}. Of these, only sources 22f, 31b, and 54d have galaxy counterparts in oMEGACat, and all three are fainter than the $B\sim 20$ limit of \cite{bellini_ground-based_2009}, and were therefore not detected in that catalog. We thus consider the previous AGN classifications unreliable, and superseded by ours. Sources 41g and 62b do not have galaxy counterparts in oMEGACat, but do have PyBDSF MeerKAT counterparts. We consider them candidate AGN, see Section \ref{meerkat_counterparts}.

\textbf{Other CSC sources matched to galaxies:}
Besides sources 22f, 31b, and 54d, 15 other sources are matched to galaxies within 3$\sigma$ separation. In Table \ref{table_pm_agn}, we show some properties of these associations, and use them to guide our analysis. For most of these associations, the separation is significantly less than the $1\sigma$ PUs. Eight galaxies have MeerKAT counterparts within $2\arcsec$, with MeerKAT flux in $\mu$Jy shown in Table \ref{table_pm_agn}. Given the lack of other associations with distinguishing features, we consider it probable that all of these galaxies are true counterparts. 

\begin{figure*}
    \centering
    \includegraphics[width=0.45\textwidth]{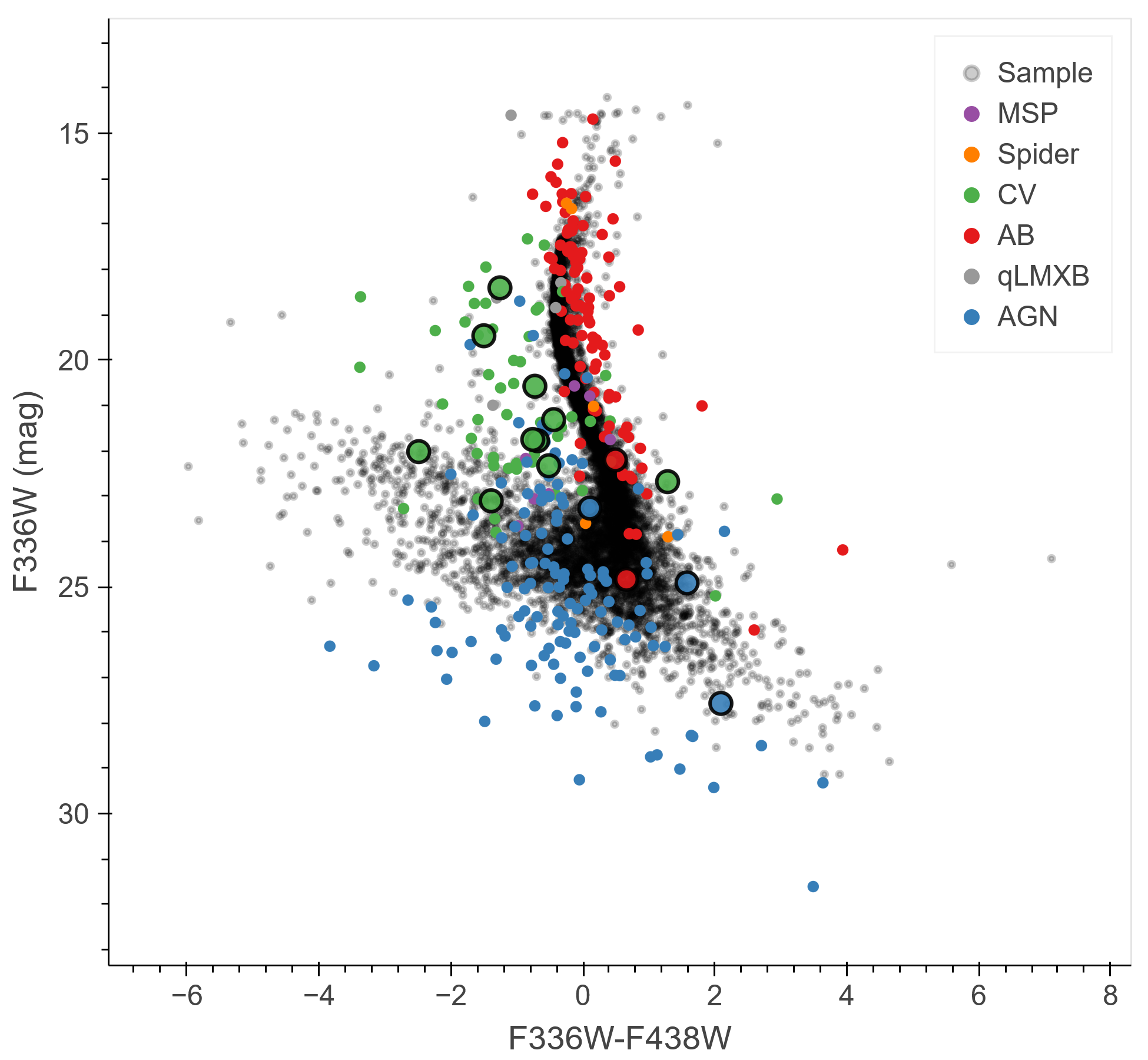}
    \includegraphics[width=0.45\textwidth]{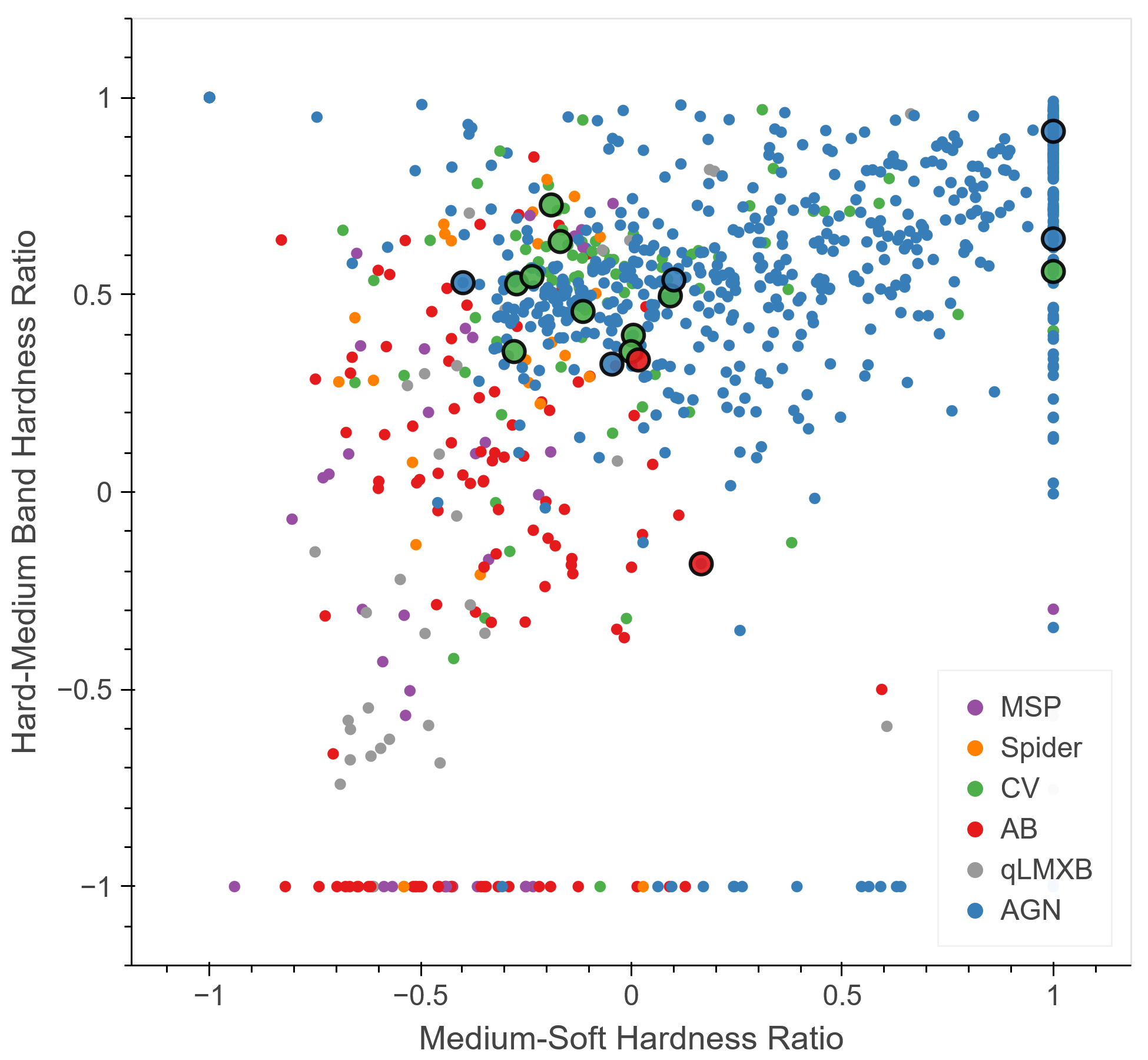}
    \includegraphics[width=0.45\textwidth]{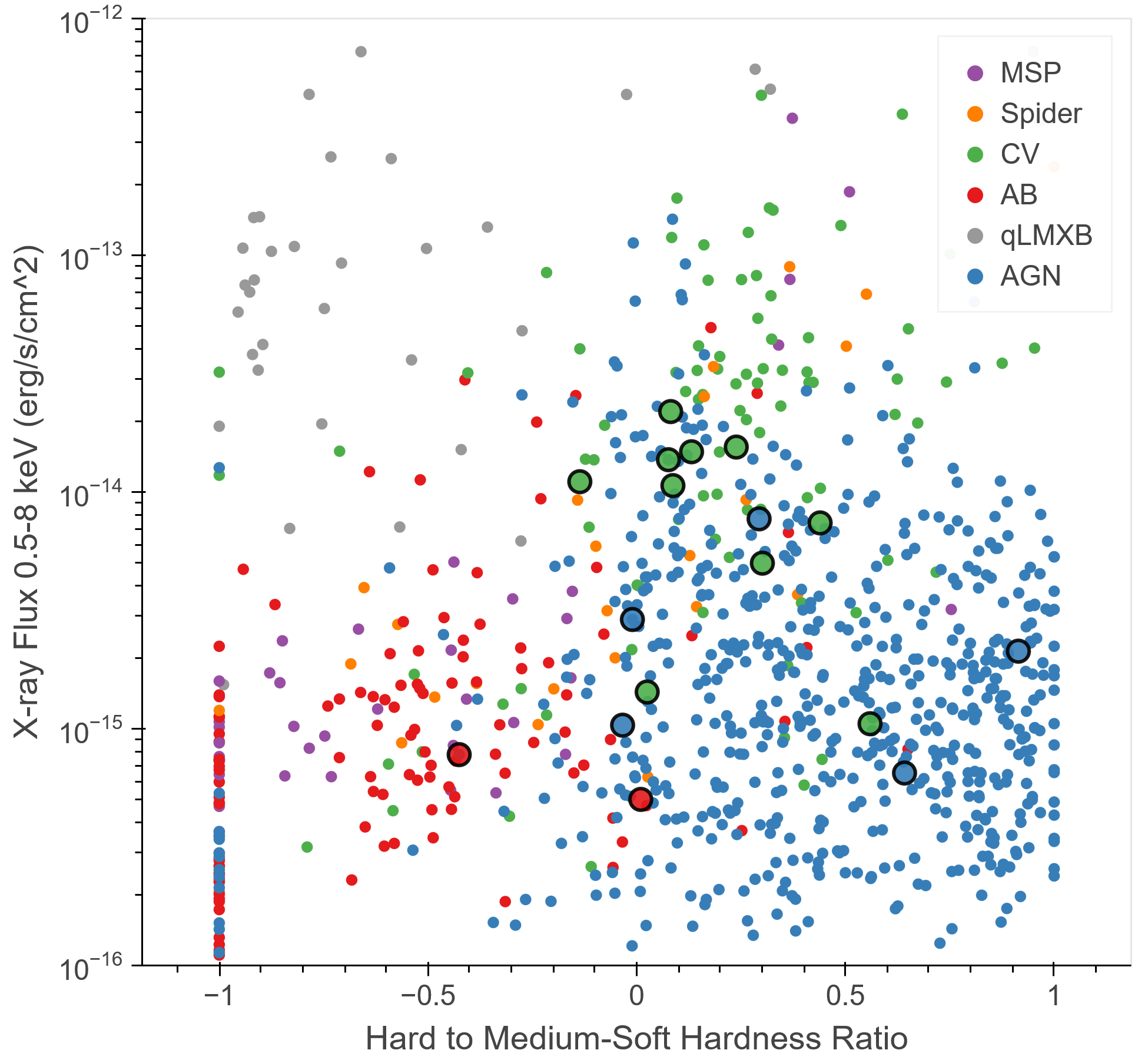}
    \caption{X-ray and optical properties of PM-selected AGN (sources outlined in black), overlayed on top of TD sources. Color corresponds to MUWCLASS classification, but as discussed in the text, most, if not all, of these sources are likely AGN. Black points in the background of the CMD are 10,000 randomly sampled stars in oMEGACat.}
    \label{fig:pm_agns}
\end{figure*}

Ten of these galaxy associations were classified by MUWCLASS as CVs, four of which were also suggested to be CV candidates in past works. They are bright and blue in the optical in at least one color relative to the MS and TD AGNs, such that their optical and X-ray features are similar to a large fraction of TD CVs, but only a small fraction of TD AGN, see Figure \ref{fig:pm_agns}. Since the local densities of sources of each class in feature space (after class balancing) are the quantities important for random forest classification, these bright background galaxy associations are favored as CVs by MUWCLASS.
Since PM measurements and radio detections are not used as features, MUWCLASS cannot reliably distinguish between AGNs and CVs in these cases. Only the spectra of very bright CVs ($>200$ counts) have visibly distinct features from those of AGN. These sources highlight the importance of PM measurements and radio observations in separating cluster members from background sources.

One additional source not named in previous works, source 2CXO J132627.9-473215, is coincident with an undetected, visibly extended source between the 2nd and 5th detected counterparts. It is hard, showing no counts in the soft and medium bands, and matches to a MeerKAT counterpart. This source is thus probably an AGN. 

We show the properties of these 17 (detected) associations, along with their MUWCLASS classifications (in color), in Fig. \ref{fig:pm_agns} together with our TD sources. There is a clear correlation between the optical and X-ray luminosities of these AGN, consistent with TD AGN and CVs. However, they are brighter in the optical than the average TD AGN, possibly due to the different depths and crowding levels of $\omega$ Cen vs. the Chandra Deep Fields.

\begin{table}
\footnotesize
\centering
\begin{tabular}{lllllllllllll}
\toprule
Assoc. Name & Lit. & Manual & Pub. & ML & P\_C & PU & Sep. & $F_{\rm CSC, b}$ & $HR_{\rm hms}$ & F814W & V-I & F\_radio \\
\midrule
 &  &  &  &  &  & (arcsec) & (arcsec) & (erg/s/cm$^2$) &  & (mag) & (mag) & ($\mu$Jy) \\
J132627.9-473215-0 &  & AGN &  & AGN & 0.87 & 0.73 &  & 2.48e-15 & 1.00 &  &  & 87.7 \\
J132640.7-472700-4 & 21g & AGN &  & AGN & 0.34 & 0.23 & 0.35 & 6.50e-16 & 0.64 & 22.07 & -0.06 & 29.7 \\
J132652.6-472713-1 & 22c & AGN & CV? & CV & 0.83 & 0.15 & 0.12 & 1.06e-14 & 0.09 & 22.30 & 0.20 &  \\
J132658.8-472820-1 & 22f & AGN & AGN & AGN & 0.59 & 0.17 & 0.07 & 2.89e-15 & -0.01 & 20.45 & 1.11 &  \\
J132642.5-473022-1 & 24i & AGN &  & AGN & 0.78 & 0.20 & 0.05 & 2.13e-15 & 0.91 & 22.19 & 1.46 &  \\
J132629.3-472813-2 & 31a & AGN & CV? & AGN & 0.74 & 0.17 & 0.15 & 7.70e-15 & 0.29 & 22.23 & 1.15 &  \\
J132631.4-472801-4 & 31b & AGN & AGN & AB & 0.80 & 0.28 & 0.34 & 7.77e-16 & -0.43 & 18.94 & 1.13 & 24.1 \\
J132700.9-473004-1 & 33e & AGN & CV? & CV & 0.33 & 0.20 & 0.08 & 1.43e-15 & 0.03 & 21.52 & 0.25 &  \\
J132659.2-473037-2 & 33f & AGN &  & AGN & 0.39 & 0.27 & 0.19 & 1.04e-15 & -0.03 & 23.05 & 1.31 &  \\
J132655.0-473113-1 & 33h & AGN &  & CV & 0.56 & 0.17 & 0.03 & 1.55e-14 & 0.24 & 18.52 & 0.73 &  \\
J132648.7-473124-1 & 33l & AGN & other & CV & 0.76 & 0.16 & 0.02 & 2.19e-14 & 0.08 & 21.01 & 0.46 &  \\
J132636.4-472528-1 & 41l &  &  & AB & 0.80 & 0.47 & 0.13 & 5.03e-16 & 0.01 & 18.77 & 1.47 &  \\
J132636.4-472528-7 & 41l & AGN &  & AB & 0.38 & 0.47 & 0.52 & 5.03e-16 & 0.01 & 20.96 & 1.13 &  \\
J132649.5-473212-1 & 43h & AGN & CV? & CV & 0.68 & 0.17 & 0.18 & 1.48e-14 & 0.13 & 21.90 & 0.62 & 19.3 \\
J132644.1-473231-1 & 44a & AGN &  & CV & 0.68 & 0.19 & 0.11 & 1.11e-14 & -0.14 & 20.34 & 0.28 & 31.2 \\
J132623.6-473044-6 & 44c & AGN & CV? & CV & 0.44 & 0.79 & 0.49 & 1.05e-15 & 0.56 & 23.10 & 0.68 &  \\
J132631.3-472438-2 & 51a & AGN &  & CV & 0.63 & 0.26 & 0.24 & 5.01e-15 & 0.30 & 20.15 & 1.38 &  \\
J132706.4-472537-1 & 52c & AGN & other & CV & 0.53 & 0.20 & 0.12 & 7.40e-15 & 0.44 & 20.68 & -0.24 &  \\
J132625.1-473227-1 & 54d & AGN & AGN & CV & 0.60 & 0.20 & 0.19 & 1.37e-14 & 0.08 &  &  & 73.3 \\
\bottomrule
\end{tabular}
\caption{Classification and multiwavelength properties of previously-known and newly classified AGN. Columns: 'Assoc. Name': Name of association with '2CXO' prefix removed, 'Lit.': Naming scheme used in \cite{henleywillis_deep_2018, cool_hstacs_2013}, 'Manual': Manual classification using all information, including those not used by MUWCLASS, 'Published': 'Previously reported classification in \cite{henleywillis_deep_2018, cool_hstacs_2013}, 'MUWCLASS': MUWCLASS classification, 'P\_C': MUWCLASS classification probability, 'PU (\arcsec)': $1\sigma$ CSC semimajor axis PU, 'Sep. (\arcsec)': Separation of CSC source to oMEGACat source, after our astrometric shifts, in arcsec, 'Member': Cluster membership as determined by proper motion cuts, '$F_{\rm CSC, b}$': CSC broadband flux, 0.2-7 keV, '$HR_{\rm hms}$': CSC hardness ratio of hard vs medium plus soft bands. 'F814W': oMEGACat F814W magnitudes, 'V-I': oMEGACat F606W-F814W colors, 'F\_radio': \textsc{PyBDSF} MeerKAT source flux. }
\label{table_pm_agn}

\end{table}

\subsection{Foreground Stars}
\label{foreground}

Out of the 87 CSC sources used for panel 4 of Figure \ref{fig:csc_omegacat_separation_density}, 16 are matched to foreground stars within $2\sigma$ PU, of which $\sim 3$ are in excess of the expected number of chance coincidences.
These 16 associations include two previously classified X-ray sources. Source 22d was previously associated with its first counterpart, a non-member blue straggler \citep{henleywillis_deep_2018}, and classified by MUWCLASS as an AB (The TD does not contain a blue straggler class). Source 31b was discussed above as being more likely matched to a galaxy. 

As coronally emitting foreground stars\footnote{not included in the TD} are soft X-ray emitters similar to ABs, they have similar hardness ratios, and low X-ray to optical flux ratios, and therefore are classified as ABs by MUWCLASS. We find 10 associations that fulfill these criteria, and have separations substantially smaller than their X-ray PU. These associations are shown in Table \ref{table_foreground}, and include sources 22d, 22a, 43f, which were assigned the foreground star counterparts in previous publications. On the other hand, not included are sources 31b, 33e, and 41l, which have foreground star associations, but are much more likely to be actually matched to their galaxy counterparts, as their X-ray spectra are too hard to be consistent with coronal sources. 

Source 43j has two foreground star counterparts. The first foreground counterpart has a large PM of $(-9.2, 5.3) $ mas relative to the cluster, and is bluer than a MS star at any foreground distance, consistent with a CV nature. Its X-ray spectrum is soft, consistent with MSPs and a small fraction of CVs in the TD, and therefore its null and UL associations were classified as MSP. Due to the lack of a MeerKAT counterpart, we suggest this source to be a foreground CV. Alternatively, the 4th association, a cluster member star near the MS turnoff, may be the true counterpart, which would make this source an AB. However, the larger separation of the third counterpart compared to the blue counterpart, and the rarity of such blue foreground stars favors the foreground CV interpretation.

Some other CSC sources are also coincident with foreground stars, but these stars are either too bright to be detected in oMEGACat (e.g., source 54f), or they are outside of its area of coverage. Using Gaia DR3 \citep{gaia_collaboration_gaia_2023}, we found 8 counterparts with PMs consistent with foreground stars, 5 of which were previously identified in \citep{haggard_deep_2013, henleywillis_deep_2018} using ground-based observations from \cite{kaluzny_cluster_2004}. We present their properties in Table \ref{table_gaia_foreground}.

\begin{table}
\footnotesize
\centering
\begin{tabular}{llllllllllll}
\toprule
Assoc. Name & Lit. & Manual & Pub. & ML & P\_C & PU & Sep. & $F_{\rm CSC, b}$ & $HR_{\rm hms}$ & F814W & V-I \\
\midrule
 &  &  &  &  &  & (arcsec) & (arcsec) & (erg/s/cm$^2$) &  & (mag) & (mag) \\
J132644.0-472856-2 &  & FG &  & AB & 0.98 & 0.17 & 0.16 & 1.85e-15 & -0.78 & 14.27 & 0.45 \\
J132648.2-472640-3 & 22a & FG & other & AB & 0.77 & 0.39 & 0.36 & 7.03e-16 & -0.56 & 17.62 & 2.01 \\
J132658.7-472728-1 & 22d & FG & AB & AB & 0.98 & 0.23 & 0.17 & 7.34e-16 & -1.00 & 15.97 & 0.76 \\
J132656.4-472815-2 & 22j & FG &  & AB & 0.86 & 0.24 & 0.20 & 4.95e-16 & -0.06 & 18.44 & 1.35 \\
J132652.3-473152-4 & 33q & FG &  & AB & 0.48 & 0.68 & 0.64 & 1.79e-16 & -1.00 & 20.39 & 0.82 \\
J132630.5-472600-1 & 41e & FG &  & AB & 0.94 & 0.41 & 0.28 & 1.76e-15 & -1.00 & 11.56 & 0.33 \\
J132649.1-472451-1 & 42d & FG &  & AB & 0.94 & 0.42 & 0.29 & 7.22e-16 & -1.00 & 11.92 & 0.31 \\
J132656.0-473202-2 & 43f & FG & other & AB & 0.96 & 0.36 & 0.31 & 6.56e-16 & -1.00 & 16.89 & 0.79 \\
J132705.2-473015-2 & 43j & FG CV? &  & CV & 0.64 & 0.40 & 0.34 & 7.10e-16 & -1.00 & 22.51 & -0.09 \\
J132705.2-473015-3 & 43j &  &  & MSP & 0.44 & 0.40 & 0.49 & 7.10e-16 & -1.00 & 22.90 & 0.92 \\
J132705.2-473015-4 & 43j &  &  & AB & 0.95 & 0.40 & 0.46 & 7.10e-16 & -1.00 & 16.96 & 0.47 \\
J132714.4-472830-1 & 52f & FG &  & AB & 0.93 & 0.36 & 0.08 & 1.50e-15 & -0.61 & 16.27 & 1.00 \\
J132617.3-472952-0 & 54n & FG &  & AB & 0.46 & 0.90 &  & 6.44e-16 & -1.00 &  &  \\
J132628.9-473437-4 & 74a & FG &  & AB & 0.93 & 1.01 & 0.81 & 1.73e-15 & -0.47 &  &  \\
\bottomrule
\end{tabular}
\caption{Classification and MW properties of foreground star associations identified from oMEGACat PMs. Columns same as in Table \ref{table_pm_agn}. }
\label{table_foreground}
\end{table}

\begin{table}
\footnotesize
\centering
\begin{tabular}{llllllllllll}
\toprule
Assoc. Name & Lit. & Manual & Pub. & ML & P\_C & PU & Sep. & $F_{\rm CSC, b}$ & $HR_{\rm hms}$ & Gmag & BP-RP \\
\midrule
 &  &  &  &  &  & (arcsec) & (arcsec) & (erg/s/cm$^2$) &  & (mag) & (mag) \\
J132644.6-472332-0 & 51e &  & Foreground & MSP & 0.35 & 0.73 & 0.39 & 7.92e-16 & -1.00 & 16.03 & 1.09 \\
J132716.2-472457-0 & 62d &  &  & MSP & 0.43 & 0.69 & 0.71 & 1.06e-15 & -1.00 & 11.08 & 0.56 \\
J132720.0-472549-0 & 62h &  &  & AB & 0.43 & 1.09 & 0.63 & 2.13e-16 & -1.00 & 17.49 & 0.92 \\
J132657.1-473351-0 & 63f &  &  & AB & 0.39 & 0.76 & 0.73 & 1.16e-15 & -0.20 & 13.55 & 0.86 \\
J132721.6-472921-0 & 63h &  & Foreground & MSP & 0.41 & 0.78 & 0.12 & 1.00e-15 & -1.00 & 14.29 & 1.04 \\
J132647.3-473559-0 & 73d &  & Foreground & AGN & 0.49 & 1.24 & 0.66 & 3.28e-15 & 0.08 & 16.64 & 1.10 \\
J132710.0-472127-0 & 82b &  & Foreground & qLMXB & 0.43 & 0.46 & 0.18 & 1.20e-14 & -1.00 & 15.60 & 1.29 \\
J132611.5-473402-0 & 84d &  & Foreground & qLMXB & 0.62 & 0.31 & 0.48 & 1.77e-14 & -0.62 & 12.95 & 1.50 \\
\bottomrule
\end{tabular}
\caption{Classification and MW properties of foreground star associations identified from Gaia data. `Sep': separation between CSC and Gaia counterparts, `Gmag': Gaia G-band magnitude, `BP-RP': Gaia BP-RP color. }
\label{table_gaia_foreground}
\end{table}

\subsection{Cluster Members}

\cite{cool_hstacs_2013, henleywillis_deep_2018} identified 33 optical counterparts to Chandra sources that are likely cluster members. These include 3 candidate BY Draconis-type ABs (24e, 43e, 14a), 8 giant stars (13b, 22e, 24f, 32f, 34b, 41h, 42c, 43c) that may belong to the most metal-rich red-giant/sub-giant branch (RGB/SGB-a), 4 optically variable stars (11b, 73a, 74d, 91a, \cite{kaluzny_cluster_2004}), 6 CVs, and 12 candidate CVs. The membership status of these counterparts was determined using limited ground-based PM information and their location on the CMD. With oMEGACat PM and metallicity measurements, and MUWCLASS classifications, we re-examine these, as well as new, associations and classifications.

\subsection{Giant and Subgiant ABs}
\label{Giant and Subgiant ABs}

The 3 BY Draconis-type ABs (main-sequence ABs) and 2 of the giant stars (41h, 42c) are faint X-ray sources that are listed in \cite{cool_hstacs_2013} but not in the \cite{henleywillis_deep_2018} catalog. They are also not detected in CSC2.1 and are therefore excluded from our analysis. The sources from \cite{kaluzny_cluster_2004} associated with the variable stars by \cite{henleywillis_deep_2018} were not in the oMEGACat footprint, except for source 11b, which we discuss below.

All known ABs in $\omega$ Cen are therefore giant or subgiant ABs. These systems can produce X-rays either via accretion (NS or BH X-ray binaries and symbiotic systems with WDs) or due to enhanced magnetic activity (e.g., RS Canum Venaticorum/RS CVn binaries). Six of these ABs identified in \cite{cool_hstacs_2013} lie redward of, or below the subgiant and giant branch of the dominant population of stars in Omega Cen, and close to an anomalous metal-rich red giant branch (RGB) and sub-giant branch (SGB), and were therefore referred to as RGB/SGB-a stars. The authors noted that their membership in that branch is uncertain, as no metallicity measurements were available for these stars at that time. We investigate these associations, and whether or not they belong to the metal rich RGB/SGB using the MUSE measurements.

The most secure of these associations are those of sources 32f and 22e, which have separations within $0.10\arcsec$. These two X-ray sources are moderately bright with luminosities at $\qty{1.0e31}{\erg\per\second}$ and $\qty{3.6e30}{\erg\per\second}$ (58 and 25 counts), respectively. Their high AB classification probability $P_{AB}>90\%$ is supported by their X-ray/optical flux ratio, although their hardness ratios are larger than for most other ABs in the TD. Their metallicities, $[M/H] = -2.49 \pm 0.05$ and $-2.4 \pm 0.03$, respectively, make them more metal-poor than the cluster average of $-1.6$ \citep{nitschai_omegacat_2023}, contradicting the metal-rich RGB/SGB-a hypothesis in \cite{cool_hstacs_2013}.

The eighth association of source 13b suggested in \cite{cool_hstacs_2013} lies on another SGB than the other 7 RGB/SGB-a stars, but was included in this group for convenience. However, it lies significantly further away than the second association with a brighter giant star lying on the most populous RGB. Its optical/X-ray flux ratio is compatible with coronal sources, and it has a below average metallicity of $[M/H]=-1.83 \pm 0.01$. Given it is significantly closer to the CSC source than the eighth association, we consider this to be the more likely counterpart.

Source 34b is relatively bright (\qty{1.3e31}{\erg\per\second}, 71 counts), and has substantially harder $\text{HR}_{\text{ms}}$ than other ABs. The giant association has an X-ray/optical flux ratio compatible with ABs, is metal rich ($[M/H]=-0.87 \pm 0.04$, but is located just outside of the $2\sigma$ PU of $0.37\arcsec$. \cite{henleywillis_deep_2018} noted that internal absorption, in addition to the cluster absorption value, was required to obtain a good spectral fit of the X-ray source. This hints at the possibility of the X-ray source being powered by mass transfer from the giant star.
The 4th counterpart of source 43c suggested in \cite{cool_hstacs_2013} is metal poor at $[M/H]=-2.2 \pm 0.05$, however, the 5th counterpart is also a faint foreground star. The X-ray source has a luminosity of \qty{2.5e30}{\erg\per\second} (24 counts) and is soft, consistent with ABs or MSPs. 
Without additional information, we cannot confirm either giant star as the true counterparts for sources 34b and 43c.

Source 24f was identified as the counterpart to a black widow spider MSP by \cite{zhao_chandra_2023}. The giant association identified in \cite{cool_hstacs_2013} is $0.5\arcsec$ away, outside the 95\% PU of $0.42\arcsec$, and is therefore unlikely to be real. We discuss possible optical counterparts to the spider in Section \ref{spiders} below. 

Source 11b was associated with the Kaluzny variable cluster member star NV371 \citep{kaluzny_cluster_2004}, which has a measured period of 2.2 days. Its location on the CMD appears above the RGB or SGB that other RGB/SGB-a giants belong to. The orbital separation implied by the period (if it is the orbital period) and stellar mass, on the order of a few solar radii, supports an AB nature.

In summary, of the 8 candidate giant star counterparts in \cite{cool_hstacs_2013}, only 2 are probable true counterparts, 1 is more likely matched to another closer giant star, 2 are marginal candidates given their large separations, 1 is confirmed to not be a counterpart, and 2 were too faint to be detected in the CSC. Furthermore, there is little evidence that these giant stars are more metal-rich than the cluster average, and therefore likely do not belong to the metal-rich RGB/SGB branch as suggested by \cite{cool_hstacs_2013}. 

The candidate giant counterparts of sources 23h, 31c suggested in \cite{henleywillis_deep_2018} lie on a different, brighter RGB than RGB/SGB-a, but the authors did not give them classifications. Source 31c has two very close giant counterparts with similar brightness, such that they both seem equally likely to be the true counterpart.

Other sources with likely giant counterparts within $2\sigma$ PU include sources 23a, 24j, 33c, and 54n, and possible giant counterparts between $2\sigma$ and $3\sigma$ PU include sources 11d, 12c, 14d, 21e, 24b, 41k, 53a, 2CXO J132659.7-472537, and 2CXO J132646.3-473038-9. 

Interestingly, source 23a is matched to a MeerKAT counterpart with flux $18 \mu$Jy. This source is then possibly a MSP accreting from the giant star, i.e. an LMXB or Huntsman-type spider system\footnote{The first Huntsman system PSR J1417–4402 was detected at $40 \mu$Jy \citep{camilo_discovery_2016} at a comparable distance}. If so, its orbital periodicity should be observable in the optical. We list this source in Table \ref{table_msp}.

The very bright counterparts to sources 53a and 2CXO J132659.7-472537 appear to be cluster member blue stragglers, and are classified by MUWCLASS as ABs. Like other blue stragglers, they have a large blue excess in all colors, except F336W-F435W. If these two associations are correct, they would be among the most luminous and the lowest X-ray to optical flux ratio ABs in the TD. The only other cluster member blue stragglers associated with Chandra sources in the TD (grouped under ABs) are in NGC 6121, identified by \cite{lugger_exotica_2023}. 

As shown in Figure \ref{fig:csc_omegacat_separation_density}, 7 (13) giant stars within $1\sigma$ ($2\sigma$) CSC PU are in excess of the chance coincidence rate. If this excess represents true associations, then these can be accounted for by the RGB/SGB-a counterparts to 32f, 22e, the closer giant counterpart to 13b, the Kaluzny variable 11b, and the proposed counterparts to sources 23h and 31c.

\begin{figure*}
    \centering
    \includegraphics[width=0.45\textwidth]{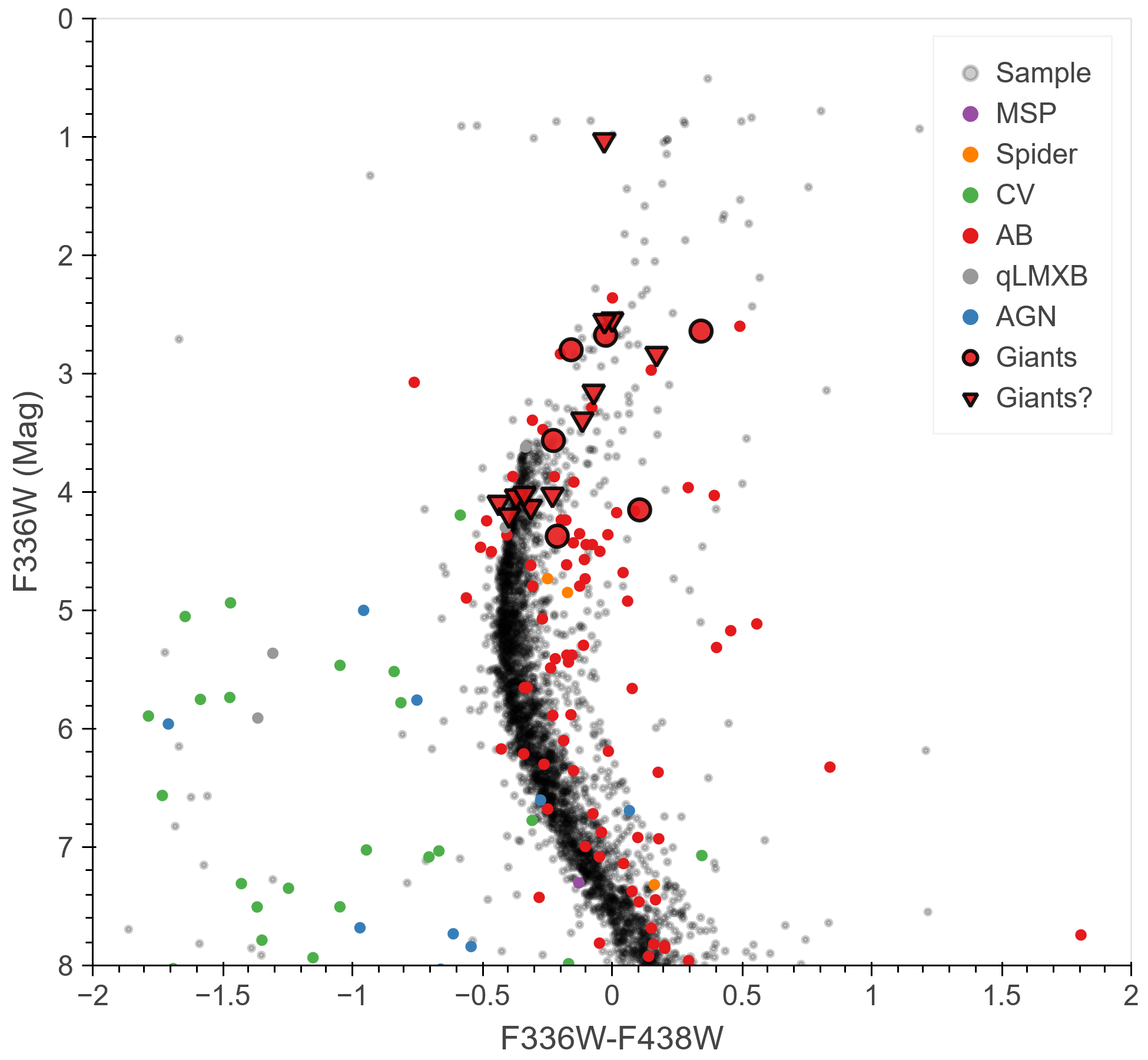}
    \includegraphics[width=0.45\textwidth]{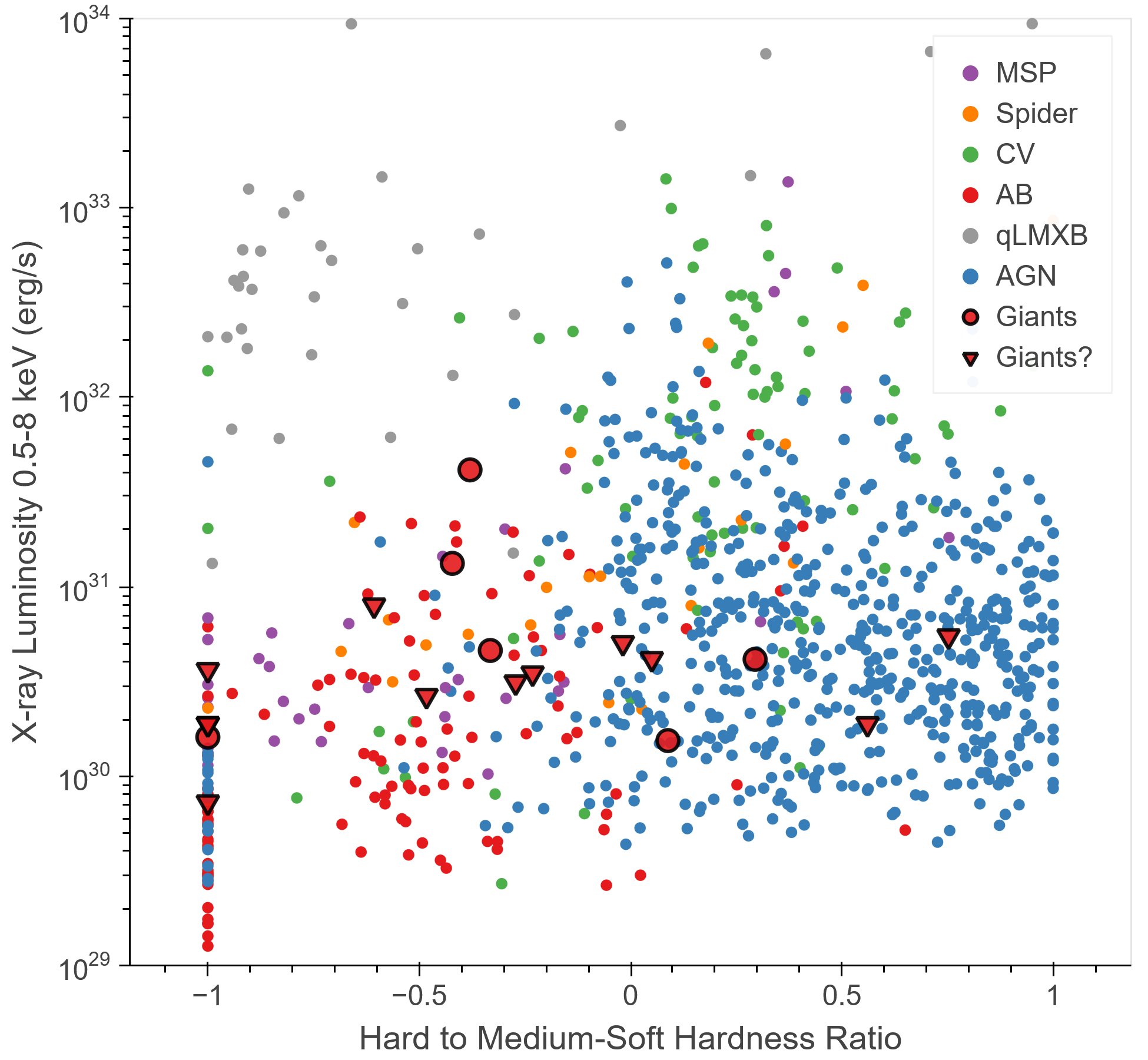}
    \includegraphics[width=0.45\textwidth]{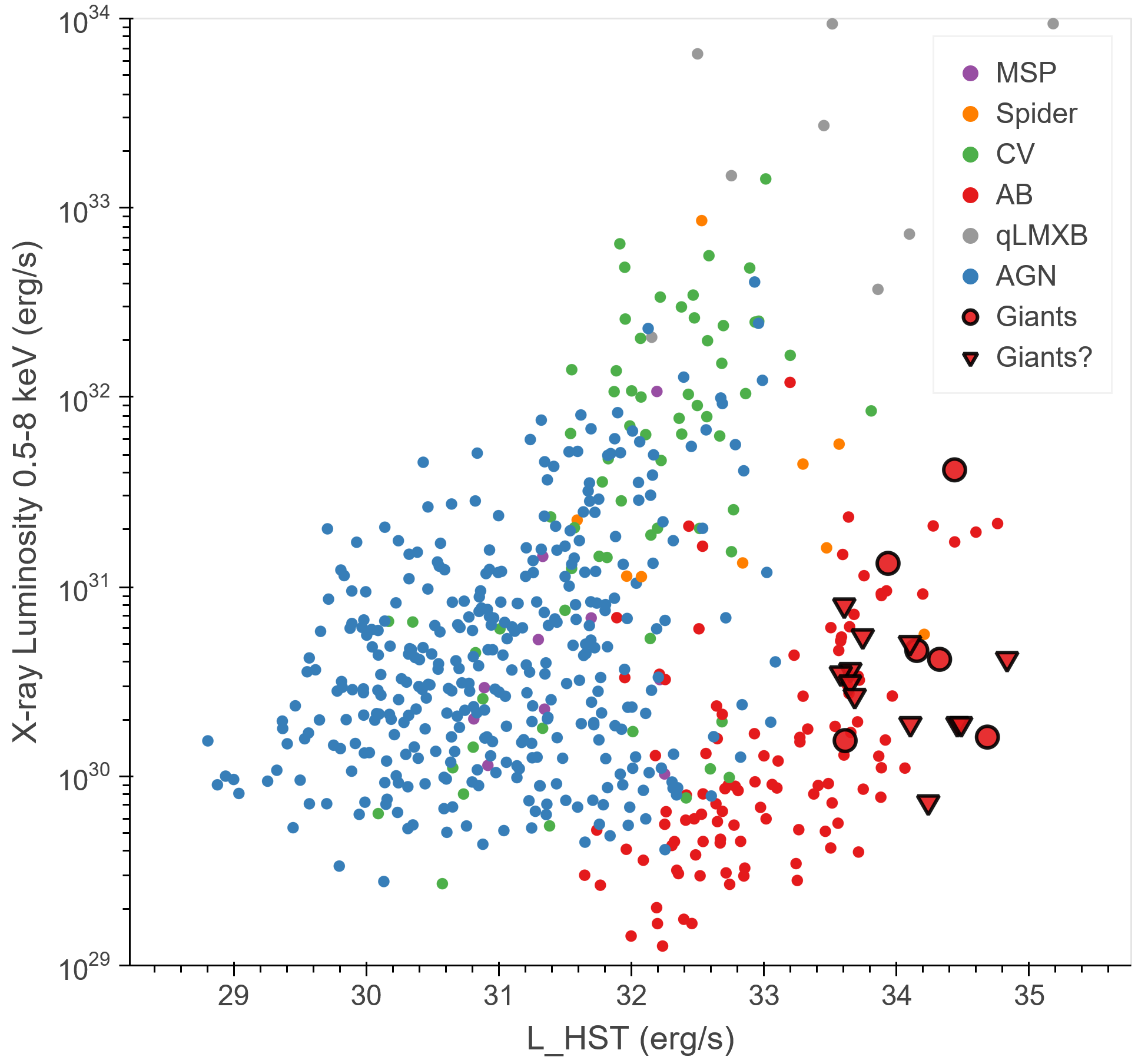}
    \caption{X-ray and optical properties of confirmed and candidate giant star ABs. Black points in the background of the CMD are 10,000 randomly sampled stars in oMEGACat. The "luminosity" and "absolute magnitudes" of TD AGN are calculated as if they were at the distance of $\omega$ Cen, to enable comparison of their properties with other classes.}
    \label{fig:NGC5139_giants}
\end{figure*}

We show the X-ray/optical properties of the confident and candidate giant star associations in Figure \ref{fig:NGC5139_giants}. 

\begin{table}
\footnotesize
\centering
\begin{tabular}{lllllllllllll}
\toprule
Assoc. Name & Lit. & Manual & Pub. & ML & P\_C & PU (\arcsec) & Sep. (\arcsec) & $L_{\rm CSC, b}$ & $HR_{\rm hms}$ & F814W & V-I & MH \\
\midrule
 &  &  &  &  &  & (arcsec) & (arcsec) & (erg/s) &  & (mag) & (mag) &  \\
J132646.3-473038-9 &  & Giant AB? &  & AB & 0.94 & 0.32 & 0.72 & 1.88e+30 & 0.56 & 14.69 & 0.68 & -1.73 \\
J132659.7-472537-1 &  & Giant AB? &  & AB & 0.83 & 0.56 & 0.11 & 7.28e+29 & -1.00 & 16.08 & 0.09 & -1.92 \\
J132641.0-472737-1 & 11b & Variable AB & Variable AB & AB & 0.77 & 0.15 & 0.07 & 4.14e+31 & -0.38 & 14.78 & 0.74 &  \\
J132640.7-472832-18 & 11d & Giant AB? &  & AB & 0.94 & 0.26 & 0.67 & 2.66e+30 & -0.48 & 16.98 & 0.48 & -1.72 \\
J132647.9-472810-6 & 12c & Giant AB? &  & AB & 0.92 & 0.19 & 0.29 & 3.13e+30 & -0.27 & 17.07 & 0.48 & -1.87 \\
J132650.5-472918-2 & 13b & Giant AB &  & AB & 0.93 & 0.29 & 0.19 & 4.14e+30 & 0.29 & 15.14 & 0.65 & -1.83 \\
J132650.5-472918-8 & 13b &  & RGB/SGB-a & AB & 0.89 & 0.29 & 0.47 & 4.14e+30 & 0.29 & 16.75 & 0.52 & -1.64 \\
J132638.0-472910-12 & 14d & Giant AB? &  & AB & 0.82 & 0.23 & 0.54 & 5.48e+30 & 0.75 & 16.75 & 0.52 & -1.49 \\
J132645.2-472652-3 & 21e & Giant AB? &  & AB & 0.90 & 0.17 & 0.39 & 7.97e+30 & -0.61 & 17.24 & 0.45 & -2.17 \\
J132659.9-472809-1 & 22e & Giant AB & RGB/SGB-a & AB & 0.96 & 0.19 & 0.05 & 4.61e+30 & -0.33 & 15.36 & 0.84 & -2.40 \\
J132653.9-473036-1 & 23h & Giant AB & None & AB & 0.97 & 0.29 & 0.16 & 1.61e+30 & -1.00 & 14.07 & 0.73 & -1.01 \\
J132639.2-473036-5 & 24b & Giant AB? &  & AB & 0.91 & 0.28 & 0.63 & 3.49e+30 & -0.23 & 17.30 & 0.45 & -2.23 \\
J132642.3-473019-2 & 24j & Giant AB? &  & AB & 0.94 & 0.25 & 0.19 & 3.64e+30 & -1.00 & 17.11 & 0.45 & -1.98 \\
J132632.3-472707-1 & 31c & Giant AB? &  & AB & 0.98 & 0.23 & 0.14 & 1.88e+30 & -1.00 & 14.79 & 0.58 & -1.76 \\
J132632.3-472707-2 & 31c & Giant AB? & None & AB & 0.99 & 0.23 & 0.22 & 1.88e+30 & -1.00 & 15.71 & 0.60 &  \\
J132705.3-472808-1 & 32f & Giant AB & RGB/SGB-a & AB & 0.98 & 0.17 & 0.06 & 1.33e+31 & -0.42 & 16.03 & 0.67 & -2.49 \\
J132703.5-472857-1 & 33c & Giant AB &  & AB & 0.91 & 0.31 & 0.15 & 1.54e+30 & 0.09 & 17.08 & 0.51 & -1.19 \\
J132637.4-473052-3 & 34b &  & RGB/SGB-a & AB & 0.70 & 0.18 & 0.40 & 1.57e+31 & 0.18 & 16.81 & 0.60 & -0.87 \\
J132630.9-472521-4 & 41k & Giant AB? &  & AB & 0.89 & 0.35 & 0.53 & 5.05e+30 & -0.02 & 15.99 & 0.40 & -1.88 \\
J132706.9-473008-4 & 43c &  & RGB/SGB-a & AB & 0.98 & 0.30 & 0.37 & 3.30e+30 & -0.35 & 16.28 & 0.67 & -2.23 \\
J132706.9-473008-5 & 43c &  &  & MSP & 0.46 & 0.30 & 0.44 & 3.30e+30 & -0.35 & 22.59 &  &  \\
J132659.8-473236-3 & 53a & Giant AB? & None & AB & 0.79 & 0.57 & 0.66 & 4.16e+30 & 0.05 & 14.74 & -0.05 &  \\
\bottomrule
\end{tabular}
\caption{Classification and MW properties of previously suggested and our giant star associations. Columns same as in Table \ref{table_pm_agn}. "MH": MUSE [M/H] metallicity measurements. }
\label{table_giant_ab}
\end{table}

\subsection{CVs}

Sources 12a, 13a, 13c, 24c, 41d, and 54h are classified as CVs in \cite{henleywillis_deep_2018, cool_hstacs_2013} based on H$\alpha$ and UV excess. All of their previously identified associations are classified as CV by MUWCLASS, and are within $1\sigma$ separation from their CSC sources. The only previously identified association lacking H-alpha detection in oMEGACat is that of source 12a, possibly due to crowding effects.

\begin{figure*}
    \centering
    \includegraphics[width=0.45\textwidth]{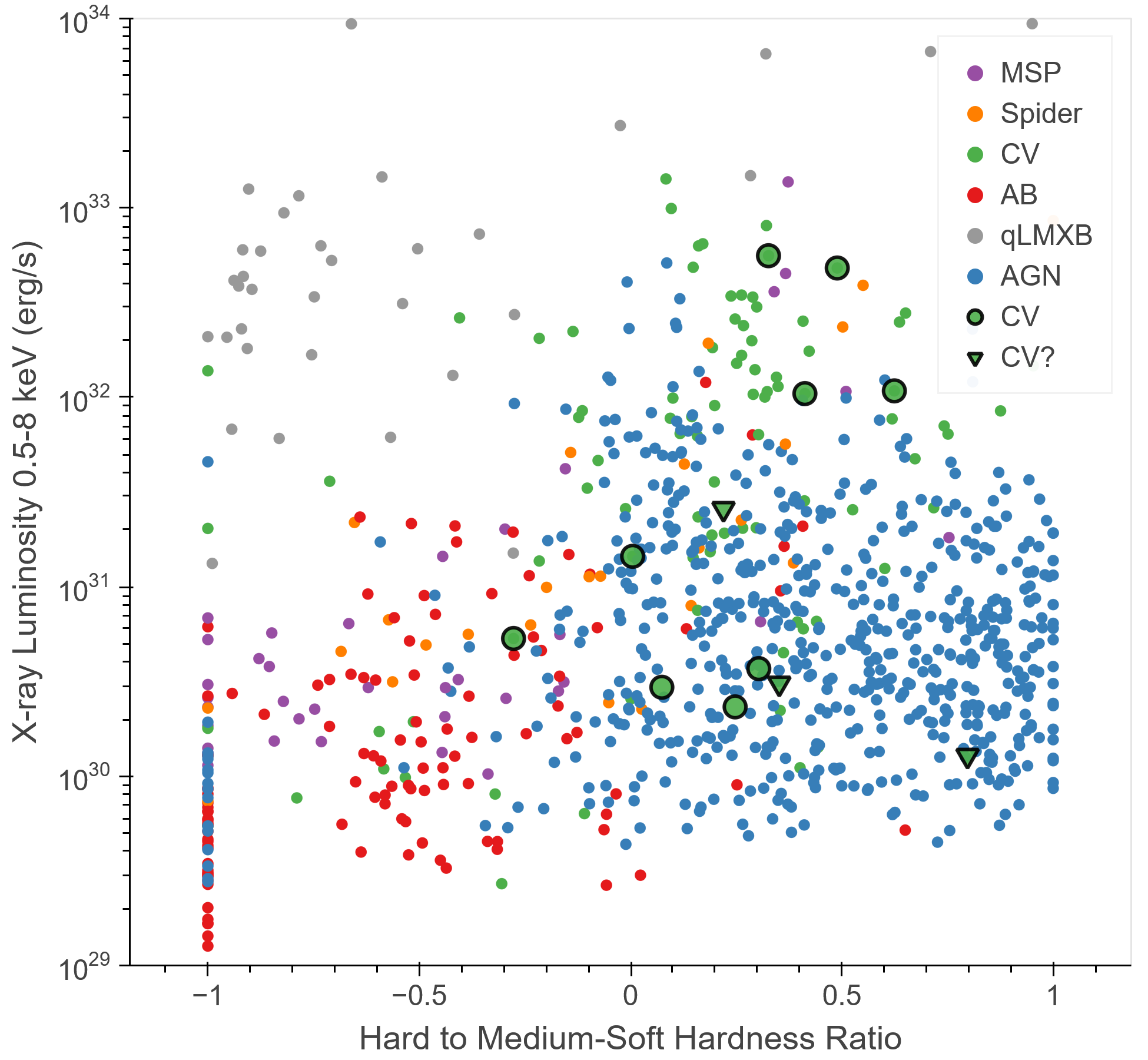}
    \includegraphics[width=0.45\textwidth]{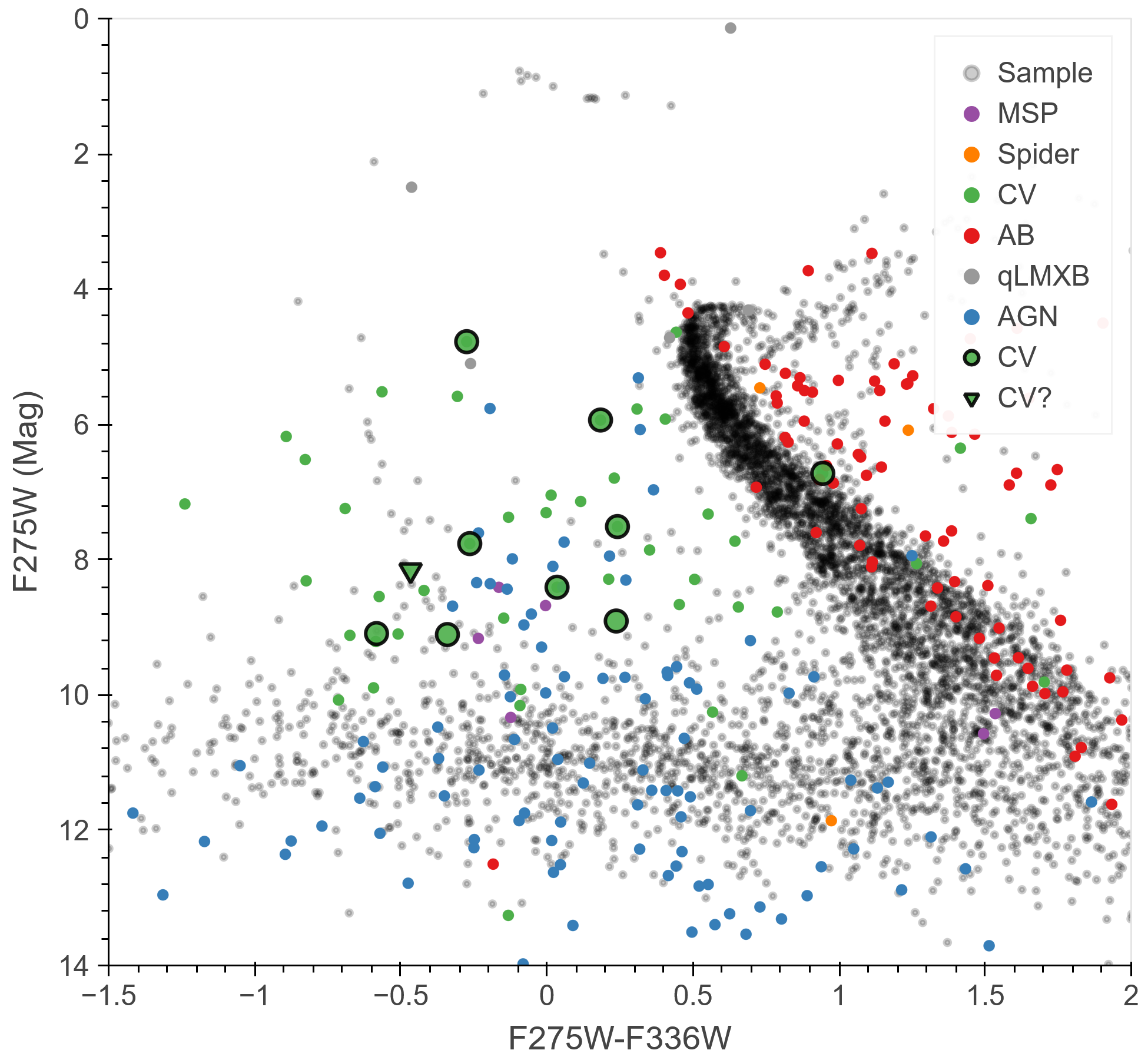}
    \includegraphics[width=0.45\textwidth]{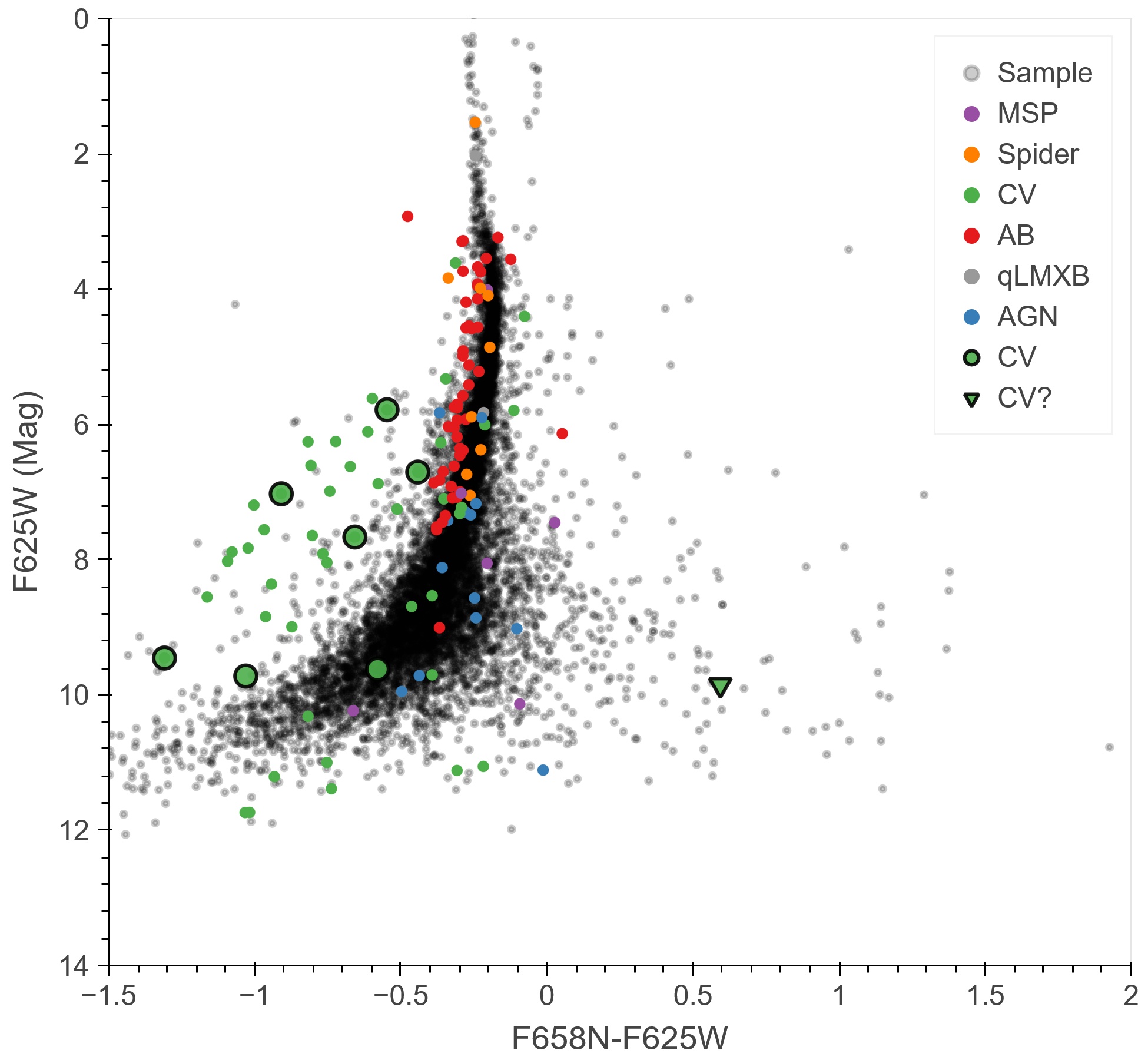}
    \caption{X-ray and optical properties of confirmed and candidate CVs. The candidate CV (Source 14e) with apparent H$\alpha$ absorption in the third panel likely has unreliable photometry due to crowding and faintness. Black points in the background of the CMD are 10,000 randomly sampled stars in oMEGACat. The "luminosity" and "absolute magnitudes" of TD AGN are calculated as if they were at the distance of $\omega$ Cen, to enable comparison of their properties with other classes.}
    \label{fig:NGC5139_halpha_excess}
\end{figure*}

As shown in panel 6 of Figure \ref{fig:csc_omegacat_separation_density}, 11 (27) sources with H$\alpha$ excess or blue excess (F336W-F435W $\leq 0.5$ and F625W $\geq 20$) are expected to be associated with CSC sources within $1\sigma$ ($2\sigma$) PU. The 6 known CVs, and some candidate CVs can account for this excess. We discuss candidate CVs and show their X-ray and optical properties along with those of confirmed CVs in Figure \ref{fig:NGC5139_halpha_excess} and Table \ref{table_cv}.

Source 12b is a bright source at 87 counts, with hardness ratios compatible with TD CVs. Its third counterpart is close to two other stars, and thus suffers from crowding, missing magnitudes in F435W, F336W, and F275W in the catalog, as well as PM measurements. However, in the UV images, this source appears blue, consistent with a CV nature. The X-ray lightcurves show variability, with the source being significantly brighter in the latter two observations (13726, 13727), as well as showing flares. Deeper UV observations could determine the nature of this counterpart, and thus the X-ray source. 

Source 13f is a faint candidate CV in \cite{cool_hstacs_2013} associated with its closest counterpart. It shows H$\alpha$ excess and appears blueward of the MS in all CMDs. Its cluster membership is confirmed by PM measurements, and X-ray and optical features are all compatible with CVs. Its SED, including data from JWST, also appears very similar to those of known CVs (e.g. 13a). Therefore, we confirm its classification as a CV. It shows strong excesses in the F212N and F470N $H_2$ filters.

Source 14e is a faint source with only 6 X-ray counts. It is missing soft band detections, and appears hard on the hard-medium scale, although with large uncertainties. Its second counterpart has a clear blue excess consistent with a WD. On the H$\alpha$ CMD it shows a deficit, however, this measurement is likely unreliable due to crowding. Given the hardness of the X-ray source, we classify this source as a candidate CV.

Source 23b is another faint candidate CV in \cite{cool_hstacs_2013} identified with its closest counterpart. Based on the clear blue excess of the closest counterpart, whose SED is similar to those of known CVs (e.g. 12a, 24c), we confirm this association as a CV. MUWCLASS slightly favored this source as a MSP (53\%) over a CV (40\%), because its X-ray features are softer and fainter than most TD CVs, and the position of the blue counterpart in several CMDs are very close to some TD MSPs. The lack of a MeerKAT counterpart disfavors a MSP classification, although if it is a MSP, it may still be detected in future radio observations.

The closest counterpart to source 23d is relatively bright in UV, but is too faint to be measured in F625W and F814W, and lacks PM measurements. In the images, this counterpart appears to be a cluster member. Its SED from F275W to F658N is power-law like, and is most similar to that of the closest counterpart to 23b, discussed above. The non-detection in F625W and F814W indicates that it has a strong H$\alpha$ excess. Because of the blue counterpart and the hardness of the X-ray emission, we classify 23d as a CV.

Source 24k is faint with 10 X-ray counts, but shows hardness ratios compatible with TD CVs. Its 6th counterpart shows a clear blue excess in the images, although it lacks F275W and F336W magnitudes in the catalog. Interestingly, there is also another source showing a clear blue excess in the images closer to the CSC source position, which is undetected in the catalog. With either of these counterparts, this X-ray source is likely to be a CV.

Except for source 23b, all of our confident and candidate CVs were classified as CVs by MUWCLASS. The known CVs have higher classification probabilities, because their X-ray features are brighter and more distinctive from other classes in the TD.

\begin{table}
\footnotesize
\centering
\begin{tabular}{llllllllllll}
\toprule
Assoc. Name & Lit. & Manual & Pub. & ML & P\_C & PU (\arcsec) & Sep. (\arcsec) & $L_{\rm CSC, b}$ & $HR_{\rm hms}$ & F275W & UV Color \\
\midrule
J132648.6-472744-1 & 12a & CV & CV & CV & 0.86 & 0.18 & 0.04 & 5.33e+30 & -0.28 & 21.21 & 0.24 \\
J132652.5-472737-3 & 12b & CV? &  & CV & 0.74 & 0.15 & 0.28 & 2.58e+31 & 0.22 &  &  \\
J132653.5-472900-1 & 13a & CV & CV & CV & 0.91 & 0.14 & 0.07 & 4.82e+32 & 0.49 & 20.42 & 0.94 \\
J132652.1-472935-1 & 13c & CV & CV & CV & 0.97 & 0.14 & 0.12 & 5.59e+32 & 0.33 & 19.63 & 0.18 \\
J132645.9-472916-1 & 13f & CV & CV? & CV & 0.62 & 0.23 & 0.04 & 3.69e+30 & 0.30 & 22.61 & 0.24 \\
J132641.8-472923-2 & 14e & CV? &  & CV & 0.59 & 0.34 & 0.13 & 1.28e+30 & 0.80 & 21.85 & -0.47 \\
J132651.6-473047-1 & 23b & CV & CV? & MSP & 0.53 & 0.20 & 0.21 & 2.95e+30 & 0.07 & 22.81 & -0.34 \\
J132658.4-472847-1 & 23d & CV &  & CV & 0.42 & 0.26 & 0.12 & 2.32e+30 & 0.25 & 22.79 & -0.58 \\
J132638.4-473036-1 & 24c & CV & CV & CV & 0.88 & 0.18 & 0.08 & 1.45e+31 & 0.01 & 21.46 & -0.26 \\
J132635.4-472838-6 & 24k & CV? &  & CV & 0.33 & 0.31 & 0.40 & 3.08e+30 & 0.35 &  &  \\
J132628.6-472627-1 & 41d & CV & CV & CV & 0.55 & 0.18 & 0.18 & 1.08e+32 & 0.62 & 22.11 & 0.03 \\
J132620.3-473002-1 & 54h & CV & CV & CV & 0.70 & 0.18 & 0.16 & 1.05e+32 & 0.41 & 18.48 & -0.27 \\
\bottomrule
\end{tabular}
\caption{Classification and MW properties of previously known and our confident and candidate CV associations. Columns same as in Table \ref{table_pm_agn}, except with F275W magnitude. "UV Color": F275W-F336W color. }
\label{table_cv}
\end{table}

\subsection{Previously Known MSPs}

The MSPs discovered in Omega Cen by \cite{dai_timing_2023, chen_meerkat_2023, bernadich_joint_2026} were categorized as either binary or isolated systems using radio timing data. All seven of them with CSC counterparts are also detected in the \textsc{PyBDSF} MeerKAT catalogs. For MSPs discovered by MeerKAT (MSPs F and after), the limited time span of the MeerKAT observations so far can only find binaries with orbital periods $\lesssim10$ days. However, \cite{chen_meerkat_2023} argues that wider binaries may exist in $\omega$ Cen due to its lower density. The more recent MeerKAT and Parkes observations by \cite{bernadich_joint_2026} provided updated timing solutions for 3 BW MSPs and discovered the isolated MSP S (not detected by Chandra), but did not discover new binary MSPs. Furthermore, MSPs I, N, and Q were updated to be in wider binaries, with N and Q having minimum companion masses $>0.2 M_\odot$.

The likely X-ray counterpart to MSP Q (21b) was reported in \cite{zhao_chandra_2023}, and it is the only known non-spider binary MSP in $\omega$ Cen with a CSC counterpart. It is not detected in the soft band, but its hardness ratio in the hard-medium scale and X-ray luminosity (8 total counts) is consistent with other TD MSPs, although with large uncertainties. Its 4th optical counterpart is blue and faint, consistent with a helium WD companion (mass $>0.205 M_\odot$) inferred from timing data by \cite{chen_meerkat_2023, bernadich_joint_2026}. The association was reported in \cite{cool_hstacs_2013} as a faint blue CV, but with the MSP discovery, we view this WD as the likely companion to MSP Q. 

The Chandra counterpart to MSP N in \cite{zhao_chandra_2023} is too faint to be detected in the CSC. Its \textsc{PyBDSF} MeerKAT counterpart is $\sim 0.8$ away from a WD in oMEGACat, compatible with its updated minimum companion mass of $>0.232 M_\odot$ in \cite{bernadich_joint_2026}. However, the separation is significantly higher than the MeerKAT PU of $0.14\arcsec$. MSP I has a \textsc{PyBDSF} MeerKAT counterpart $\sim 4\arcsec$ away from the published positions in \cite{chen_meerkat_2023}, within its $3\sigma$ PU, and which is also coincident with an undetected CSC source. Updated timing positions are necessary to confirm the counterparts to these binary MSPs.

Of the isolated MSPs, only MSP C and MSP E have X-ray counterparts from \cite{zhao_chandra_2023} that are cross-matched to the CSC. The \textsc{PyBDSF} MeerKAT position of MSP J is coincident with a possible Chandra counterpart, undetected in the CSC due to its location in the gap between ACIS chips.
They are not classified correctly by MUWCLASS in their null associations (i.e., using X-ray only features), being slightly more favored as ABs rather than MSPs, due to large uncertainties in their CSC fluxes causing their CSC features to be more consistent with those of ABs after uncertainty sampling. However, the upper-limit associations are classified correctly. This demonstrates that upper limits of current observations become important when classifying sources, if there is reason to believe that true counterparts are below these detection limits. 

Despite their reported isolated nature, we find relatively bright WDs $1.3\arcsec$ and $0.2\arcsec$ away from MSPs C and E, respectively. The WD near MSP E has a large optical variability of $\sim 1$ mag in the F606W filter. To estimate the chance coincidence probability of WDs to occur near MSPs, we select  WDs based on several cuts in the UV-optical CMDs, see Figure \ref{fig:ngc5139_msp_wd_cmd}. Qualitatively, these criteria select cluster members that are blue-wards of the MS in F275W-F336W, F336W-F438W, and F625W-F814W, and fainter than blue-stragglers, with cuts on the \textsc{phot\_hq\_flag}, \textsc{n\_measurements}, \textsc{\_RMS} parameters to limit spurious detections. However, some likely WDs, such as the probable counterpart to MSP Q, are not selected due to faintness.

\begin{figure*}
    \centering
    \includegraphics[width=0.45\textwidth]{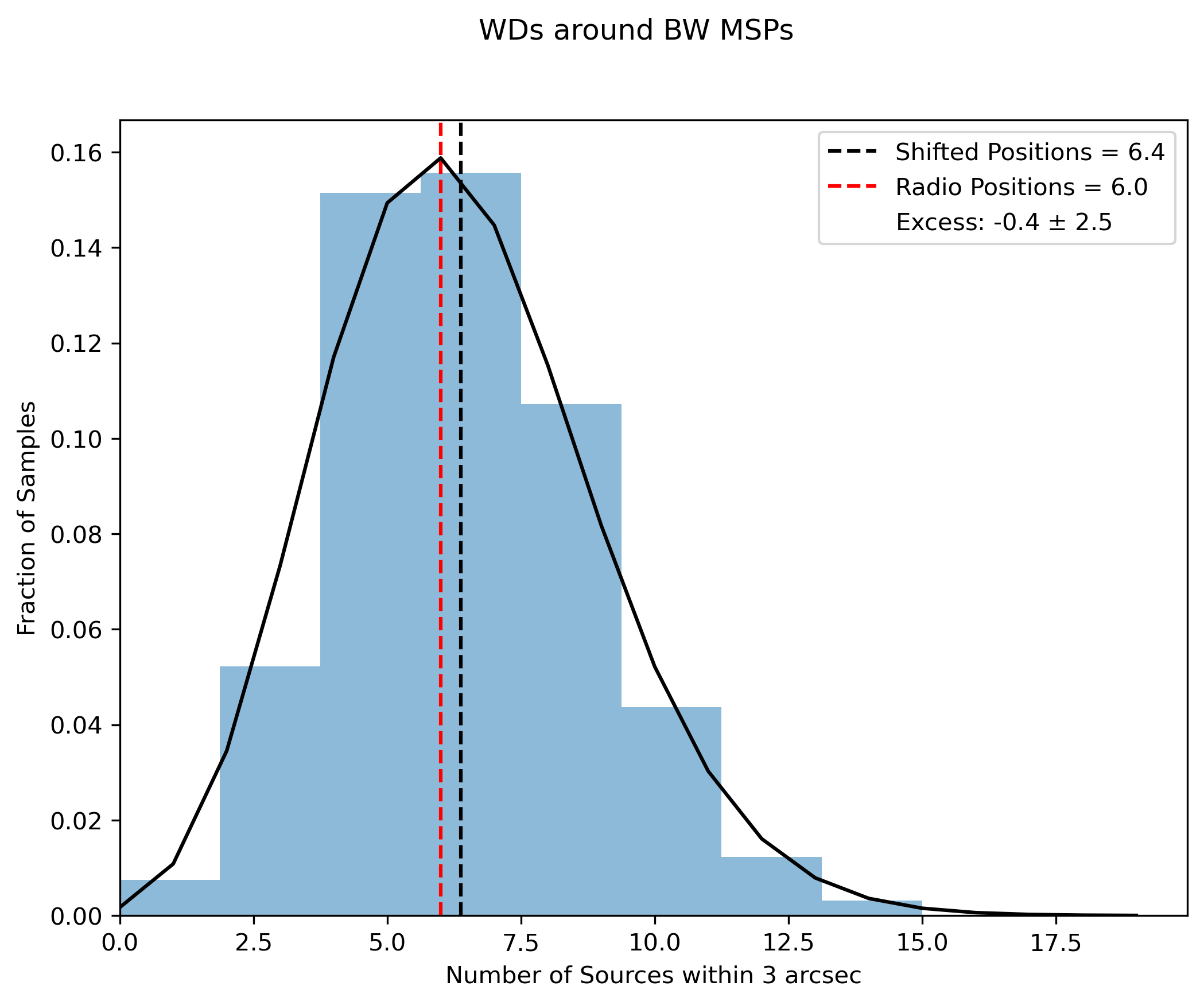}
    \includegraphics[width=0.45\textwidth]{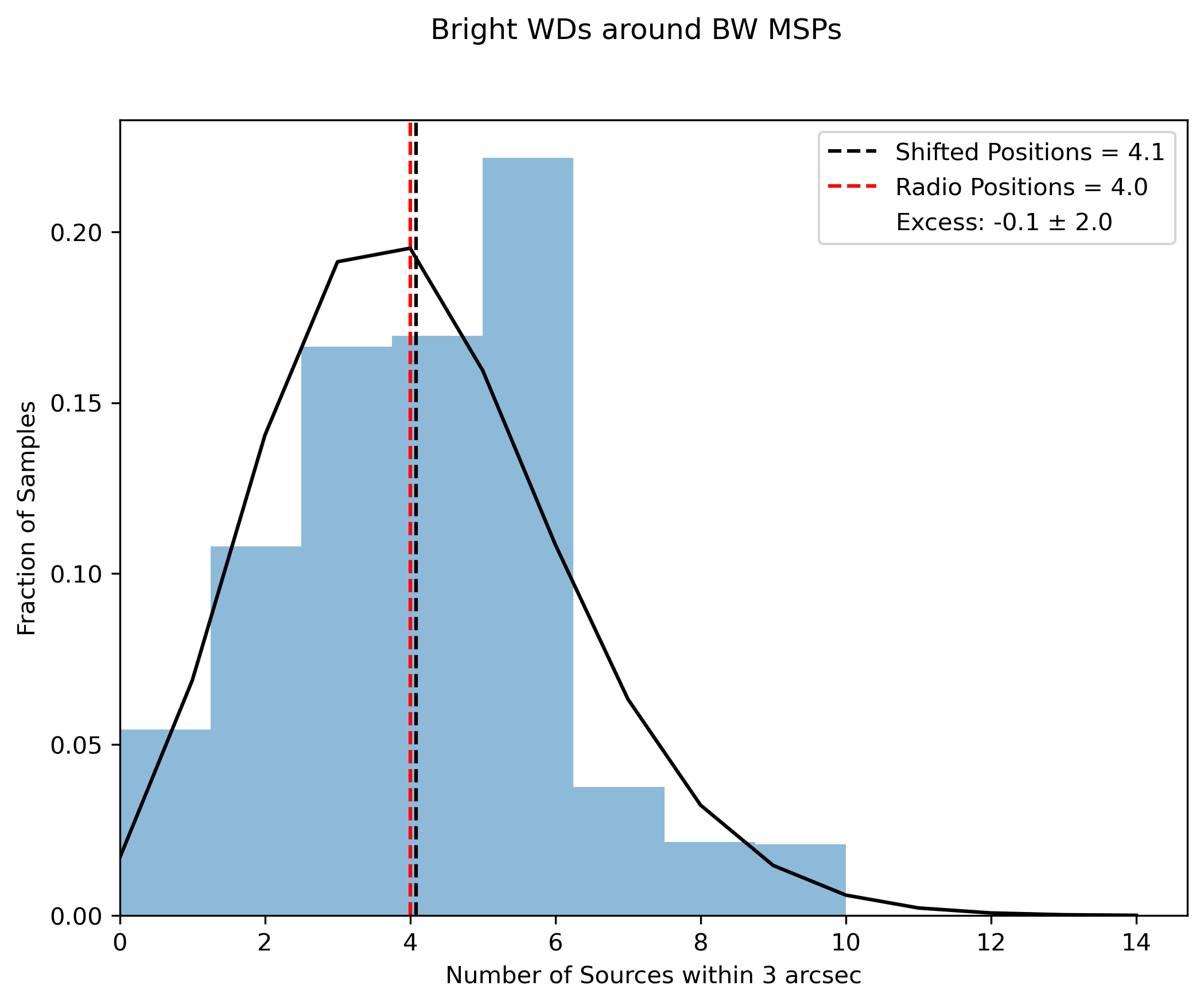}
    \includegraphics[width=0.45\textwidth]{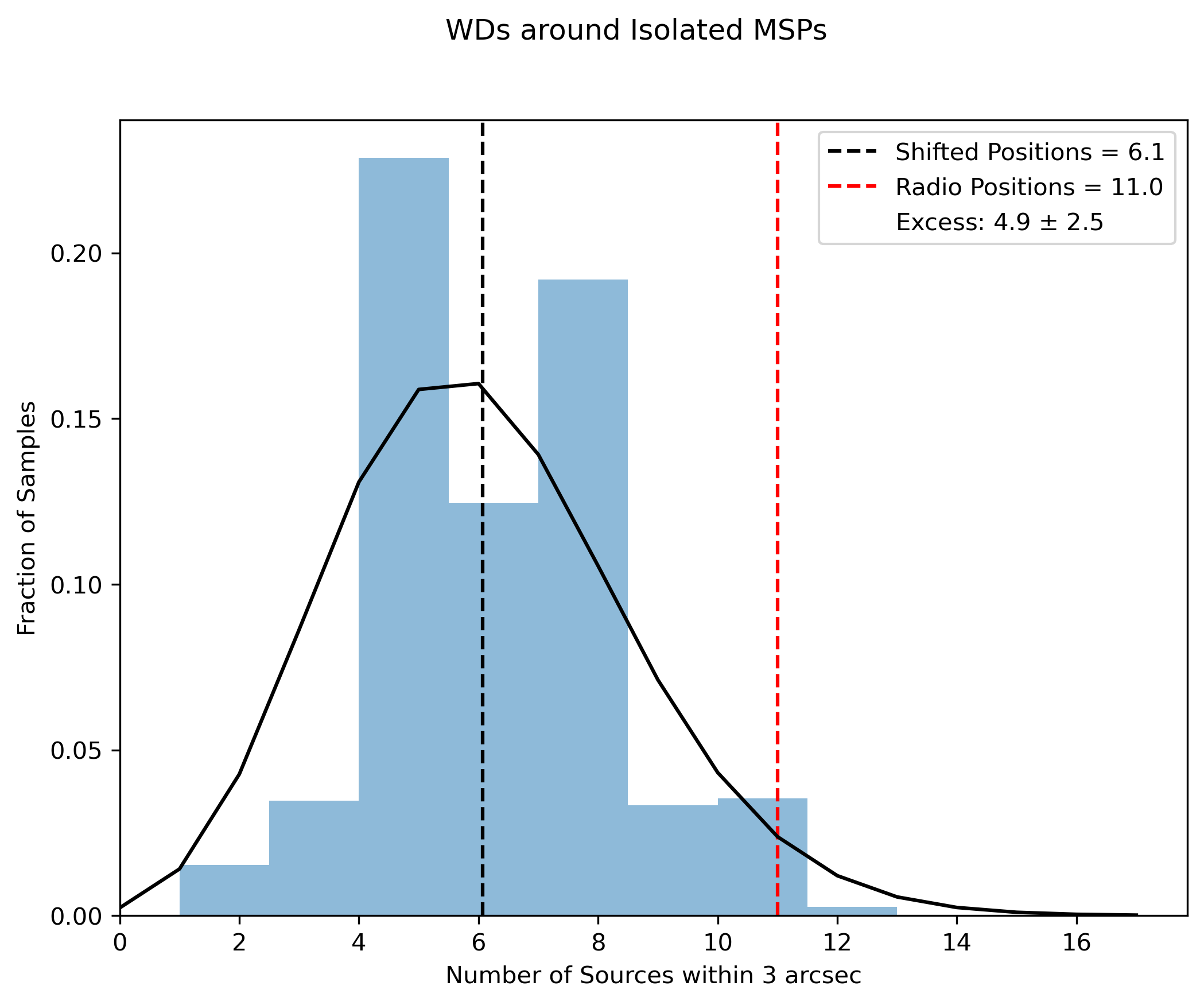}
    \includegraphics[width=0.45\textwidth]{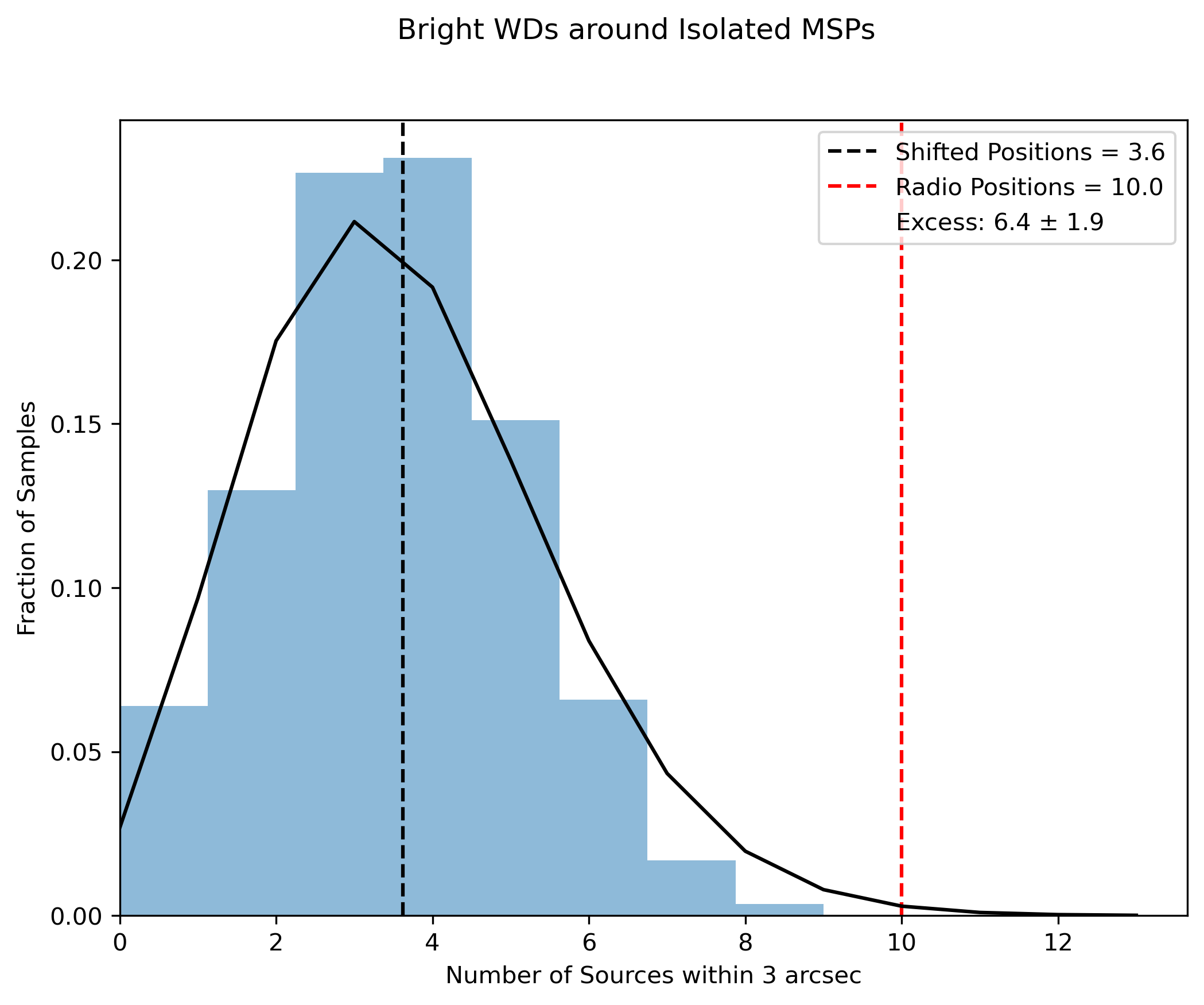}
    \caption{Distribution of the number of WDs within $3\arcsec$ of random positions close to \textsc{PyBDSF} MeerKAT positions of BW (panels 1 and 2) and isolated (panels 3 and 4) MSPs, similar to Figure \ref{fig:csc_omegacat_counterparts_hist}. Panels 2 and 4 show only WDs brighter than 23 mag in F275W.}
    \label{fig:msp_wd_excess}
\end{figure*}

\begin{figure*}
    \centering
    \includegraphics[width=0.5\textwidth]{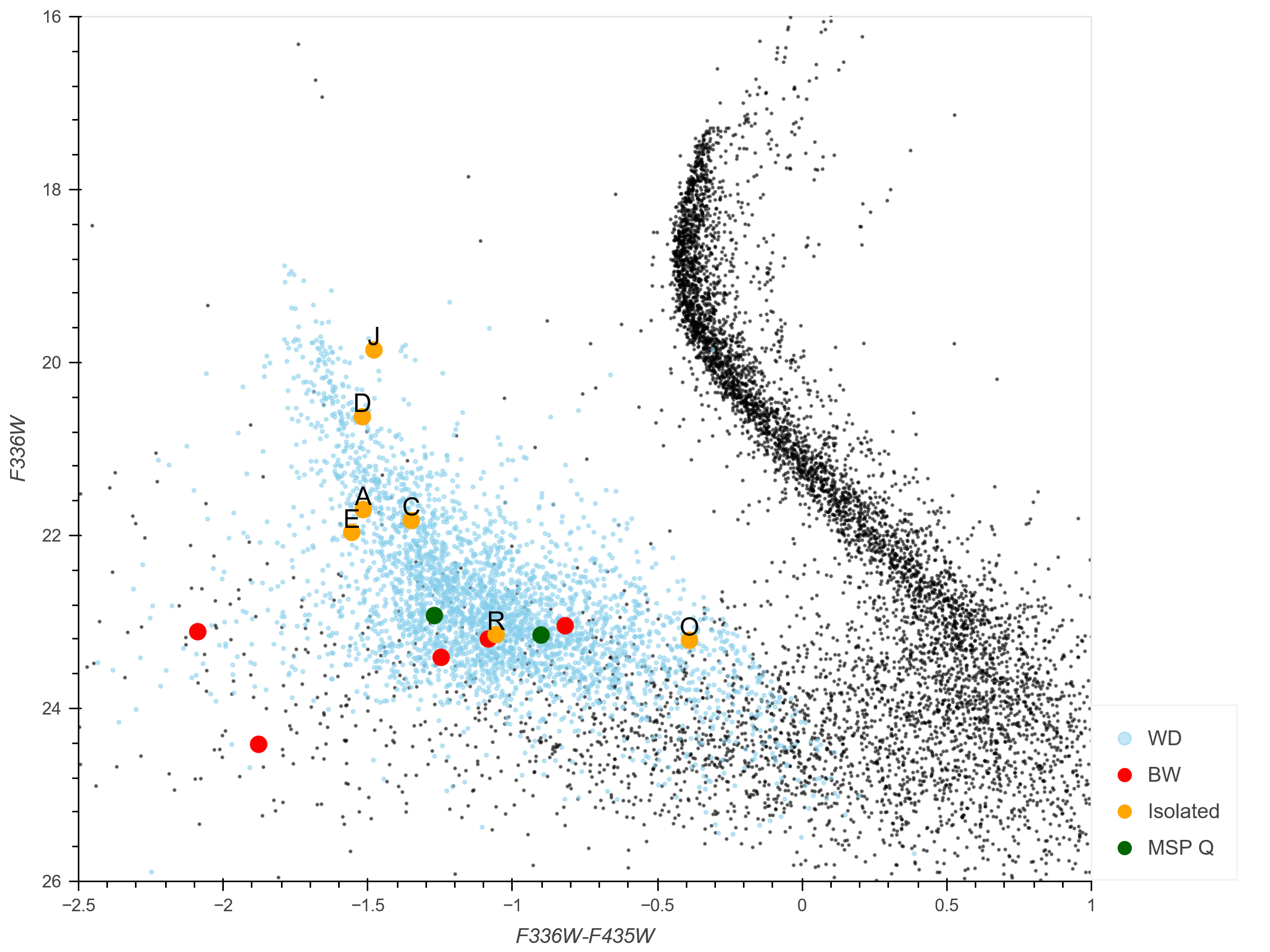}
    \caption{F336W-F435W vs. F435W CMD of the nearest WDs to \textsc{PyBDSF} MeerKAT positions of known BW and isolated MSPs, along with the probable Helium WD counterpart to MSP Q. A sample of 10,000 stars within 3\arcmin of the cluster center, as well as all WDs within that radius, are plotted in the background.}
    \label{fig:ngc5139_msp_wd_cmd}
\end{figure*}

We find $\approx 1650$ WD candidates within 2\arcmin of the cluster center. The chance coincidence probability within a $1\arcsec$ radius to find one WD is thus 10.8\%\footnote{As the WD density decreases with distance from the cluster center, sources further away will have smaller chance coincidence probabilities}. To improve the sample size, we search for WDs around BW and isolated MSPs not expected to have WD companions, including those not crossmatched to the CSC. For the MeerKAT discovered MSPs, the timing positions in \cite{chen_meerkat_2023} are too inaccurate to match to HST, except the BW MSPs G, H, and K, which have updated timing solutions from \cite{bernadich_joint_2026}. However, for all of these MSPs except MSPs F, M, and P, there is an obvious counterpart in the \textsc{PyBDSF} MeerKAT catalog \citep{smirnov_mining_2025}. We plot the separation distribution of WDs around randomly shifted points between 5 to $10\arcsec$ away from the \textsc{PyBDSF} MeerKAT positions, similar to Figure \ref{fig:csc_omegacat_sigma_excess}, in Figure \ref{fig:msp_wd_excess}.

Within $3\arcsec$ separation of the 5 BW MSPs, the number of WDs closely matches the chance coincidences within $3\arcsec$ of the randomly shifted positions. However, within $3\arcsec$ separation of the 7 reported isolated MSPs, there are $\sim 5$ excess WDs above the expected number of chance coincidences, at a nominal significance of $\sim 2\sigma$. Interestingly, the 3 WDs within $1\arcsec$ separation from MSPs D, E, and J are also among the brighter WDs in the cluster, at 20.8, 22.3, and 20.0 mag in F275W respectively. The isolated MSPs A (21.9 mag, $3.7\arcsec$) and C (22.1 mag, $1.3\arcsec$) are also matched to bright WDs outside $1\arcsec$ separation.

Thus, if we select only WDs brighter than 23 mag in F275W, the significance of the excess around isolated MSPs is enhanced to  $>3\sigma$ (see panel 4 of Figure \ref{fig:msp_wd_excess}), before accounting for the problem of running multiple statistical tests on the same data. For another view of the unusual nature of these WDs, we plot the WDs found near the isolated and known WD binary MSPs on the UV F336W-F435W vs. F435W CMD in Figure \ref{fig:ngc5139_msp_wd_cmd}, along with the general population of WDs. The closest WDs to the BW MSPs are also shown, as examples of chance coincidences. Many of these WDs lack F658N detections in oMEGACat due to faintness. Of those with F658N detections, only the WDs matched to MSPs C and E show signs of H$\alpha$ excess.

\cite{chen_meerkat_2023} notes that due to the low stellar density of $\omega$ Cen, some of these reported isolated MSPs may turn out to be in wide binaries with future observations, which would explain this apparent excess. However, the separations between some WDs and the \textsc{PyBDSF} position of the MSPs, e.g. MSPs C and D, seem too high for the WDs to be the companions. Future timing observations may still correct these positions to be significantly closer to the WDs. Otherwise, it is possible that the progenitor of these WDs were giant stars from which the MSPs were accreting, and after the giant stars evolved to WDs, the binaries were disrupted by cluster dynamical interactions relatively recently, which may explain their brightness and concentration near MSPs compared to the average WD. 

\subsection{Previously Known Spider-Type MSPs}
\label{spiders}

Five spider MSPs in Omega Centauri are classified and crossmatched to Chandra sources by \cite{zhao_chandra_2023}: MSP B (13d, 2CXO J132649.5-472924), MSP G (24f, 2CXO J132637.2-472942), MSP H (14c, 2CXO J132644.0-472854), MSP K (21d, 2CXO J132638.2-472740), and MSP L (32d, 2CXO J132702.4-472646). Their orbital parameters from \cite{chen_meerkat_2023} and \cite{bernadich_joint_2026} are consistent with black widow (BW) systems, with the highest minimum companion mass being $>0.042 M_\odot$ for MSP K. Additionally, the updated coordinates of MSPs G, H, and K from \cite{bernadich_joint_2026} are consistent with the CSC counterpart coordinates within $1\sigma$ PU. 

MS stars of the same mass as these BW companions are below the detectable limit of oMEGACat, which goes down to roughly $0.2 M_\odot$. However, spider companions are ablated and heated by the pulsar wind, enhancing their optical emission. \cite{koljonen_gaia_2023} compiled known galactic field spider MSPs with their Gaia optical counterparts, and found a correlation between the orbital periods and optical luminosities of redback spiders. On the other hand, due to their faintness, only three BWs have confirmed companions, detected with larger ground-based telescopes. These BW companions are fainter than the correlation for redbacks would suggest. Therefore, we do a simple estimate of BW companion detectability based on the brightest BW companion reported (for J2256-1024, 21.4 mag in $i$-band, at 2.0 kpc \cite{breton_discovery_2013}). This gives an upper limit to the magnitudes of BW companions at the distance of $\omega$ Cen of $\sim 23.6$ mag in F814W. Since this is near the sensitivity limit in oMEGACat for reliably detected sources, some BW companions may be detectable.

Of the 5 BW systems, MSP K had two, and MSP G had one candidate counterparts in \cite{cool_hstacs_2013}, before their MSP nature was revealed. However, the candidate counterpart to MSP G is a giant or subgiant, and those of MSP K appear to be MS stars inconsistent with a BW companion, so no counterparts were used in the TD for these MSPs. \cite{zhao_chandra_2023} notes that the proposed counterparts of MSP K have blue or H$\alpha$ excess, but any such excess could also be explained by contamination from nearby stars. 

For each of these BWs, a few associations are classified by MUWCLASS as MSP\footnote{As described in Section \ref{counterparts}, when classifying one of these known BWs using MUWCLASS, that source is removed from the TD to prevent overfitting.} due to unusual optical features, but these features could mostly be attributable to faintness and contamination. After inspecting the actual images, we discuss some associations where the unusual features appear to be real.

There is a hint of a faint, blue counterpart undetected in oMEGACat just south of the 3rd counterpart of MSP B in the F275W and F336W images, which is also visible in the NIRCam F200W, F444W images. 

The 6th counterpart to MSP G is faint, but has a large apparent F625W excess (22.7 mag) compared to the neighboring F606W magnitude (23.6 mag), as well as a NIRCam F212N deficit that indicates variability. However, this source is $0.45\arcsec$ away from the updated timing position of MSP G from \cite{bernadich_joint_2026}, while the nearest counterpart from that position is too bright to be a BW companion. 

The closest counterpart to MSP K is very faint, being only detected in F435W, F606W, F625W, and F658N from 25.3 to 23.6 mag. If this counterpart is a real source, the lack of detection in F814W is hard to explain other than significant optical variability, potentially compatible with a BW companion. Notably, this source is within $0.1\arcsec$ of the updated timing position of MSP K from \cite{bernadich_joint_2026}. 

The closest counterpart to MSP H lies close to several other stars. In the oMEGACat catalog it is only detected in F606W (23.4 mag). In the JWST data it is very faint in most bands, but is clearly visible in the F212N image (19.7 mag), with no contamination. This could also be a variable source. The second closest counterpart is the closest counterpart to the updated timing position of MSP H from \cite{bernadich_joint_2026}. The SED of this source is typical of a faint MS star, although with fluctuations, and an apparent H$\alpha$ deficit, indicative of variability.

The 6th counterpart to MSP L is faint and shows significant H$\alpha$ excess. Compared to other stars of similar brightness in F814W (22.8 mag), it appears to have a deficit in F435W (27 mag) and an excess in F336W (25.2 mag), which may be due to variability. The relative isolation of this source increases the reliability of these measurements, despite the low flux.

All 5 of these MSPs are classified correctly with the upper-limit association, but only MSPs K, G and L are classified correctly with the null association, as MSPs B and H are harder than any other MSPs in the TD on the medium-soft hardness ratio scale. The unusual optical counterparts for each BW are all faint, close to the oMEGACat detection limit ($\sim 25$ mag), consistent with the upper limit derived from \cite{koljonen_gaia_2023}. All of these faint counterpart associations, except for MSP G, were classified incorrectly, due to the large fraction (73\%) of Spider, and especially BW, sources without optical counterparts in the TD. If some of these associations are real, they can significantly increase in the number of known BW counterparts in GCs. Detection of orbital periodicity in the optical would be the most secure way to confirm these associations, since the orbital period for these systems are known. 

\subsection{Candidate MSPs and Spider-Type MSPs}
\label{Candidate MSPs}

There are around 20 X-ray sources which are soft in the X-rays ($\text{HR}_{\text{HMS}}\lesssim0$), but appear brighter than the majority of ABs in the TD. Given the low source density of sources from other classes in this region of X-ray flux vs.\ hardness ratio plot, these may be unidentified MSPs or spiders.

Interestingly, 6 of these sources have faint radio counterparts in MeerKAT, 4 of which are within $1\arcsec$ separation (chance coincidence probability $<2\%$). The MeerKAT counterparts have fluxes in the range of tens of $\mu$Jy, a factor of 3 fainter than the MeerKAT counterparts of the known MSPs in Omega Cen. We discuss each of these sources below. 

Source 13e (2CXO J132646.2-472948) has 10 counts and is soft. The 3rd and 6th closest counterparts are similarly bright and have significant $H\alpha$ excess, but the SED of the former appears to deviate from a thermal SED, possibly due to variability. If the association is true, this SED could indicate that it is a spider system. 

Source 23f (2CXO J132657.2-473012, 13 counts) is soft in the X-rays, compatible with MSPs and ABs. The closest counterpart is relatively bright (21.7 mag in F275W) and shows a clear, strong blue excess in all CMDs. It is classified as an MSP by MUWCLASS; however, it is not detected in F658N due to a streak from a bright star. The location of the counterpart on the CMD is strongly compatible with WD companions of MSPs in the TD.
The 7th counterpart also shows blue excess compatible with WDs, but it is fainter and further away ($0.4\arcsec$), and lacks H$\alpha$ measurements due to crowding. The low density of bright WDs (628 with F275W$<$22 mag within 4\arcmin of cluster center) implies a chance coincidence probability of $\sim 0.3\%$. Combined with the 2\% chance coincidence probability of the radio counterpart, we classify the association with the brighter WD as a strong binary MSP candidate. 

Source 24g is bright (78 counts). The counterparts to this source lack F275W measurements due to a gap in the F275W coverage, except the 4th counterpart. The two closest counterparts are separated by $\sim 0.1\arcsec$. However, the SED of the second closest counterpart has a large blue excess in F336W and F435W 
, and the infrared tail, measured in NIRCam F115W and F277W, drops faster than a thermal source (see panel 3 of Figure \ref{fig:NGC5139_msp_seds}. This association may be a binary WD-MSP system, with the X-ray source being brighter than most other TD MSPs.

Source 24d has 17 counts. There is a faint source to the north of the first counterpart undetected in oMEGACat, that has an excess in F625W and F658N compared to sources similarly bright in F814W. The second closest counterpart is faint, with a significant $H\alpha$ and JWST F212N excess, but the crowding may affect the photometry, although the measured errors are small (0.04 for F658N). If one of these counterparts is real, then the system may be an interacting binary MSP system.

Source 32h has 32 counts, and is crossmatched to a background galaxy at $1.2\arcsec$ separation, but this is well outside its $1\sigma$ PU of $0.25\arcsec$. Its X-ray properties are consistent with MSPs and spiders. Its closest counterpart is faint, but has a significant H$\alpha$ excess. This association is classified as a CV due to proximity to some TD CVs in the CMDs. The presence of a MeerKAT counterpart adds weight to a MSP or spider interpretation, but we still consider this classification as very tentative.

Source 33k (2CXO J132649.7-473147) is soft and bright at 60 counts. In the F336W image, there is a source to the east of the first counterpart undetected in oMEGACat, which is not visible in the other images. Another undetected source in the F814W image is just to the north of the first undetected source, but these sources should not be the same source, because the observations in the F814W and F336W images are at different epochs. Regardless, the F336W source is very blue compared to MS stars. This source may be an MSP with a WD companion or a spider.

In summary, we find 6 MSP candidates based on their X-ray properties, and the presence of radio counterparts, in addition to source 23a discussed in Section \ref{Giant and Subgiant ABs}. Sources 24g and 33k are bright and soft, while sources 13e, 24d, and 23f are fainter, and sources 23a and 32h are harder. Source 23f and 24g are strong candidate MSPs with WD companions, and source 23a is possibly a huntsman-type system. The other sources lack associations with properties clearly consistent with MSP systems. Given the apparent faintness of the companions of the known BWs in the optical, it is possible that these sources are other spider systems, isolated MSPs not detectable in the optical, or in wide, non-interacting, non-WD binary systems. We discuss the implications of these candidates in Section \ref{discussion}.  

\begin{figure*}
    \centering
    \includegraphics[width=0.45\textwidth]{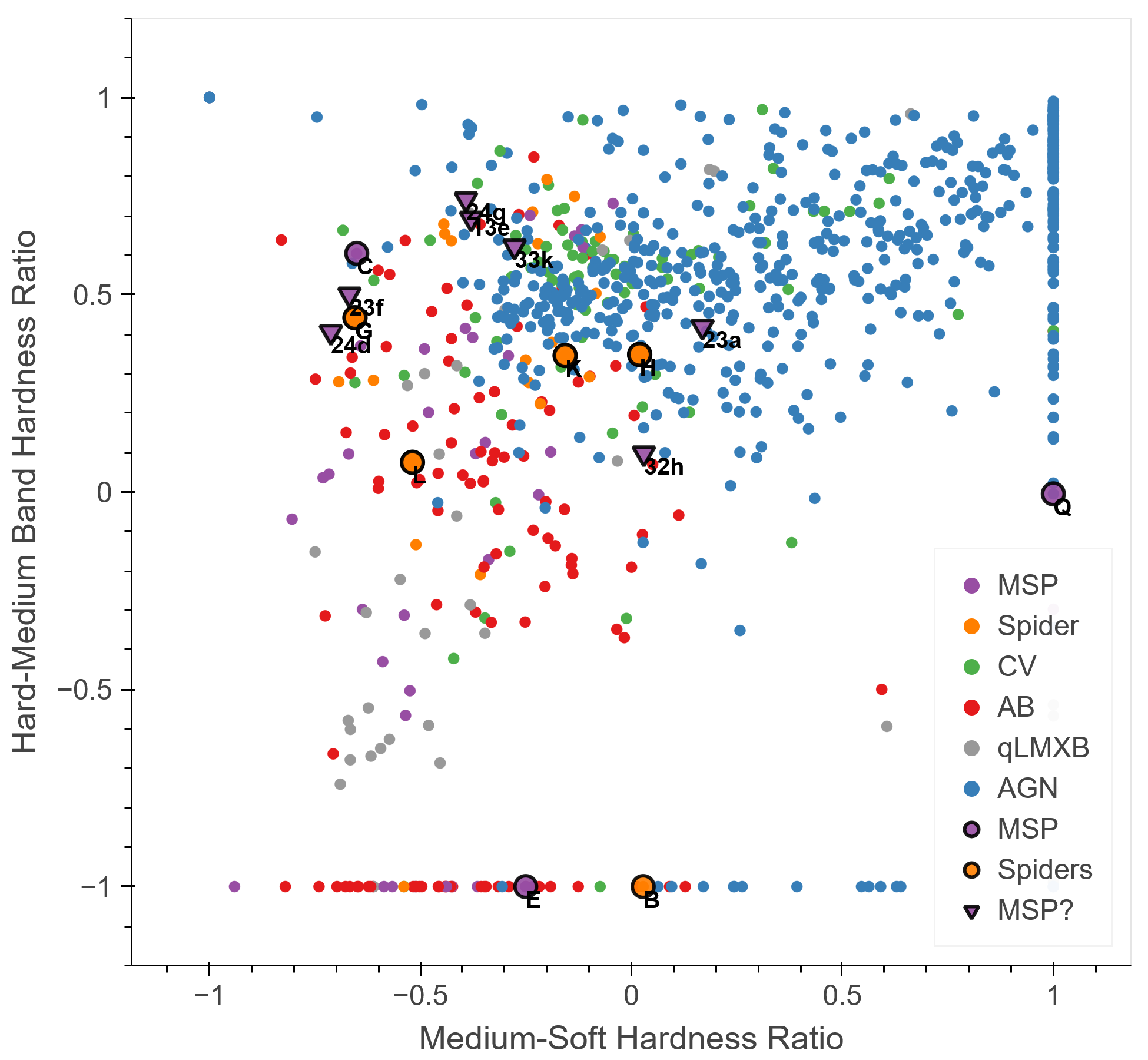}
    \includegraphics[width=0.45\textwidth]{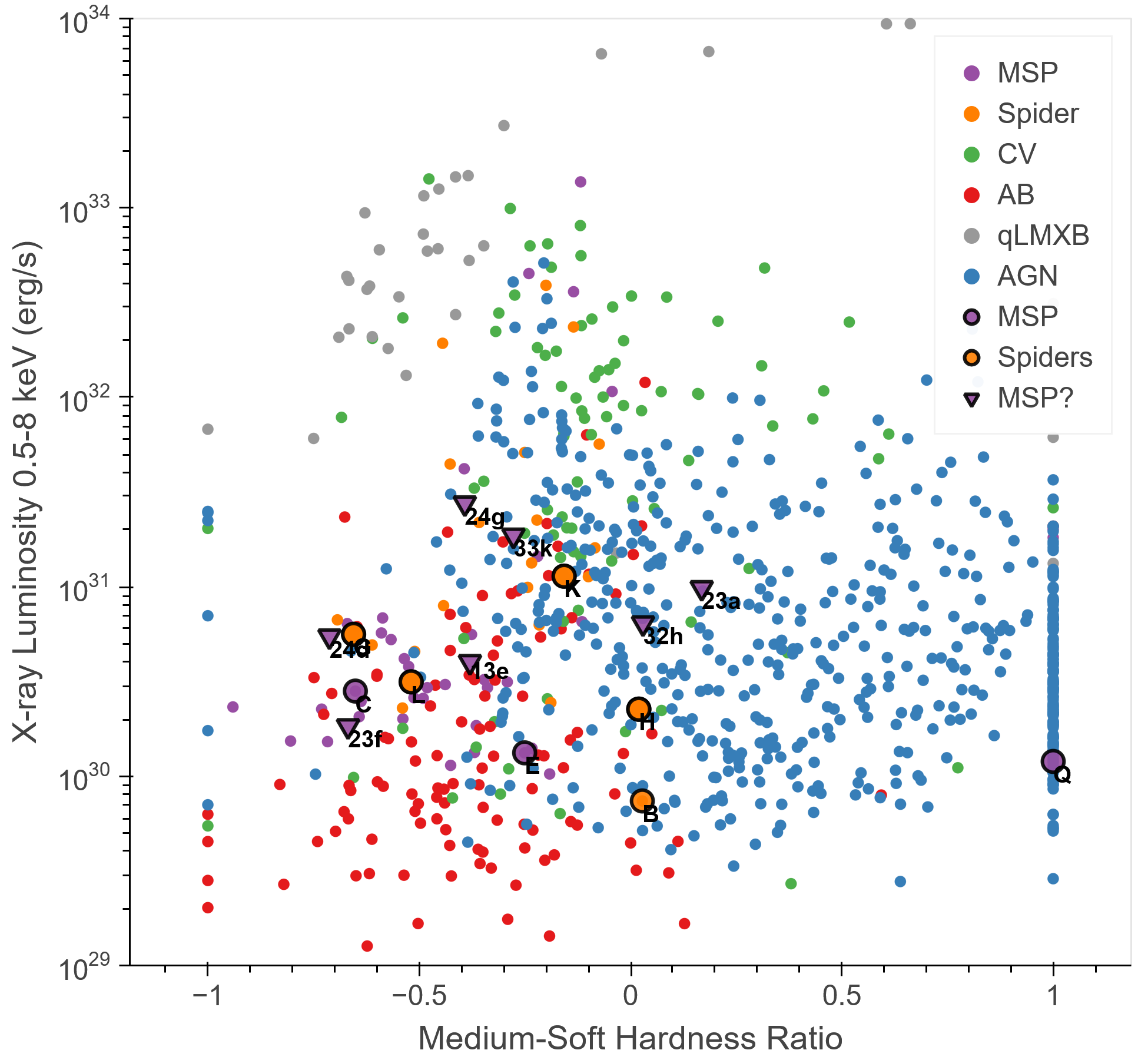}
    \includegraphics[width=0.45\textwidth]{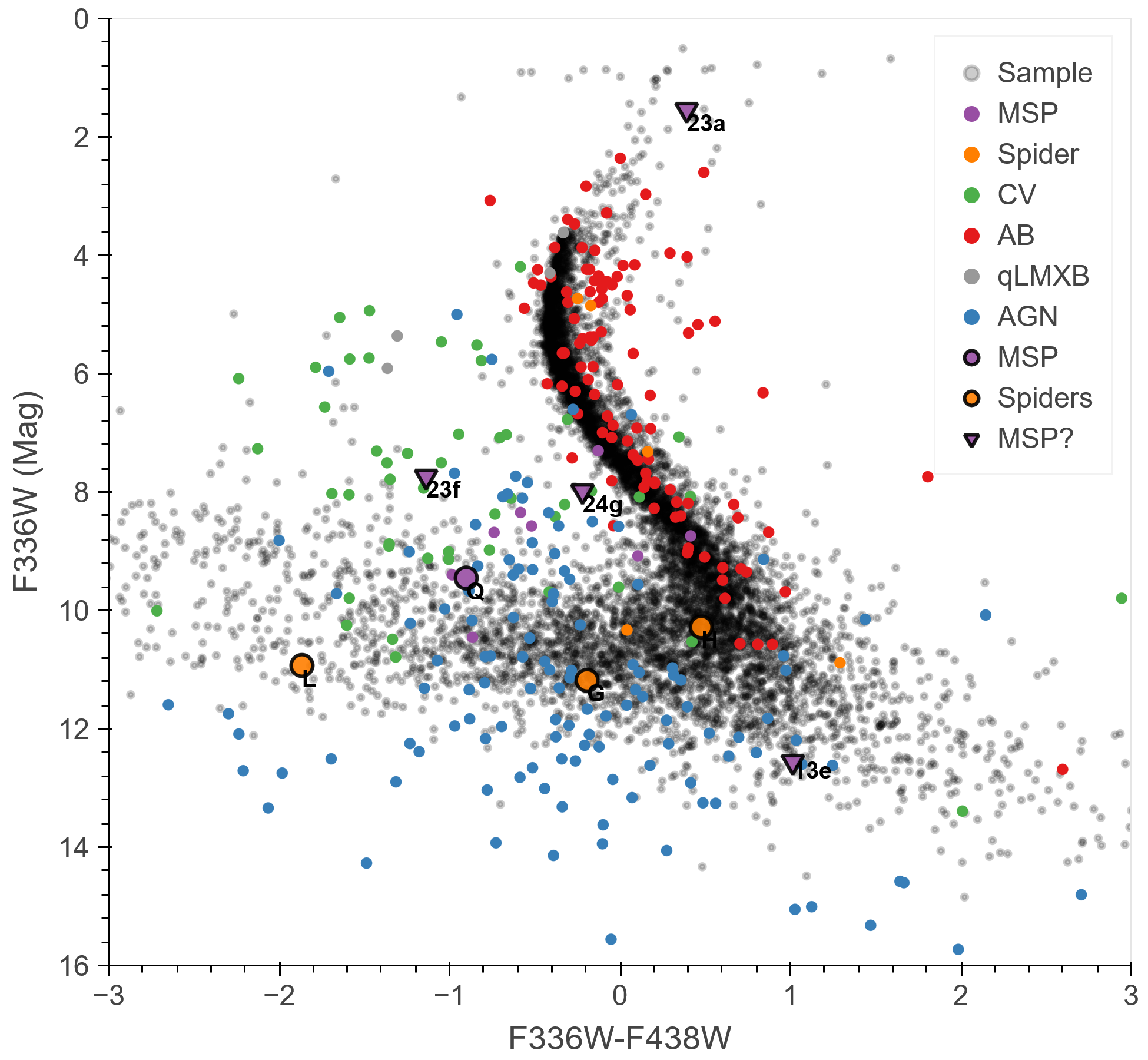}
    \caption{X-ray and optical properties of confirmed MSPs, spider MSPs, and MSP candidates. Black points in the background of the CMD are 10,000 randomly sampled stars in oMEGACat. The "luminosity" and "absolute magnitudes" of TD AGN are calculated as if they were at the distance of $\omega$ Cen, to enable comparison of their properties with other classes.}
    \label{fig:NGC5139_msps}
\end{figure*}

\begin{figure*}
    \centering
    \includegraphics[width=0.43\textwidth]{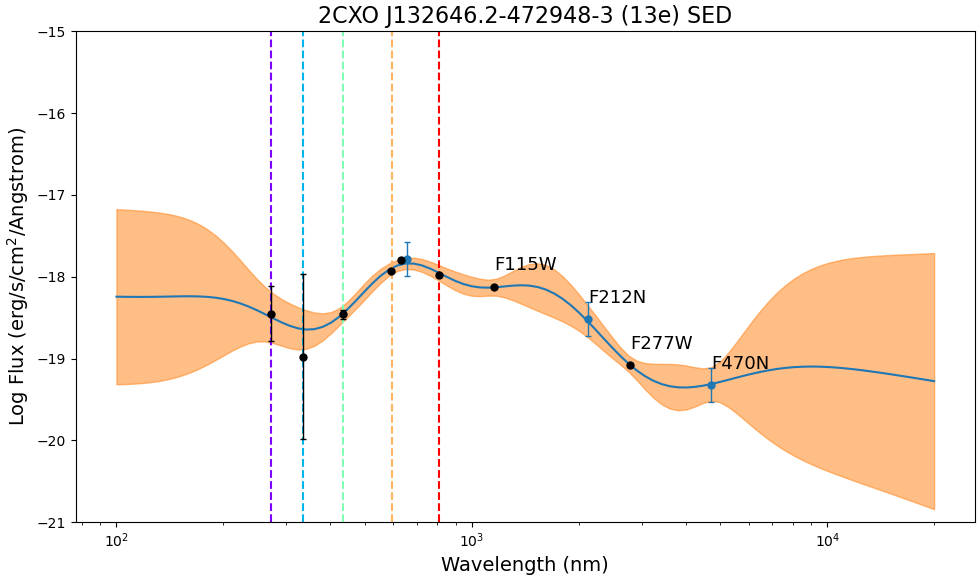}
    \includegraphics[width=0.43\textwidth]{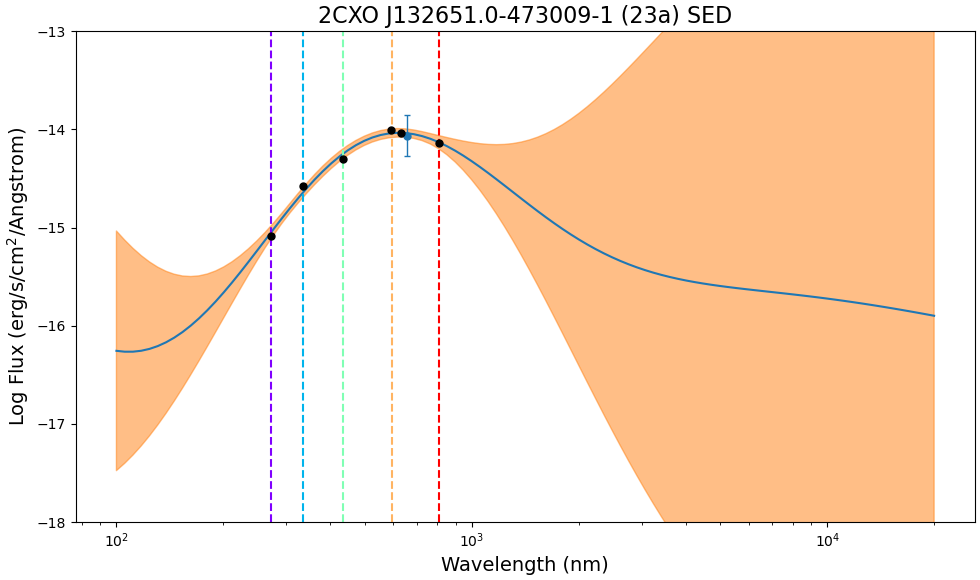}
    \includegraphics[width=0.43\textwidth]{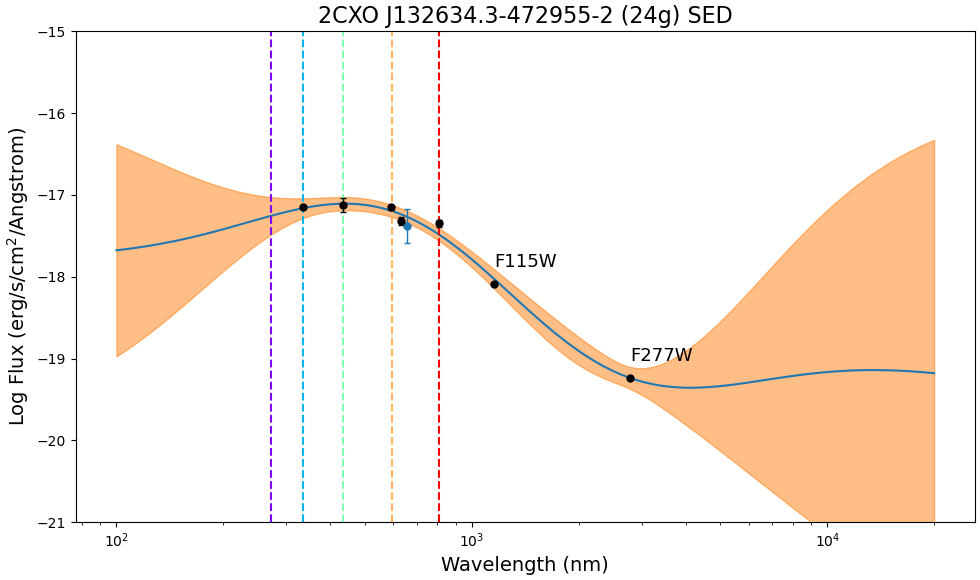}
    \includegraphics[width=0.43\textwidth]{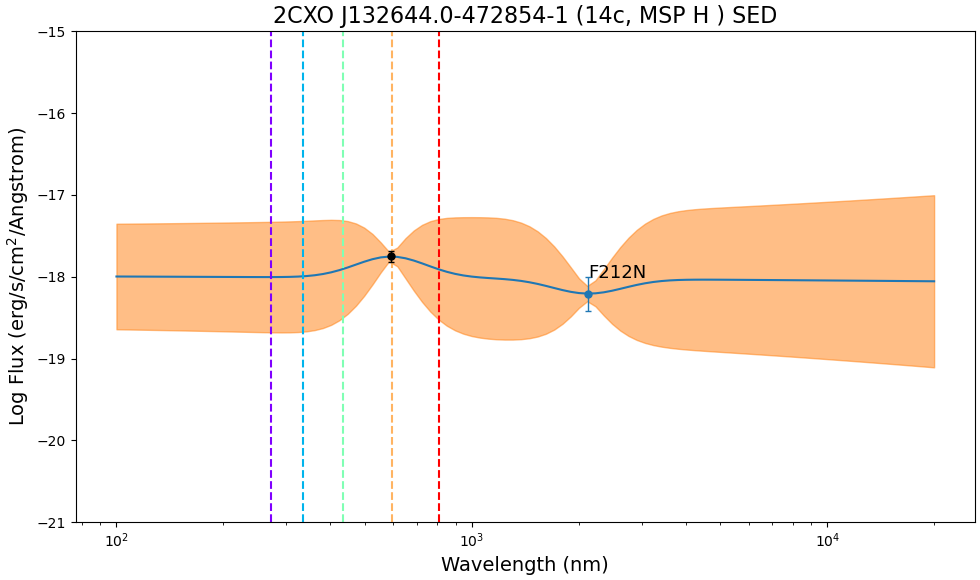}
    \includegraphics[width=0.43\textwidth]{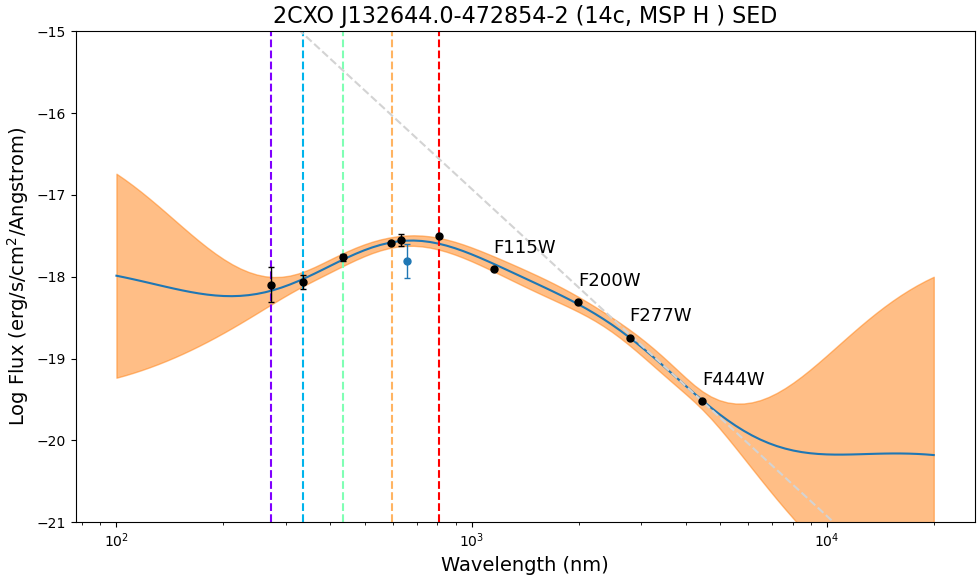}
    \includegraphics[width=0.43\textwidth]{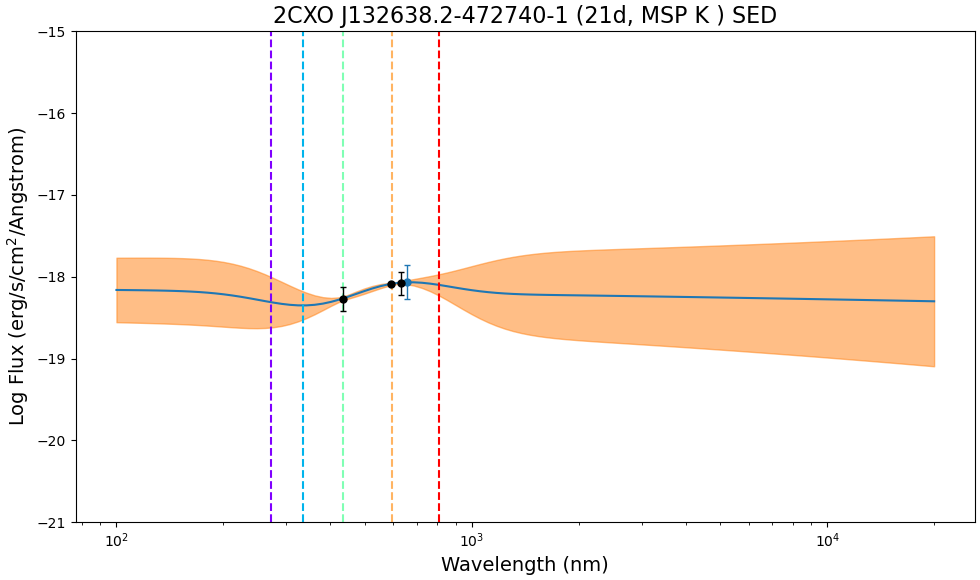}
    \includegraphics[width=0.43\textwidth]{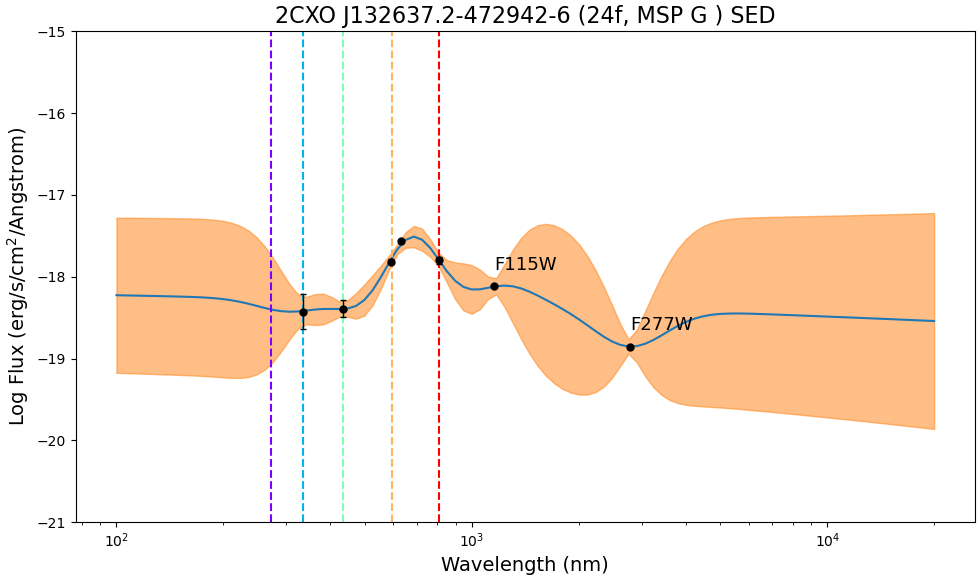}
    \includegraphics[width=0.43\textwidth]{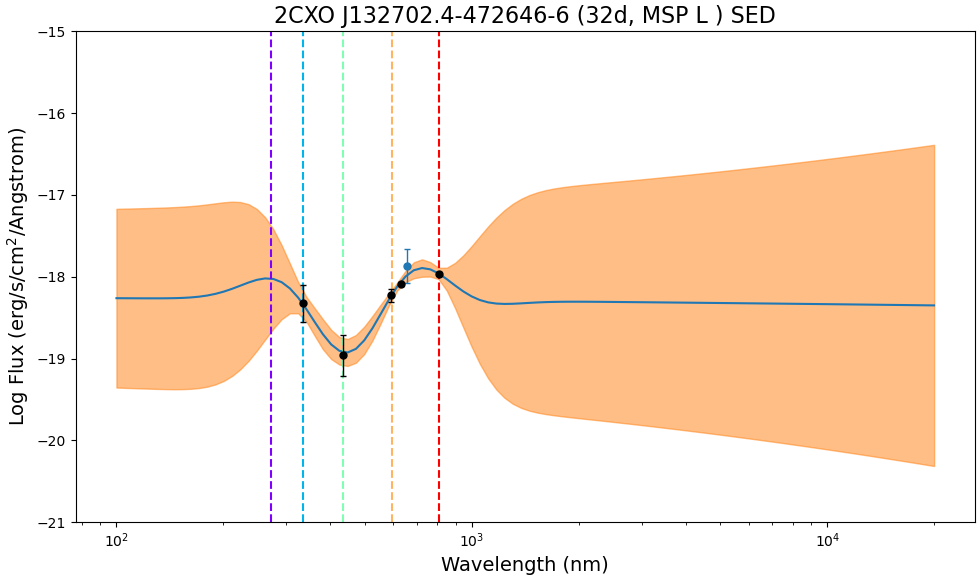}
    \includegraphics[width=0.43\textwidth]{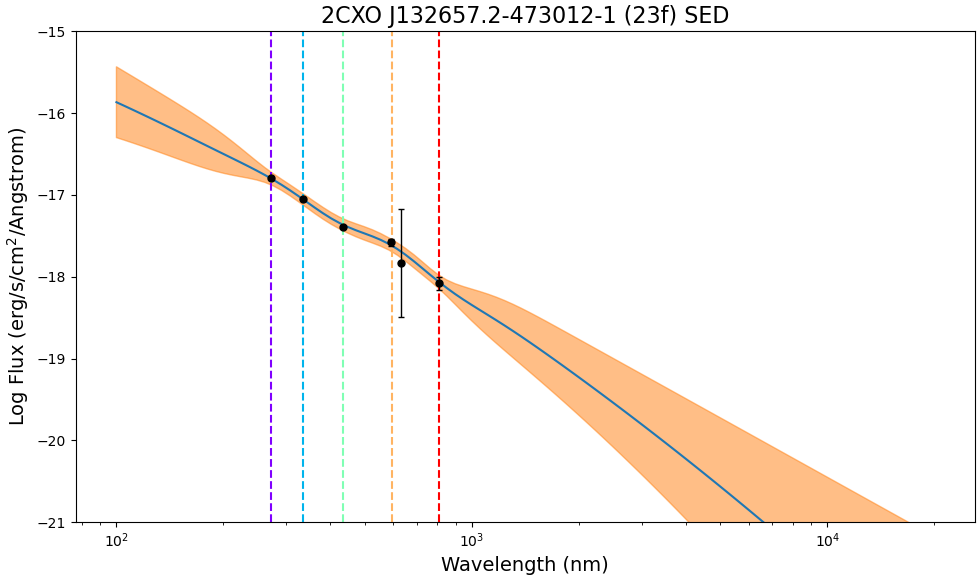}
    \includegraphics[width=0.43\textwidth]{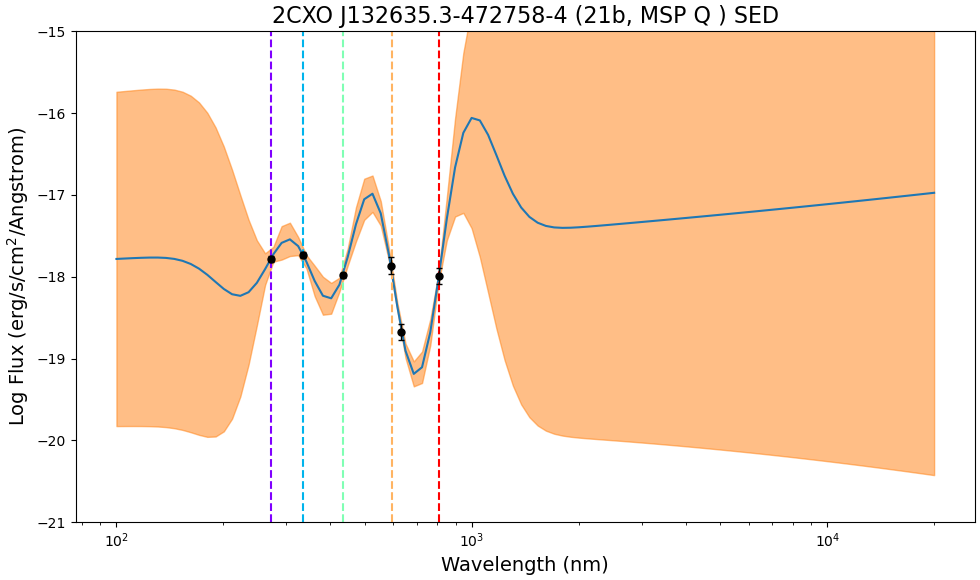}
    \caption{HST to JWST SEDs of possible counterparts to known spider MSPs and candidate MSPs, shown in same order as counterpart associations in Table \ref{table_msp}.}
    \label{fig:NGC5139_msp_seds}
\end{figure*}

We show the X-ray and optical properties of known and candidate MSPs and Spider MSPs in Figure \ref{fig:NGC5139_msps} and Table \ref{table_msp}. We additionally present the HST-JWST SEDs of the unusual sources that may be counterparts of the known spider MSPs and our candidate MSPs in Figure \ref{fig:NGC5139_msp_seds} 

\begin{table}
\scriptsize
\centering
\begin{tabular}{llllllllllllll}
\toprule
Assoc. Name & Lit. & MSP & Published & Manual & ML & P\_C & PU (\arcsec) & Sep. (\arcsec) & $L_{\rm CSC, b}$ & $HR_{\rm hms}$ & F275W & UV Color & F\_radio ($\mu$Jy) \\
\midrule
J132646.2-472948-3 & 13e &  &  & MSP? & AGN & 0.32 & 0.46 & 0.25 & 4.01e+30 & 0.26 & 25.10 & -1.16 & 28.05 \\
J132651.0-473009-1 & 23a &  &  & MSP? & AB & 0.91 & 0.35 & 0.15 & 9.86e+30 & 0.18 & 16.68 & 1.44 & 23.73 \\
J132637.4-473006\_ul & 24d &  &  & MSP? & MSP & 0.96 & 0.50 &  & 5.47e+30 & -0.49 &  &  & 21.64 \\
J132634.3-472955-2 & 24g &  &  & MSP? & CV & 0.66 & 0.34 & 0.16 & 2.76e+31 & 0.34 &  &  & 34.72 \\
J132650.6-472512\_ul & 32h &  &  & MSP? & MSP & 0.89 & 0.50 &  & 6.42e+30 & -0.23 &  &  & 34.08 \\
J132649.7-473147\_ul & 33k &  &  & MSP? & MSP & 0.81 & 0.39 &  & 1.86e+31 & 0.22 &  &  & 26.58 \\
J132649.5-472924\_ul & 13d & MSP B & BW & Spider & MSP & 0.89 & 0.72 &  & 7.39e+29 & -1.00 &  &  & 132.65 \\
J132644.0-472854-1 & 14c & MSP H & BW & Ctpt? & CV & 0.39 & 0.42 & 0.18 & 2.26e+30 & 0.03 &  &  & 35.59 \\
J132644.0-472854-2 & 14c & MSP H & BW & Ctpt? & AGN & 0.30 & 0.42 & 0.22 & 2.26e+30 & 0.03 & 24.21 & 0.23 & 35.59 \\
J132644.0-472854\_ul & 14c & MSP H & BW & Spider & MSP & 0.88 & 0.42 &  & 2.26e+30 & 0.03 &  &  & 35.59 \\
J132638.2-472740-1 & 21d & MSP K & BW & Ctpt? & AGN & 0.46 & 0.34 & 0.09 & 1.14e+31 & -0.07 &  &  & 44.55 \\
J132638.2-472740\_ul & 21d & MSP K & BW & Spider & MSP & 0.87 & 0.34 &  & 1.14e+31 & -0.07 &  &  & 44.55 \\
J132637.2-472942-6 & 24f & MSP G & BW & Ctpt? & MSP & 0.33 & 0.42 & 0.41 & 5.61e+30 & -0.38 &  &  & 100.78 \\
J132637.2-472942\_ul & 24f & MSP G & BW & Spider & MSP & 0.94 & 0.42 &  & 5.61e+30 & -0.38 &  &  & 100.78 \\
J132702.4-472646-6 & 32d & MSP L & BW & Ctpt? & CV & 0.41 & 0.47 & 0.53 & 3.15e+30 & -0.56 &  &  & 31.52 \\
J132702.4-472646\_ul & 32d & MSP L & BW & Spider & MSP & 0.91 & 0.47 &  & 3.15e+30 & -0.56 &  &  & 31.52 \\
J132642.6-472723\_ul & 11f & MSP E & Isolated & MSP & MSP & 0.91 & 0.82 &  & 1.33e+30 & -1.00 &  &  & 80.55 \\
J132655.2-473011\_ul & 23g & MSP C & Isolated & MSP & MSP & 0.90 & 0.63 &  & 2.81e+30 & -0.17 &  &  & 50.47 \\
J132657.2-473012-1 & 23f & MSP 23f & MSP? & MSP? & CV & 0.62 & 0.83 & 0.21 & 1.84e+30 & -0.33 & 20.94 & -0.49 & 17.46 \\
J132635.3-472758-4 & 21b & MSP Q & WD & MSP & MSP & 0.48 & 1.06 & 0.28 & 1.20e+30 & -0.00 & 23.43 & 0.28 & 36.11 \\
\bottomrule
\end{tabular}
\caption{Classification and MW properties of previously known and our confident and candidate MSP and Spider-type MSPs. Columns same as in Table \ref{table_pm_agn}, except with F275W magnitude. "UV Color": F275W-F336W color. }
\label{table_msp}
\end{table}

\subsection{qLMXB}

The only known qLMXB in Omega Centauri, which does not have a known counterpart, 2E 3049 (44e), is classified correctly in the null and upper limit associations, while the other associations ($>0.2\arcsec$ separation) are classified as CV. In the images, there is a faint source undetected in oMEGACat, close to the first counterpart, that appears to have a F336W excess. If this source is the real counterpart, this implies the mass of the companion is $\lesssim 0.1 M_\odot$. There is no MeerKAT counterpart.

No other X-ray source seems likely to be a qLMXB. The X-ray sources sufficiently bright and soft to be qLMXBs are all crossmatched to foreground stars, discussed in Section \ref{foreground}.

\subsection{Lower Mass ABs}

\begin{figure*}
    \centering
    \includegraphics[width=0.45\textwidth]{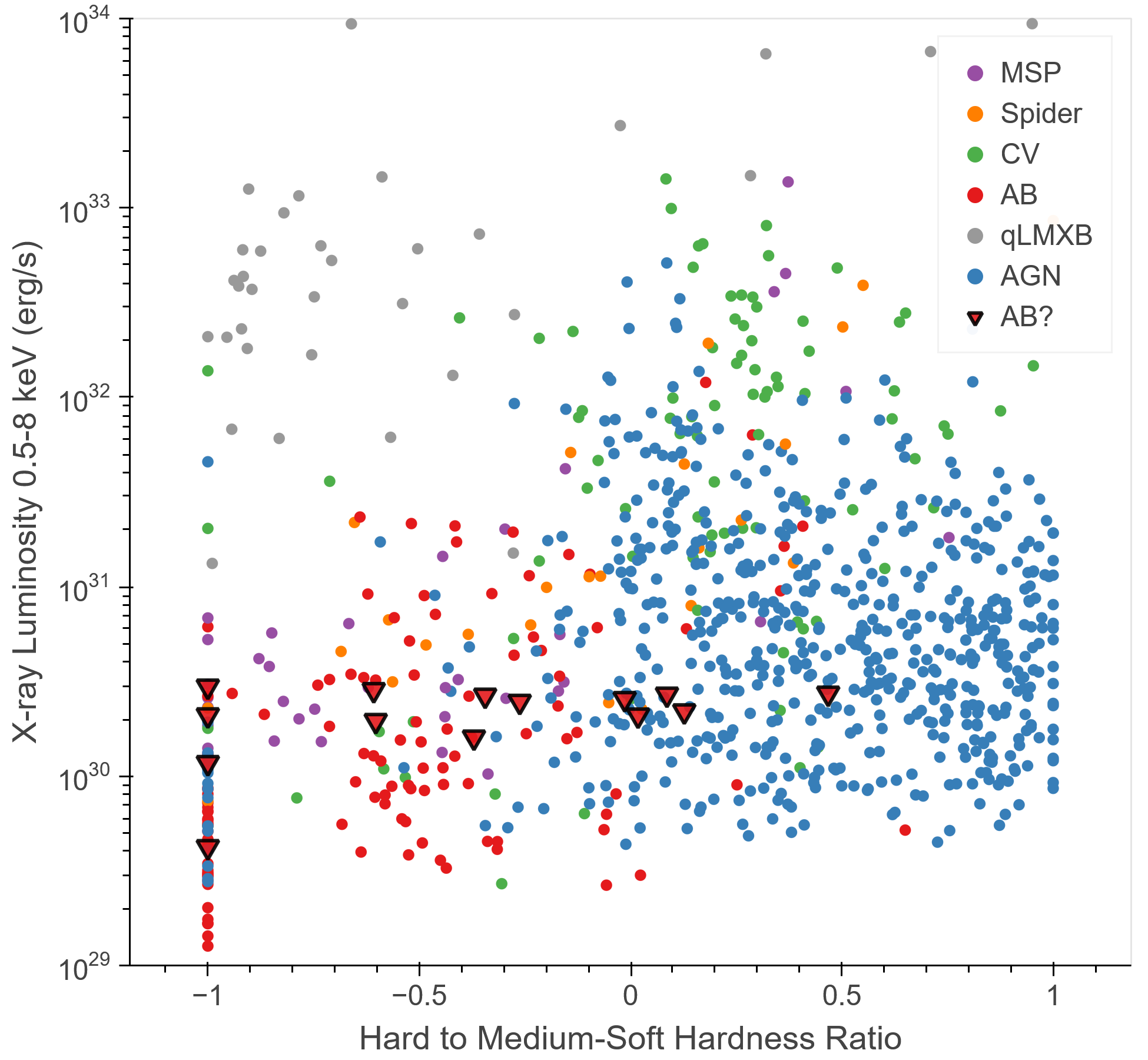}
    \includegraphics[width=0.45\textwidth]{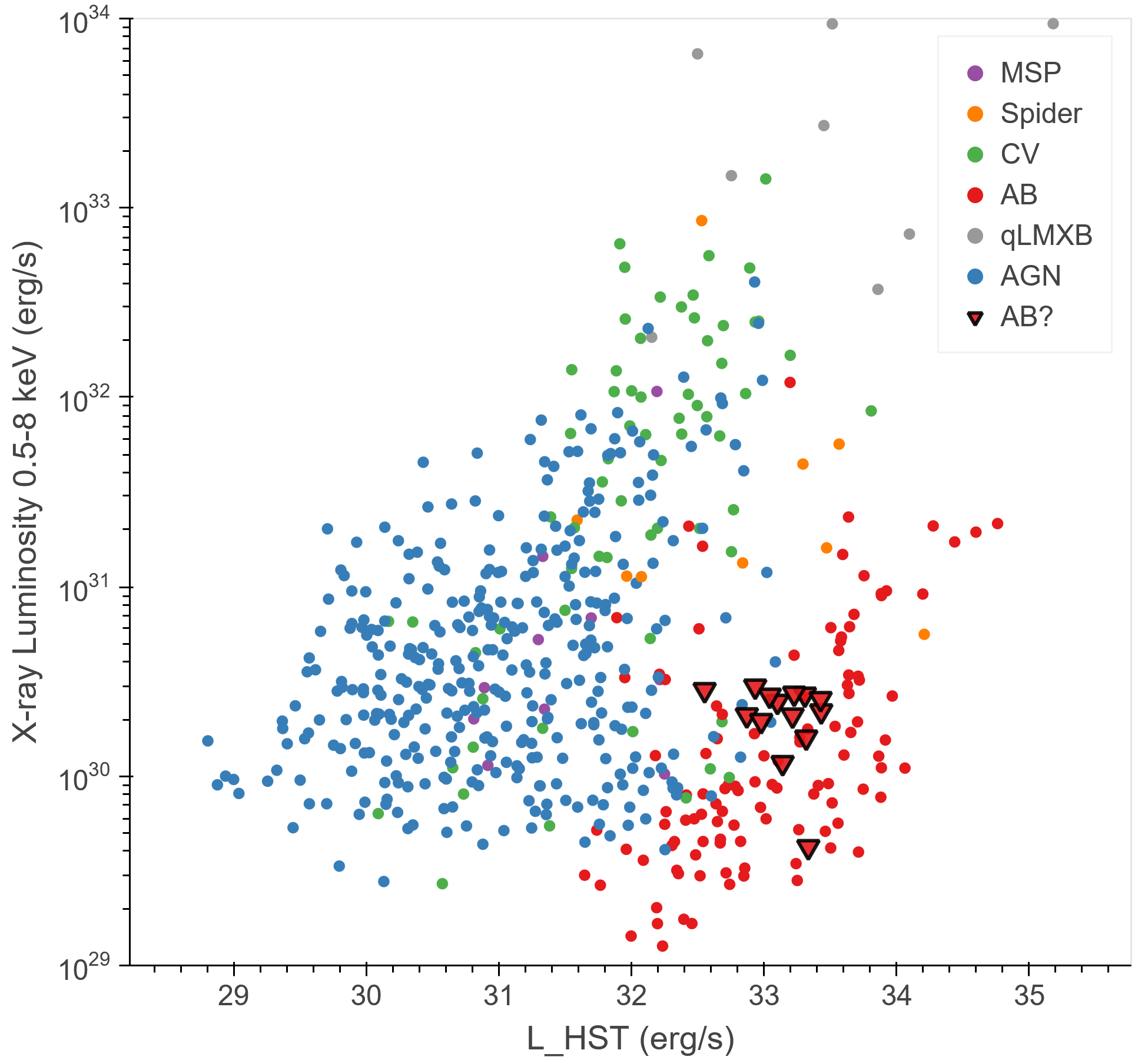}
    \includegraphics[width=0.45\textwidth]{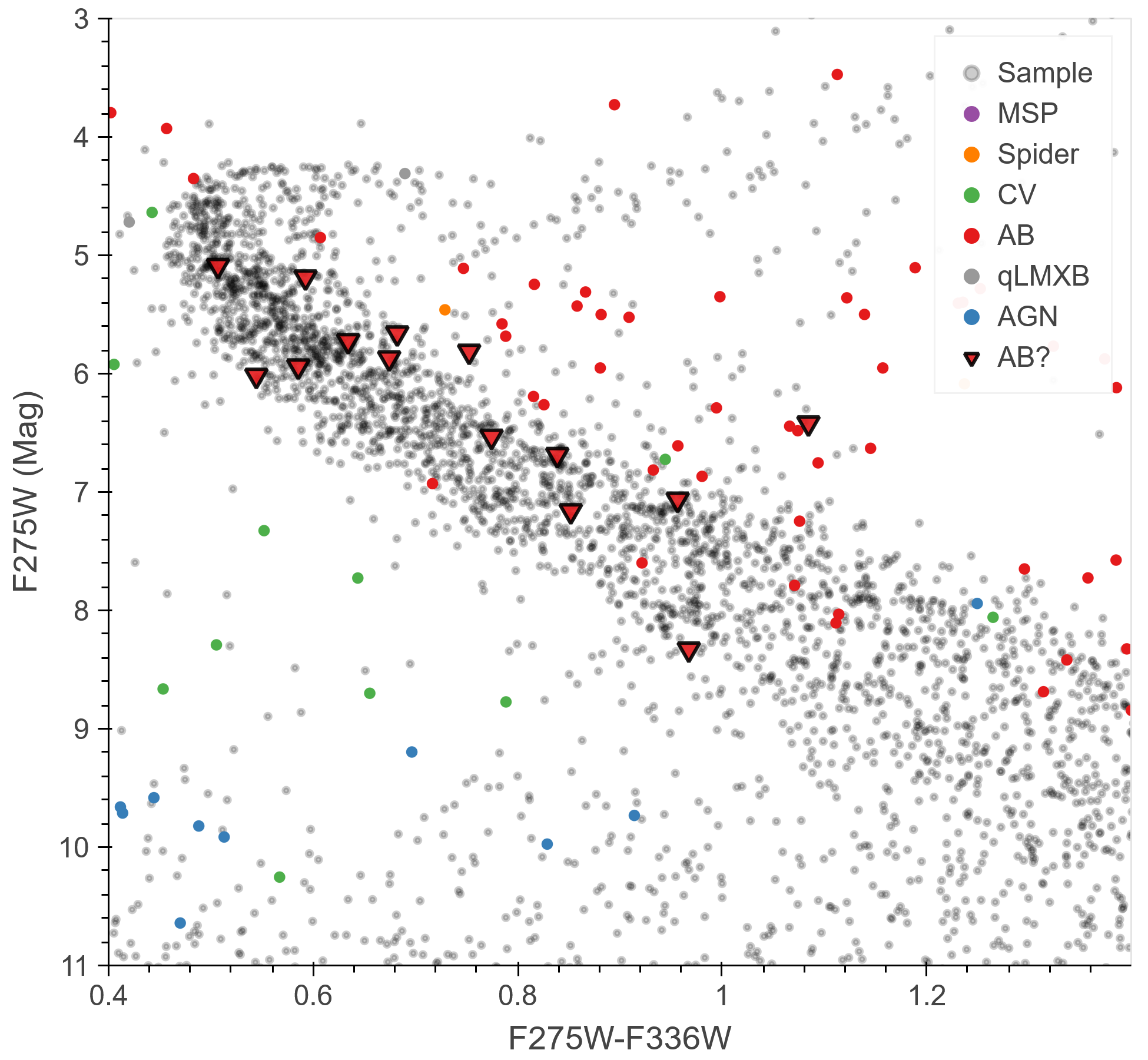}
    \includegraphics[width=0.45\textwidth]{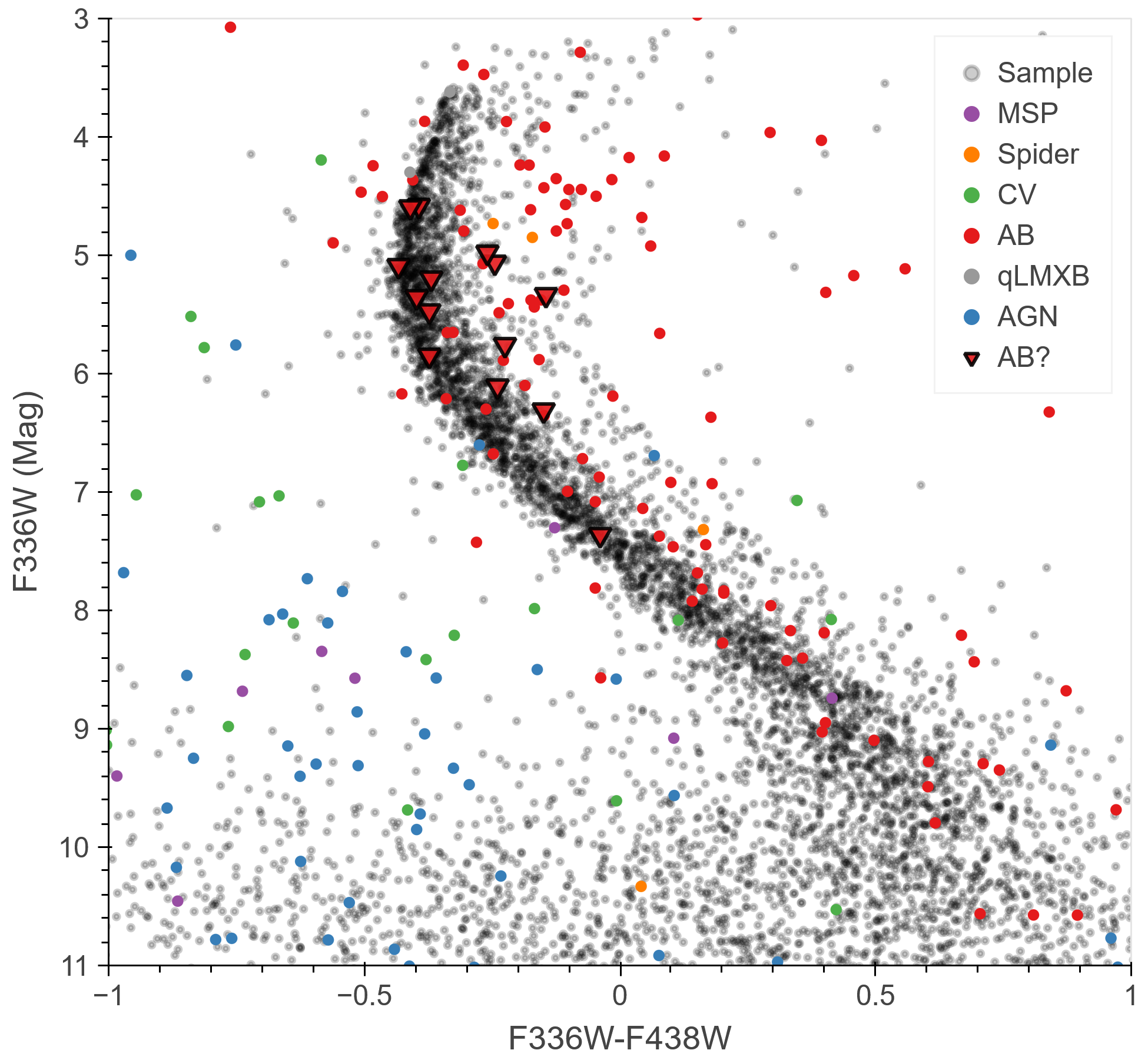}
    \includegraphics[width=0.45\textwidth]{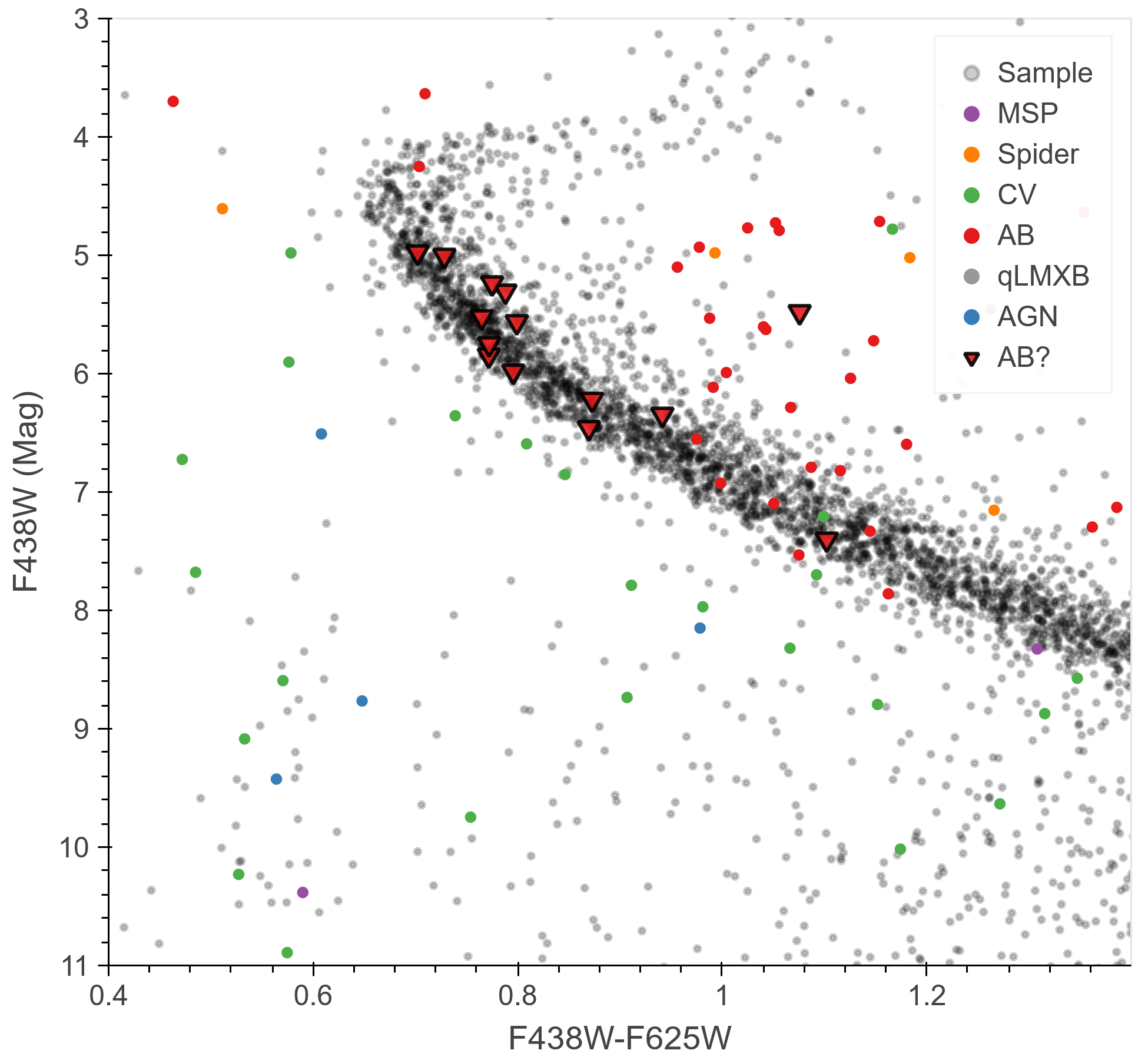}
    \includegraphics[width=0.45\textwidth]{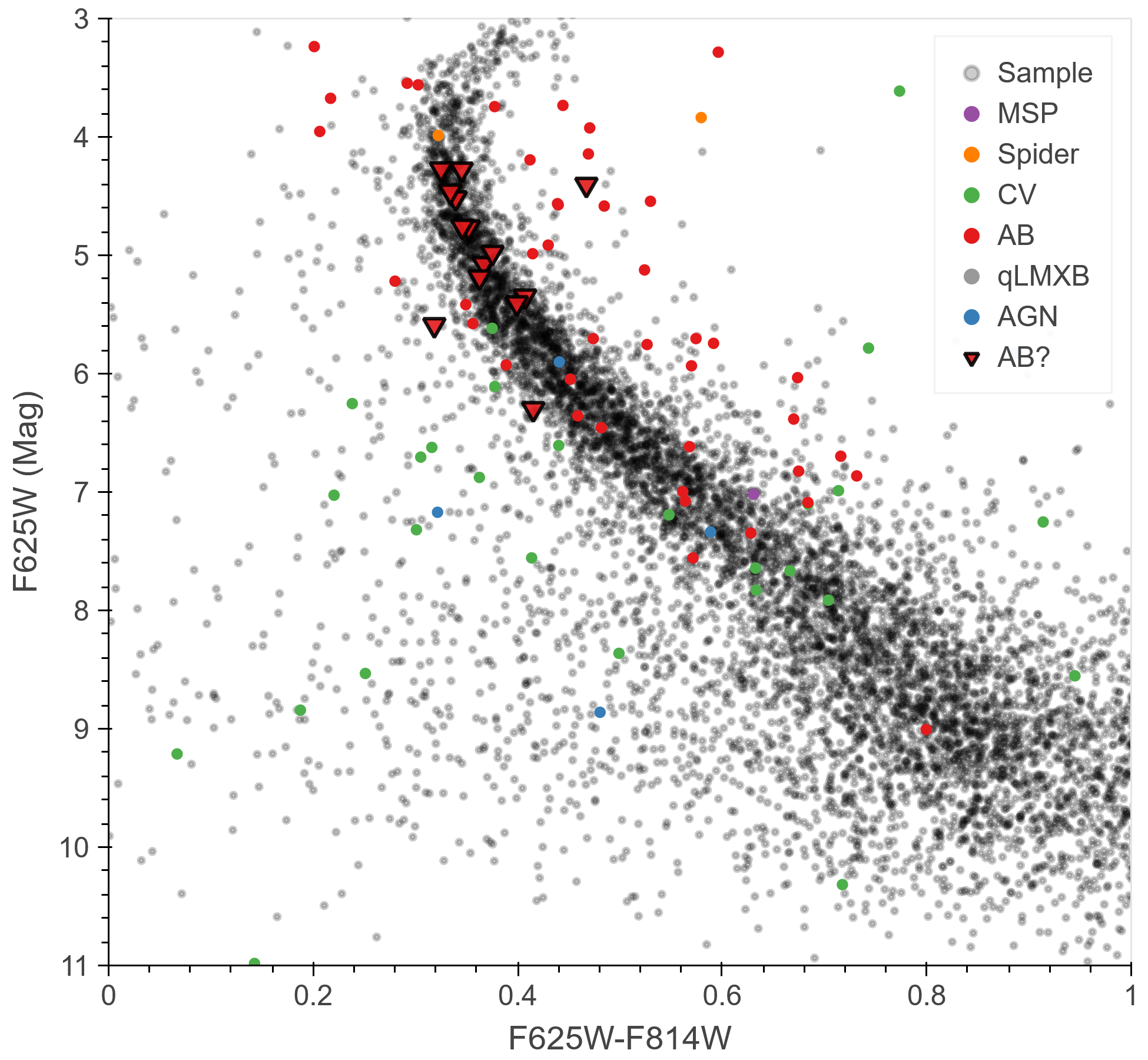}
    \caption{X-ray and optical properties of candidate ABs. Black points in the background of the CMD are 10,000 randomly sampled stars in oMEGACat. The "luminosity" and "absolute magnitudes" of TD AGN are calculated as if they were at the distance of $\omega$ Cen, to enable comparison of their properties with other classes.}
    \label{fig:NGC5139_abs}
\end{figure*}

In panel 7 of Figure \ref{fig:csc_omegacat_separation_density}, 21 CSC sources are tested for an excess of MS cluster member stars, to reveal a possible population of non-giant ABs. These 21 sources were selected by filtering out confident or candidate associations of other classes discussed above, by selecting sources soft in at least one hardness ratio ($\text{HR}_{\text{ms}}\leq0, \text{HR}_{\text{hm}}\leq0.5, \text{HR}_{\text{hms}}\leq0$), and by selecting sources not matched to a MeerKAT counterpart. Even though some ABs in the galactic field are detected in the radio, they are much fainter compared to MSPs and AGNs, and are undetectable at the distance of $\omega$ Cen. An excess of 9 optical sources is seen within $2\sigma$ PU. We manually select some candidate ABs based on MUWCLASS probabilities, X-ray vs optical luminosities, and source positions in the optical and X-ray images. Their X-ray and optical properties are shown in Figure \ref{fig:NGC5139_abs} and Table \ref{table_ab_ms}. Since we did not select these candidates based on their positions on the optical CMDs, this information can be used to independently check our selection.

These candidates fall within the brighter end of the MS, with magnitudes in F814W $\lesssim 19.5$. On the bluer CMDs, around half of these candidates lie close to the top of, or above the MS, indicating their binarity. The fraction is higher than what chance coincidence would suggest, lending support to their AB classification. However, on the redder CMDs, these sources are much more difficult to separate from the MS. The one exception is source 11e, which is located to the red of the MS on all CMDs, suggesting a subgiant AB. Its X-ray lightcurve shows evidence of flaring, as most counts in observation 13726 are clustered in a 10 ks period. 

Given the multiple stellar populations in $\omega$ Cen, the binarity of most of these candidates is hard to confidently determine from the CMDs alone. Precise spectroscopic radial velocity measurements of the orbital periods of these fainter stars are needed to definitively uncover the population of fainter ABs in this cluster.

\begin{table}
\footnotesize
\centering
\begin{tabular}{lllllllllllll}
\toprule
Assoc. Name & Lit. & Manual & Pub. & ML & P\_C & PU (\arcsec) & Sep. (\arcsec) & $L_{\rm CSC, b}$ & $HR_{\rm hms}$ & F814W & V-I & MH \\
\midrule
J132625.4-472742-5 &  & AB? &  & AB & 0.88 & 0.79 & 0.62 & 2.98e+30 & -1.00 & 18.70 & 0.57 & -1.49 \\
J132651.2-472931-2 &  & AB? &  & AB & 0.87 & 0.30 & 0.21 & 1.96e+30 & -0.60 & 18.64 & 0.55 & -1.41 \\
J132641.5-472832-2 & 11a & AB? &  & AB & 0.88 & 0.27 & 0.27 & 2.20e+30 & 0.13 & 17.64 & 0.47 & -1.73 \\
J132643.2-472816-5 & 11e & AB? &  & AB & 0.97 & 0.24 & 0.32 & 1.60e+30 & -0.37 & 17.63 & 0.67 & -2.87 \\
J132637.8-472715-3 & 21f & AB? &  & AB & 0.87 & 1.88 & 0.34 & 2.10e+30 & -1.00 & 18.97 & 0.41 & -1.43 \\
J132654.8-472832-4 & 22l & AB? &  & AB & 0.90 & 0.20 & 0.51 & 2.67e+30 & -0.34 & 18.52 & 0.50 & -0.69 \\
J132658.2-472957-5 & 23e & AB? &  & AB & 0.82 & 0.34 & 0.49 & 2.09e+30 & 0.02 & 18.11 & 0.51 & -1.40 \\
J132634.2-472641-1 & 31g & AB? &  & AB & 0.93 & 0.55 & 0.21 & 2.70e+30 & 0.09 & 17.88 & 0.49 & -1.34 \\
J132657.7-473021-3 & 33p & AB? &  & AB & 0.91 & 0.39 & 0.36 & 4.23e+29 & -1.00 & 17.82 & 0.48 & -1.12 \\
J132631.6-473105-1 & 34f & AB? &  & AB & 0.81 & 0.58 & 0.11 & 2.73e+30 & 0.47 & 18.11 & 0.51 & -1.49 \\
J132631.4-472932-1 & 34h & AB? &  & AB & 0.75 & 0.35 & 0.29 & 2.85e+30 & -0.61 & 19.58 & 0.59 & -1.46 \\
J132623.9-472730-2 & 41i & AB? &  & AB & 0.78 & 0.58 & 0.28 & 1.17e+30 & -1.00 & 18.30 & 0.54 & -1.62 \\
J132624.3-472646-2 & 41j & AB? &  & AB & 0.89 & 0.58 & 0.39 & 2.56e+30 & -0.01 & 17.63 & 0.50 & -2.04 \\
J132707.9-472944-1 & 43a & AB? &  & AB & 0.76 & 0.34 & 0.08 & 2.47e+30 & -0.26 & 18.40 & 0.52 & -1.54 \\
\bottomrule
\end{tabular}
\caption{Classification and MW properties of our candidate main sequence ABs associations. Columns same as in Table \ref{table_pm_agn}.}
\label{table_ab_ms}
\end{table}

\subsection{Other CSC Sources with MeerKAT Counterparts}
\label{meerkat_counterparts}

Aside from the AGNs and MSPs discussed above, there are 64 other CSC sources matched with MeerKAT sources within $3\sigma$ combined CSC and MeerKAT PU. Since the chance coincidence rate is $\sim 2\%$, we believe almost all of these MeerKAT associations are real. Most of these sources are relatively hard, with medium-soft hardness ratio greater than -0.5. Therefore, most of these are likely unidentified AGN, as MSPs and spider MSPs, the only other classes with detectable radio flux, are typically soft in the X-rays. Out of these 64 sources, 15 have oMEGACat coverage, and of those, none have optical counterparts with PM consistent with background galaxies, indicating that they may be analogous to the TD AGN without optical counterparts.

\begin{figure*}
    \centering
    \includegraphics[width=0.45\textwidth]{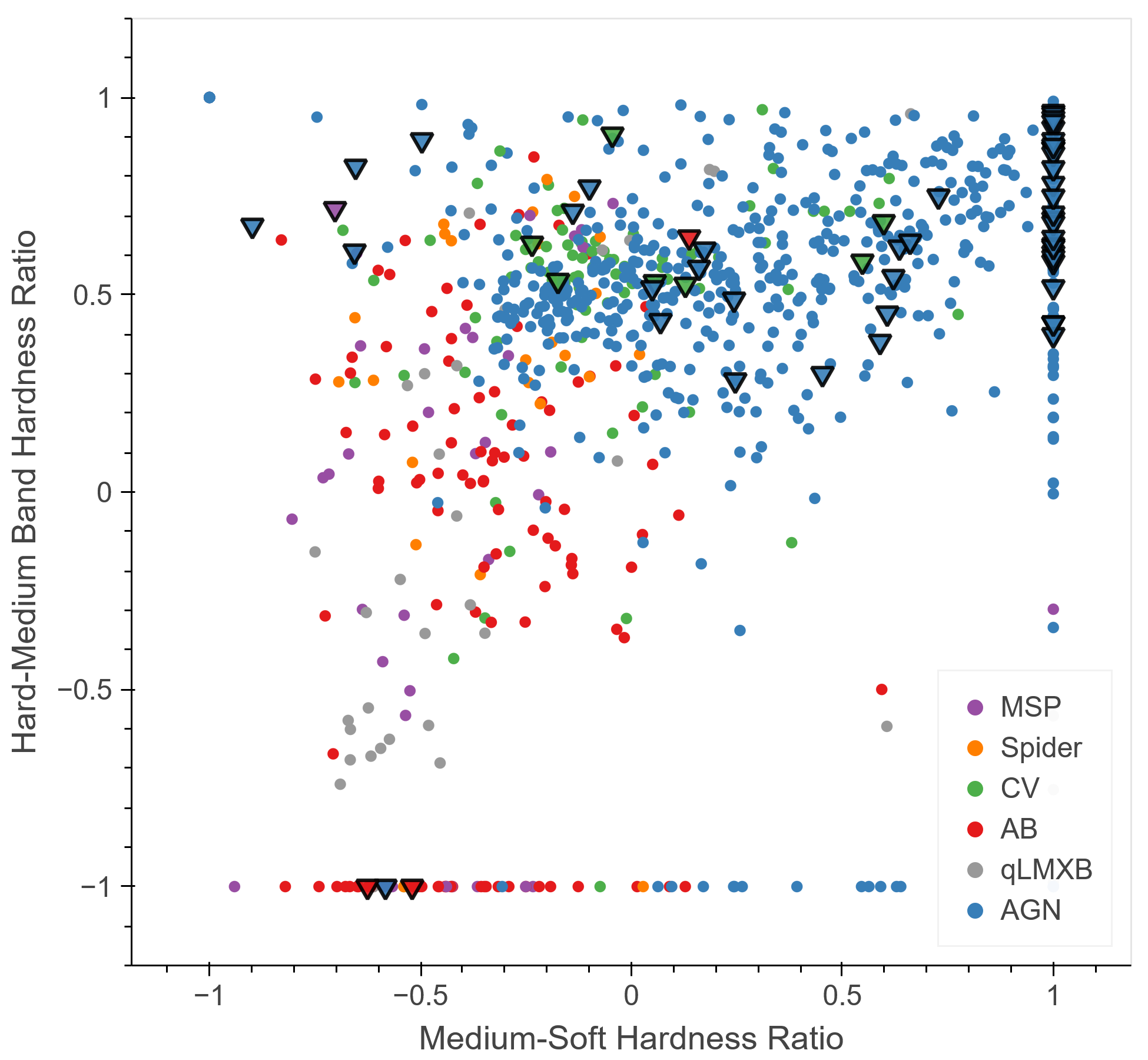}
    \includegraphics[width=0.45\textwidth]{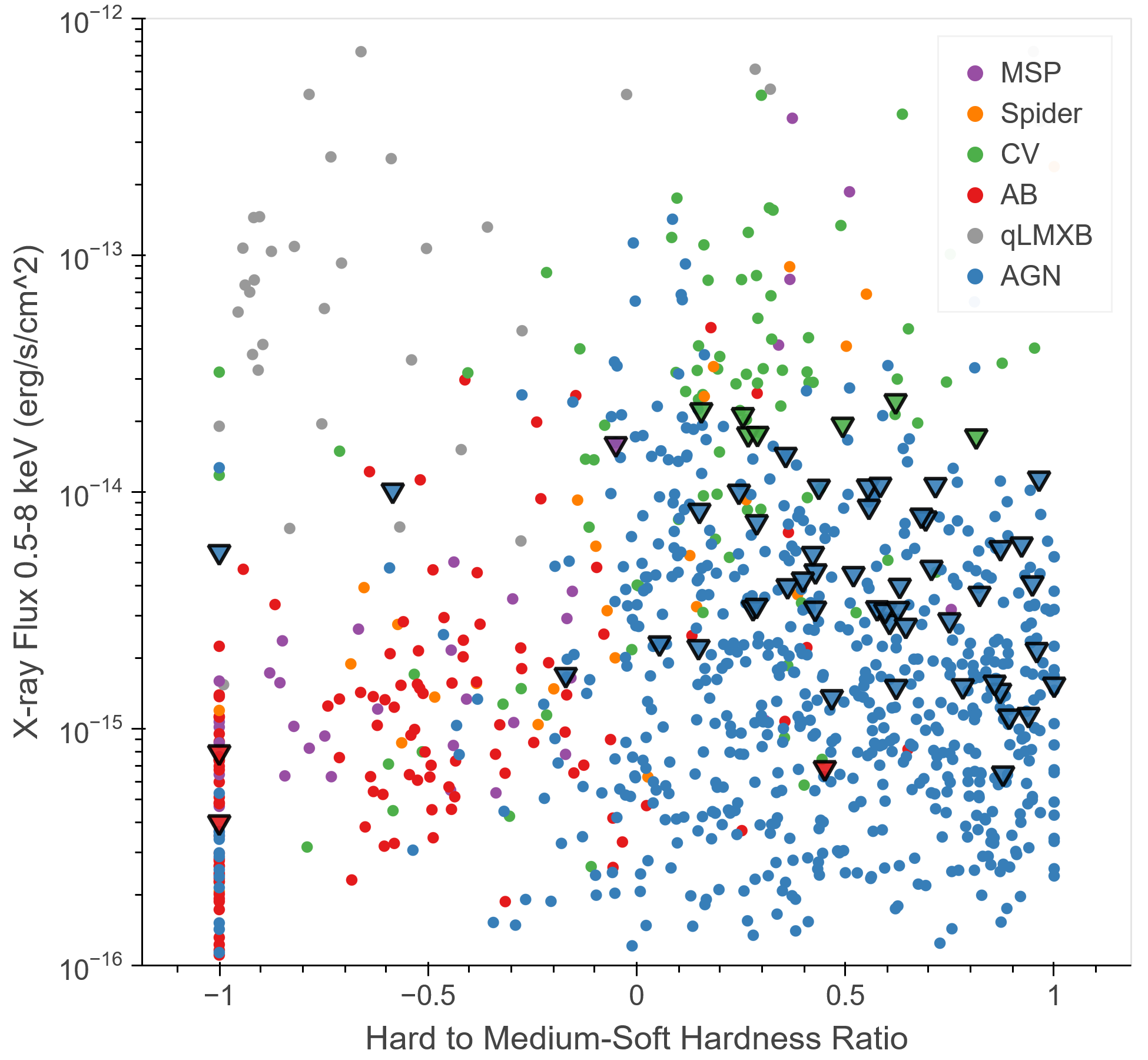}
    \caption{X-ray properties of candidate AGNs. Marker colors show their MUWCLASS classifications. }
    \label{fig:candidate_agns}
\end{figure*}

In Figure \ref{fig:candidate_agns}, we present their X-ray properties along with their MUWCLASS classification, which only relies on their X-ray properties. Most sources are classified as AGN, as expected, with some brighter ones classified as CV, and some softer ones classified as AB. 

Sources 2CXO J132635.1-472051 and 2CXO J132722.8-471813 are the two bright, soft sources (see panel 2 of Figure \ref{fig:candidate_agns}), in the region of the HR-flux diagram sparsely populated with TD sources of most classes, except qLMXBs. The hard-medium HR of 2CXO J132635.1-472051 is harder and more typical of a TD AGN, while 2CXO J132722.8-471813 lacks detection in the hard band. Both these sources are outside the field of view of HST observations, and, for this reason, were  not previously classified. Their X-ray properties suggest they may not be typical AGN, and the lack of matching foreground stars in Gaia, and the MeerKAT counterpart disfavors a foreground star classification. If these are cluster members, then it is possible that they are MSP or qLMXB systems outside the half-light radius, although their presumed luminosity is lower than most TD qLMXBs, while being higher than most TD MSPs.

Source 54j is soft. Near the CSC source position there is a hint of an extended source with PM consistent with galaxies, supporting the AGN classification. Although it is undetected in oMEGACat, this source contaminates nearby faint stars such as the first and fourth counterparts. 

Source 43l is a hard source with 11 photons. It was classified as an MSP in the UL-association because a few TD MSPs also lack soft-band photons. It has two faint, MS candidate counterparts with no obvious distinguishing features. The MeerKAT counterpart is relatively bright at $\SI{37}{\micro\jansky}$. It is thus likely to be an optically faint but radio-loud AGN, although an interpretation as an MSP with a faint optical counterpart is also possible.

\subsection{CSC Sources Not Matched to Counterparts}

There are 64 CSC sources remaining without candidate or confident optical counterparts assigned and also without MeerKAT counterparts. These include 38 sources without oMEGACat coverage, and 26 sources with coverage, but where a counterpart determination could not be made. The last panel in Figure \ref{fig:csc_omegacat_separation_density} shows the HST source density around 22 of these sources with PU$<0.5\arcsec$. Aside from one narrow peak, there is no significant excess, indicating most of the optical counterparts of these sources are not detectable in oMEGACat.

\begin{figure*}
    \centering
    \includegraphics[width=0.45\textwidth]{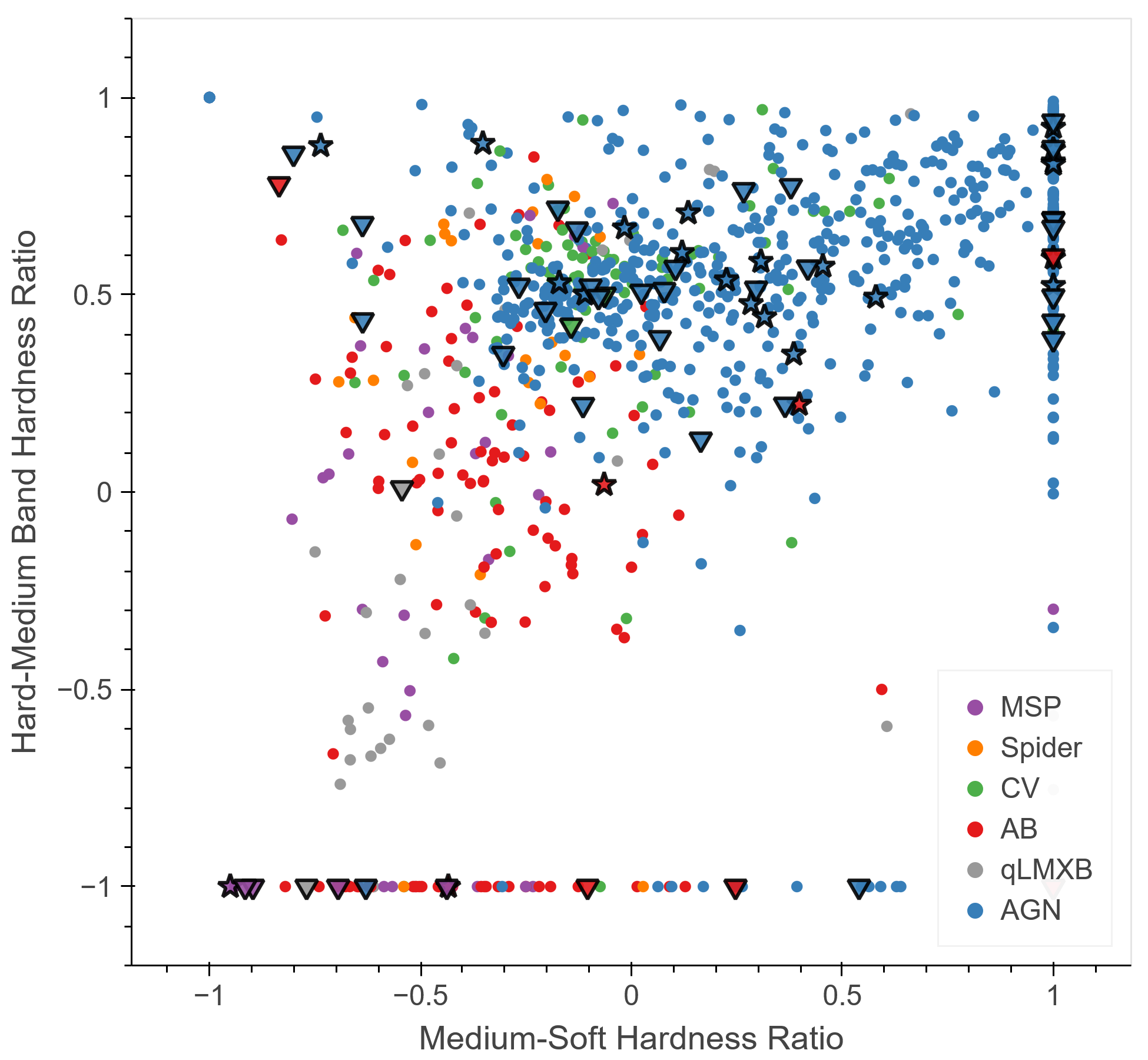}
    \includegraphics[width=0.45\textwidth]{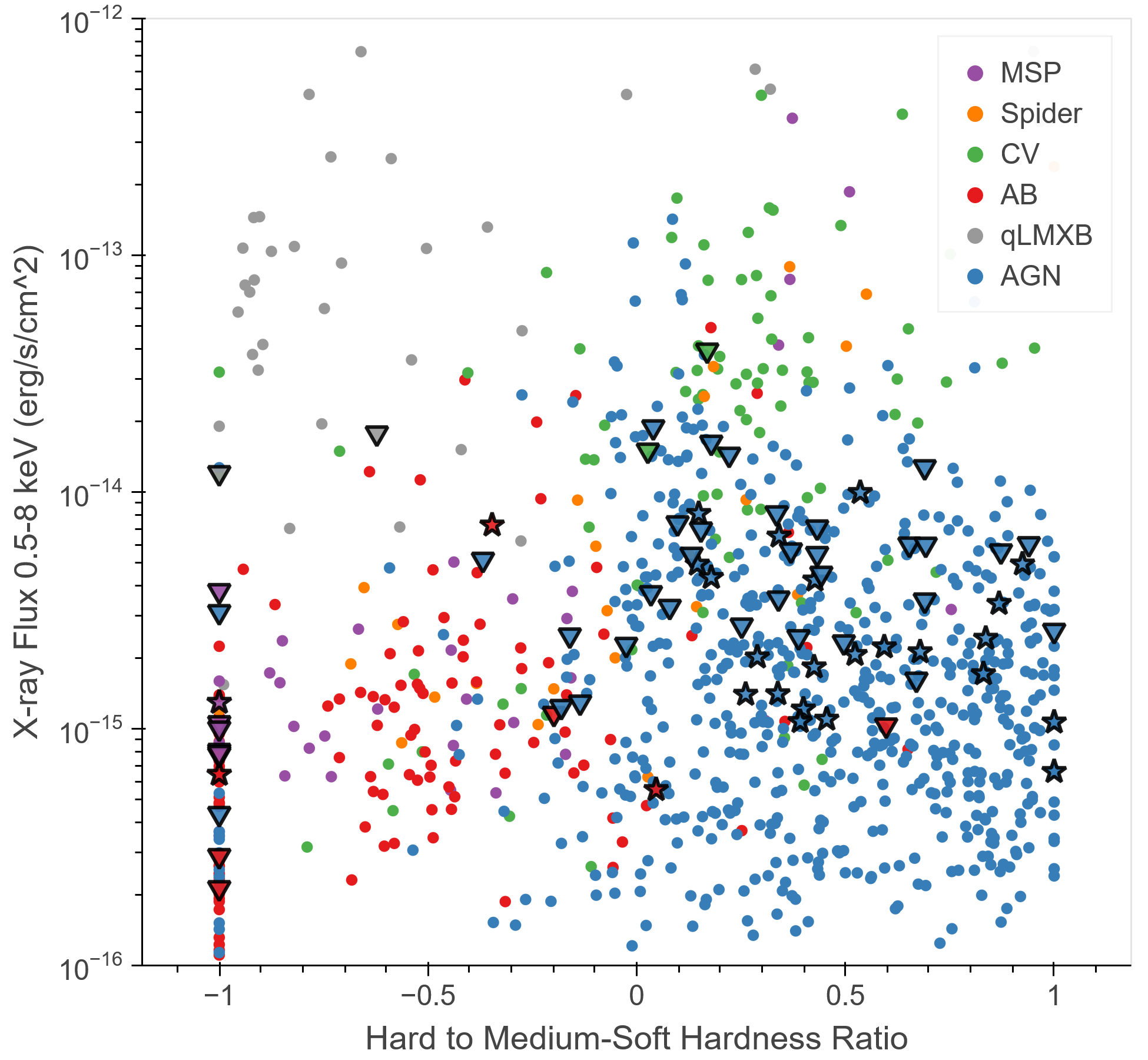}
    \caption{X-ray properties of other CSC sources not matched to a counterpart. Star markers are sources inside oMEGACat field of coverage, inverted triangle markers are sources outside. }
    \label{fig:other_csc}
\end{figure*}

\begin{figure*}
    \centering
    \includegraphics[width=0.43\textwidth]{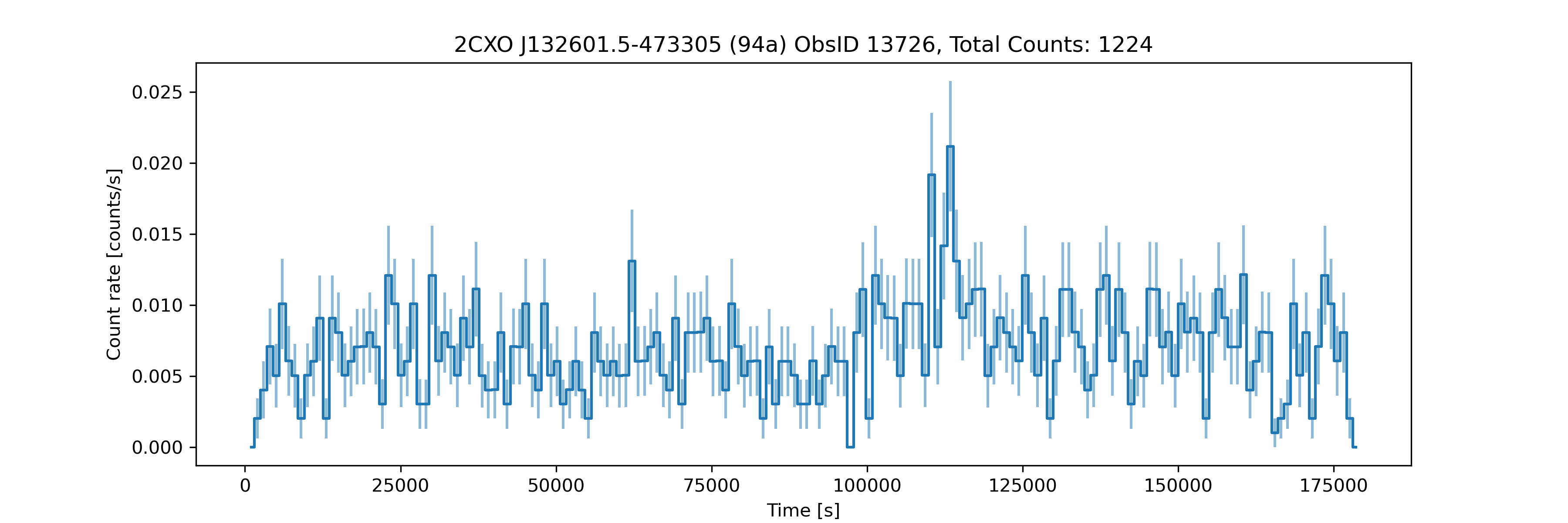}
    \includegraphics[width=0.43\textwidth]{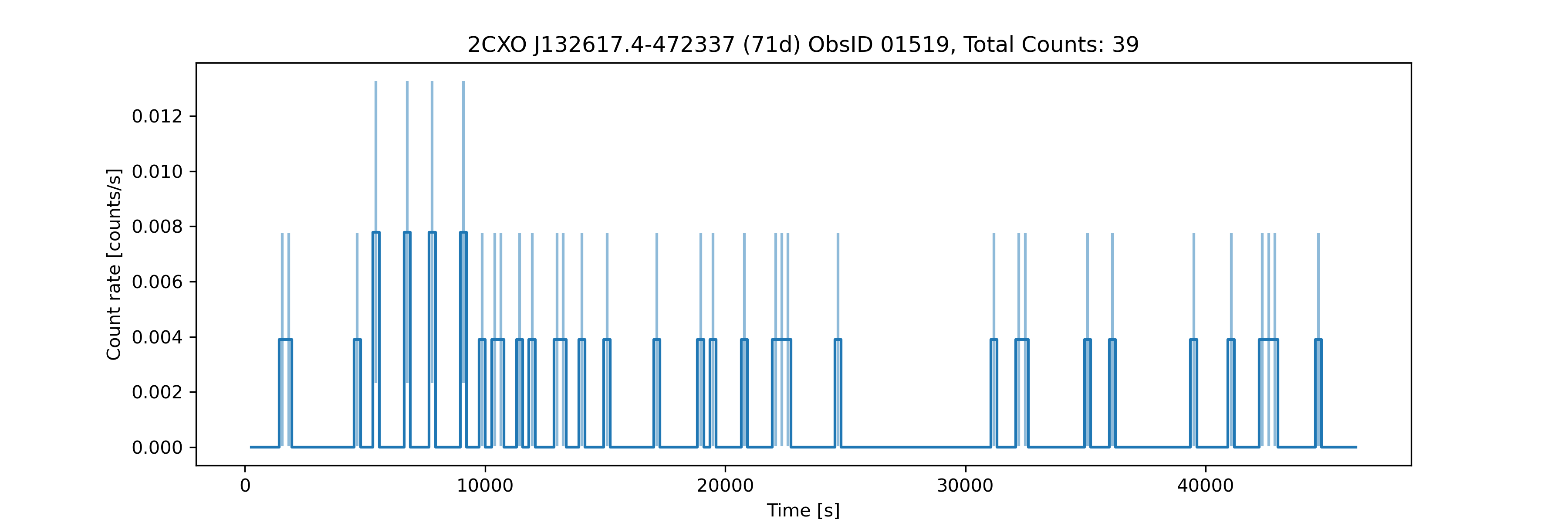}
    \includegraphics[width=0.43\textwidth]{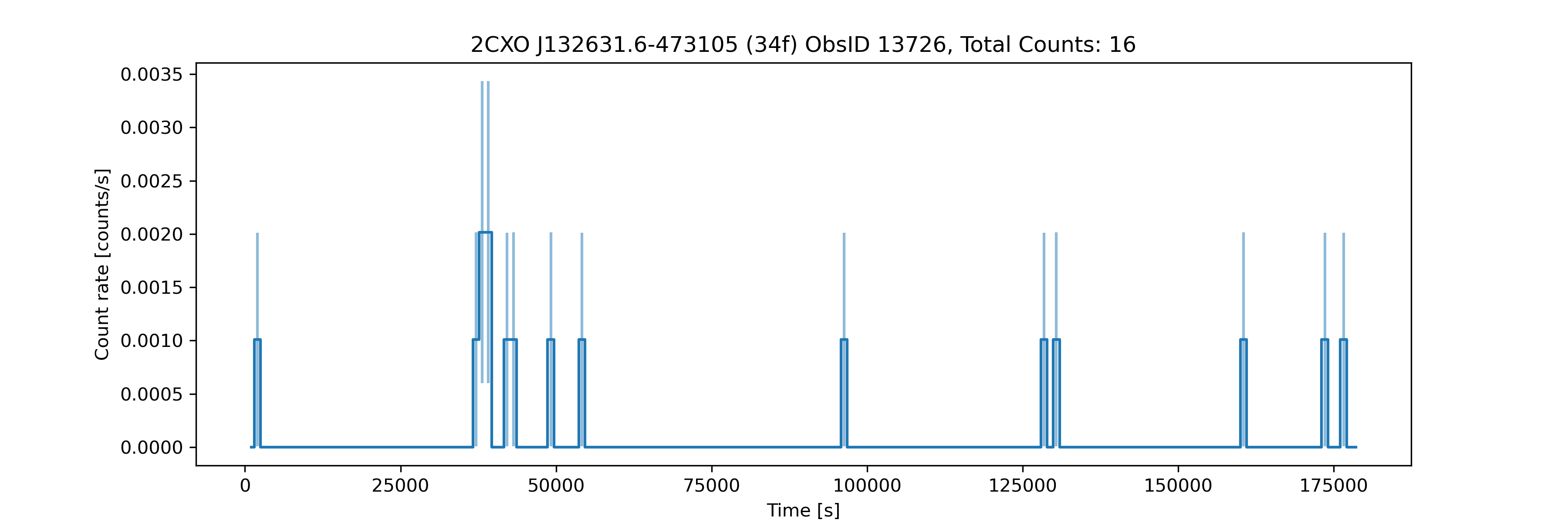}
    \includegraphics[width=0.43\textwidth]{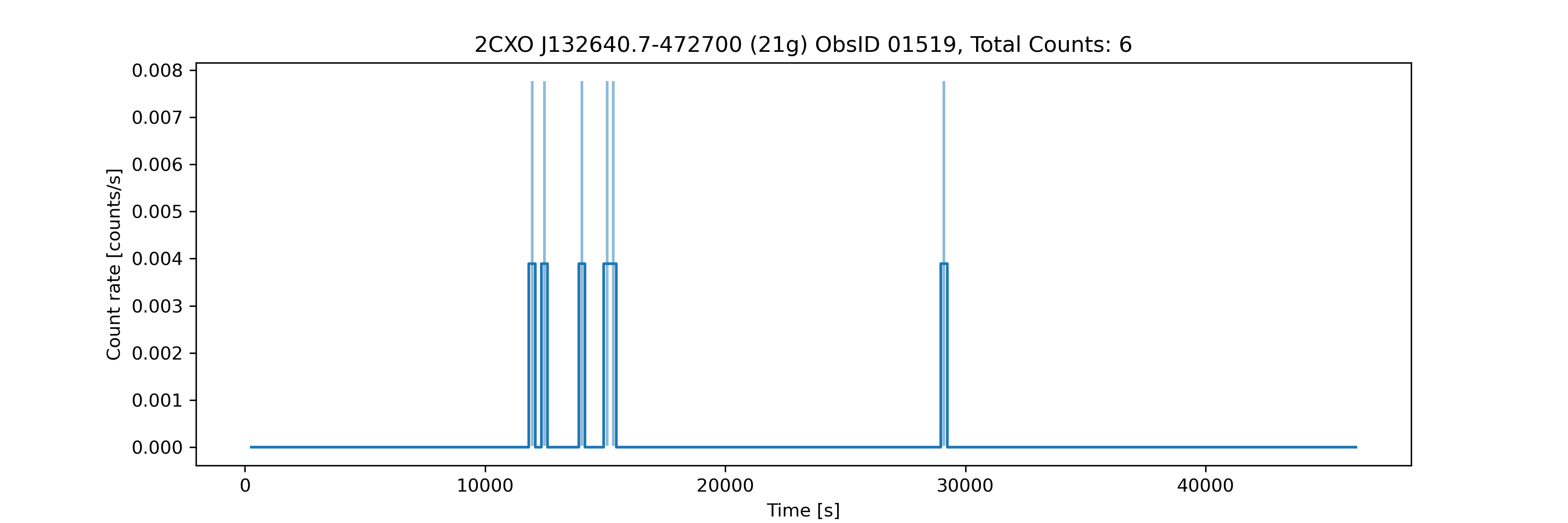}
    \includegraphics[width=0.43\textwidth]{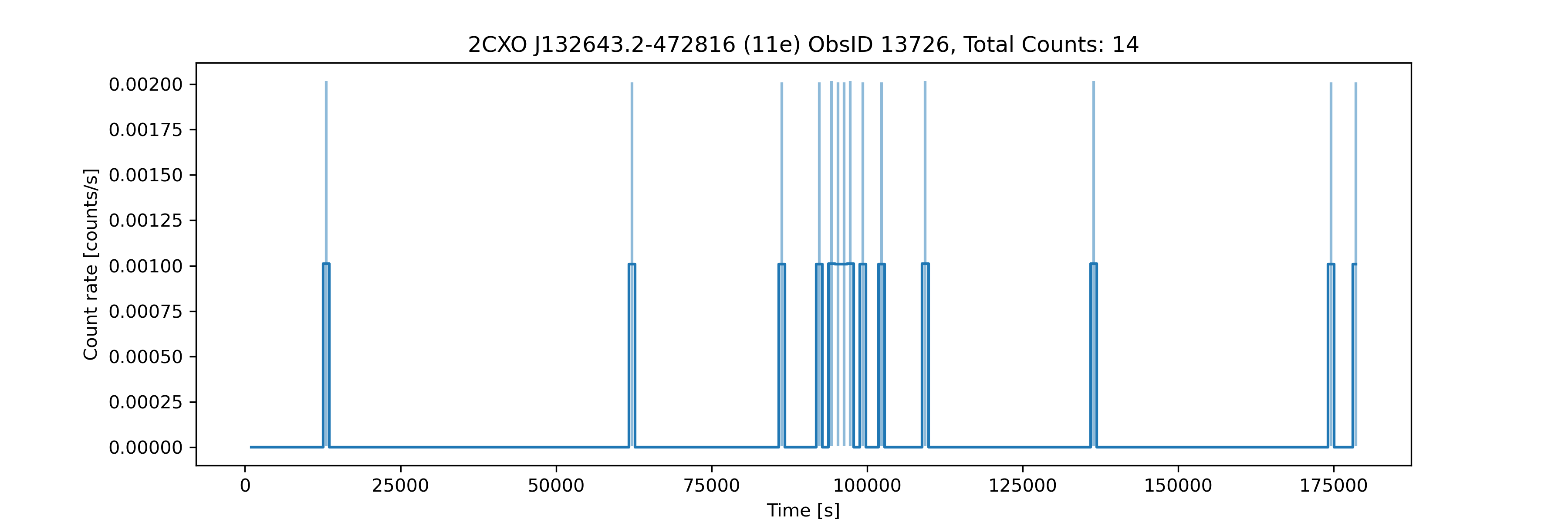}
    \includegraphics[width=0.43\textwidth]{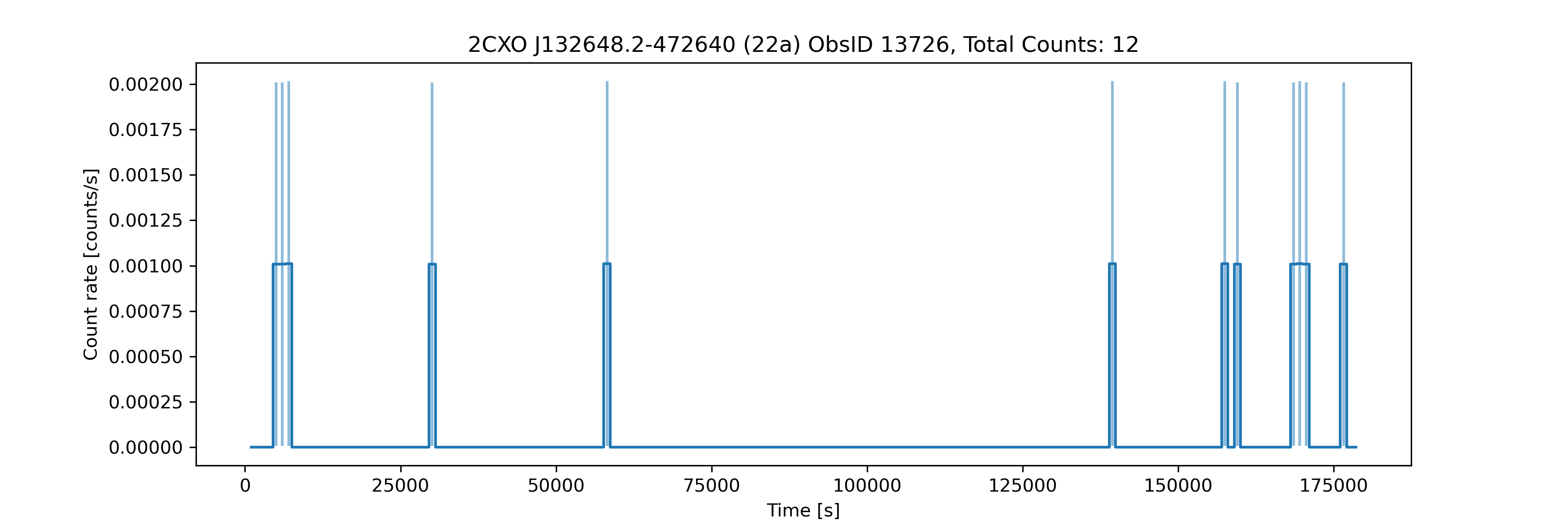}
    \includegraphics[width=0.43\textwidth]{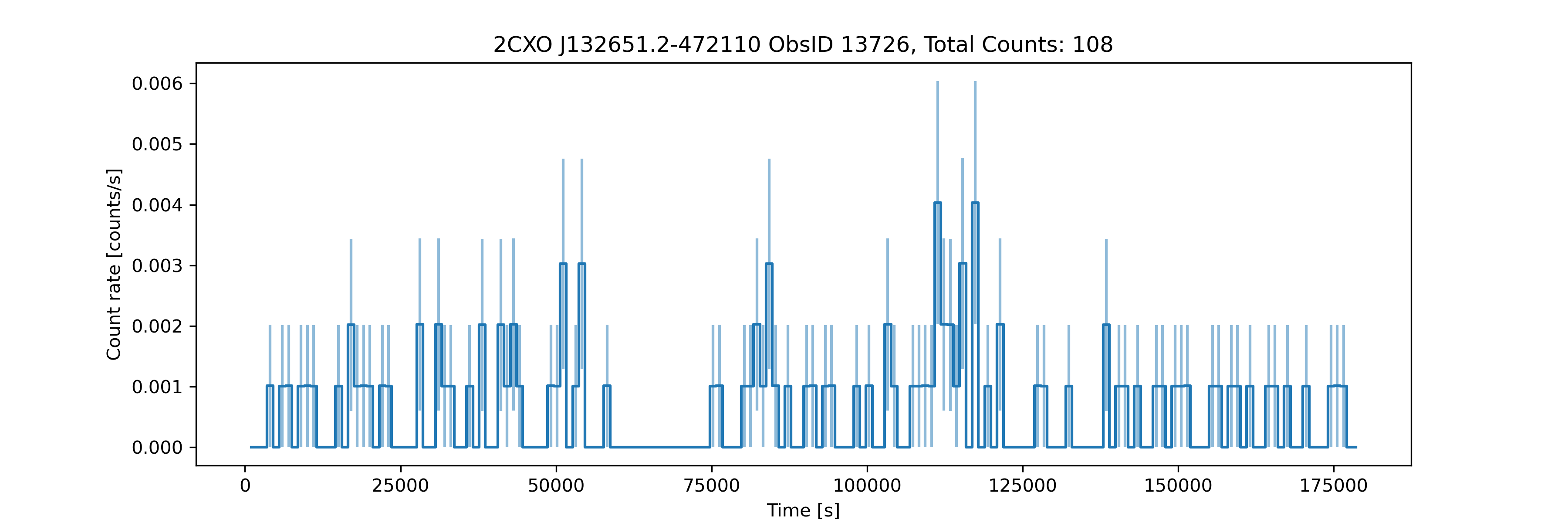}
    \includegraphics[width=0.43\textwidth]{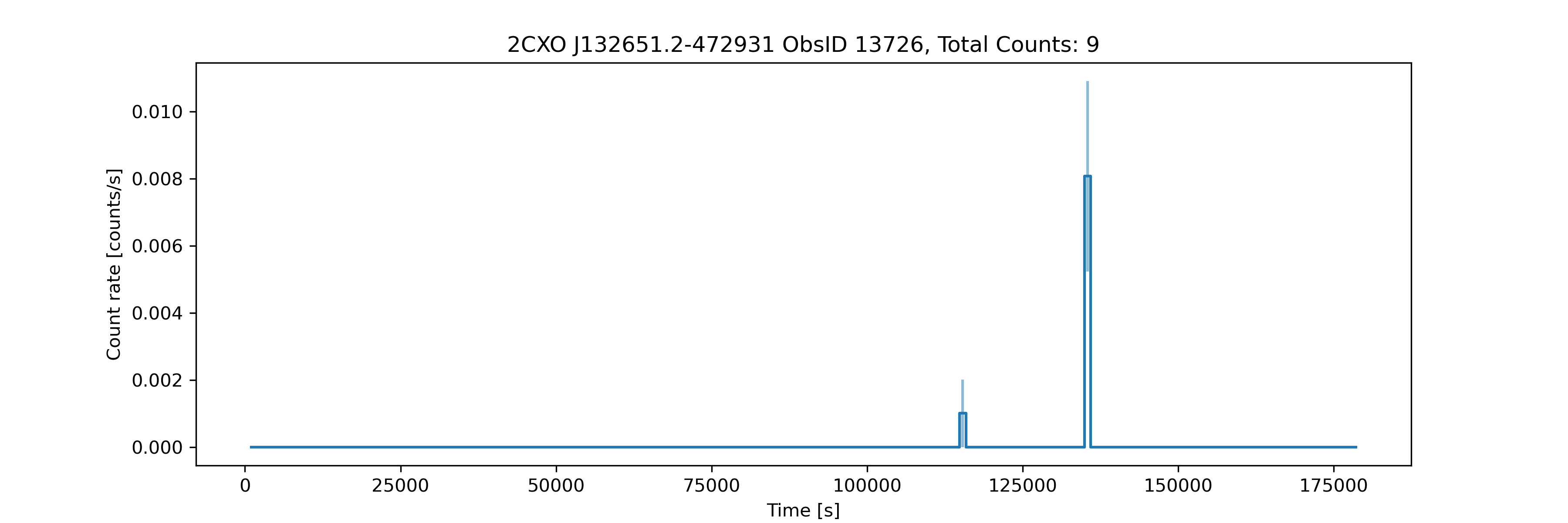}
    \includegraphics[width=0.43\textwidth]{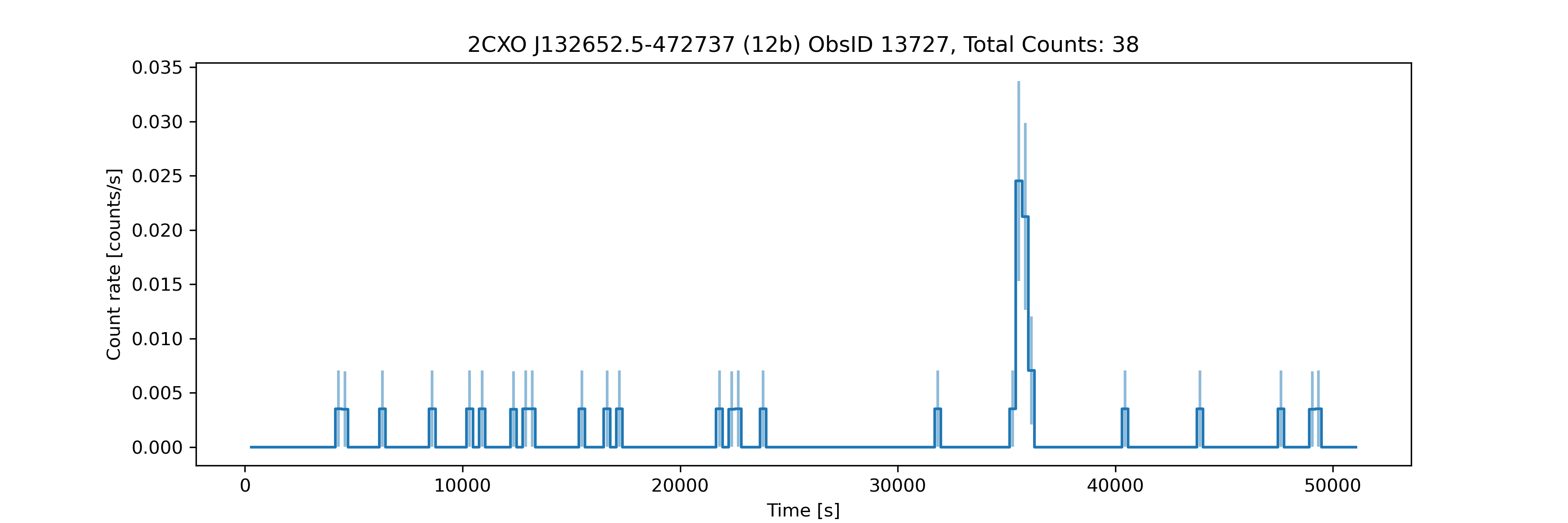}
    \includegraphics[width=0.43\textwidth]{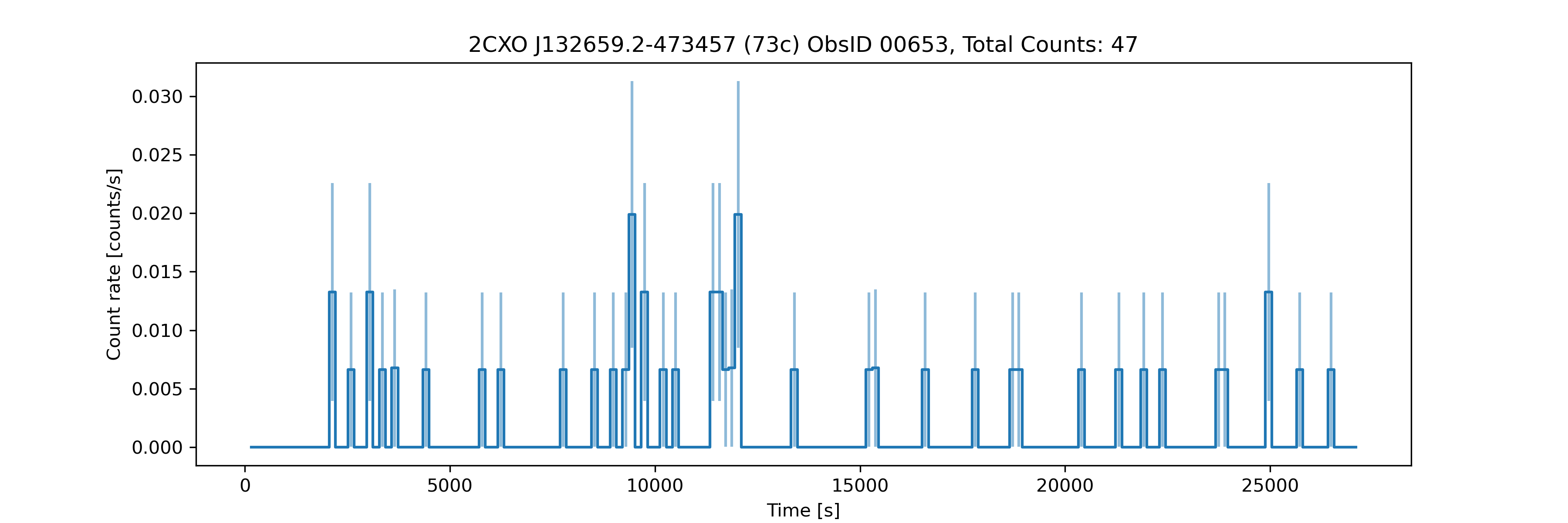}
    \includegraphics[width=0.43\textwidth]{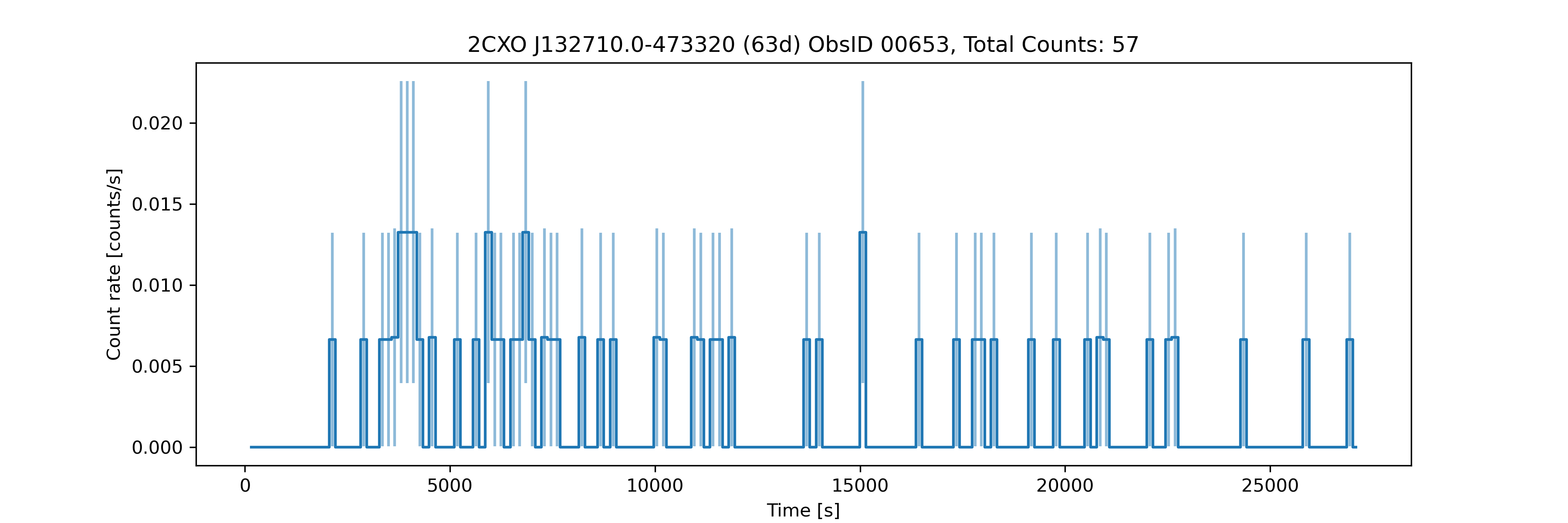}
    \includegraphics[width=0.43\textwidth]{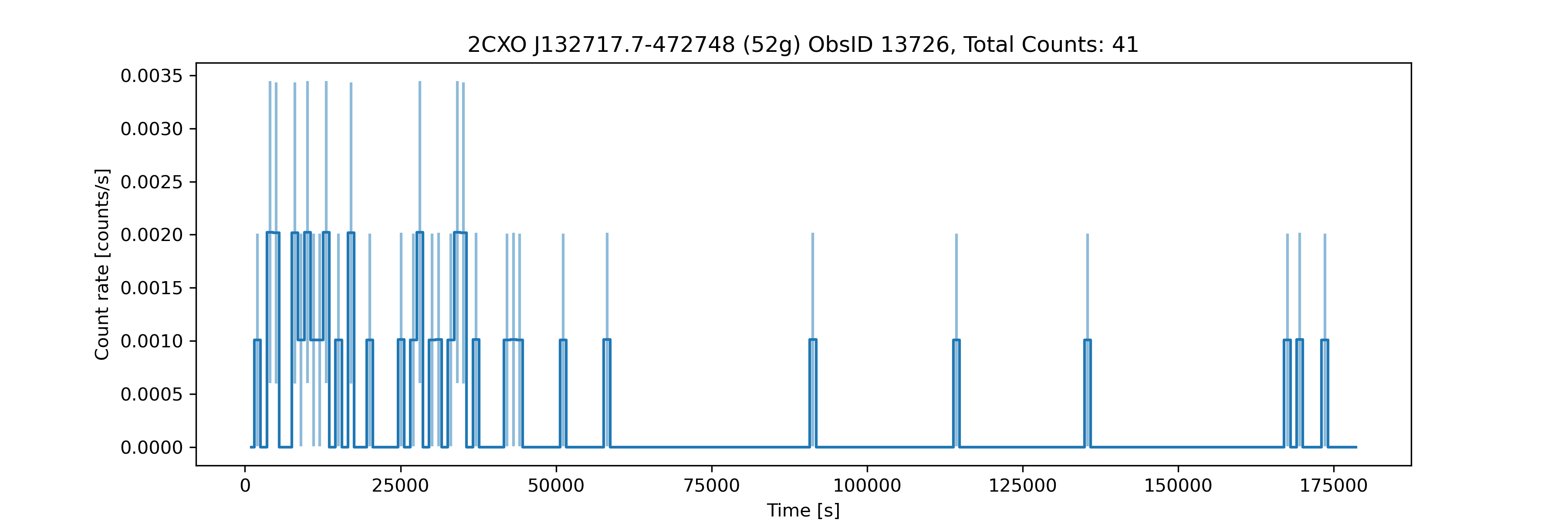}
    \caption{Light curves of variable sources at the observation listed in Table \ref{tab:variable}. Time bins are 1 ks.}
    \label{fig:variable}
\end{figure*}

In Figure \ref{fig:other_csc}, we show their X-ray properties along with their MUWCLASS classification, only using X-ray properties. As expected, most sources are classified as AGN.

Source 94a (2CXO J132601.5-473305) is a very bright source with  $>$1,000 counts. It was classified in \cite{henleywillis_deep_2018} as a symbiotic binary system, where a WD is accreting from the wind of a known carbon star. Since it is far from the cluster center, there is no oMEGACat coverage, and the X-ray source was classified as a CV. Its X-ray spectrum appears similar to those of other bright CVs, especially sources 13a and 13c. Its X-ray lightcurves show possibly periodic variability on the timescale of 10 ks. 

Source 83e (2CXO J132712.8-473456) has 530 counts and was not classified previously because it is 7.5\arcmin from the cluster center without any HST coverage. MUWCLASS classified the null association as a CV at 86\% probability. There is a cluster-member Gaia DR3 source (ID 6083509776780150528) $1.6\arcsec$ away, significantly outside the CSC PU. However, it is bright and blue, with Gmag of 19 and BP-RP color of 0.3, consistent with a WD. Given the presence of these possible cluster-member X-ray sources far from the cluster center, future HST observations to cover these sources are encouraged.

\begin{deluxetable*}{cllllllll}
\tablecaption{CSC Sources showing variability in lightcurves}
\label{tab:variable}
\tablehead{ 
\colhead{Name} & \colhead{Lit.} & \colhead{Manual} & \colhead{ML} & \colhead{P\_C} & \colhead{ObsID} & \colhead{T} & \colhead{Counts}
\\
\colhead{} & \colhead{} & \colhead{} & \colhead{} & \colhead{} & \colhead{} & \colhead{ks} & \colhead{} & \colhead{}
}

\startdata
J132601.5-473305-0 & 94a & CH star & CV & 0.87 & 13726 & 20 & 1019 \\
J132617.4-472337-0 & 71d &  & AGN & 0.82 & 1519 & 7 & 6 \\
J132631.6-473105-1 & 34f & AB? & AB & 0.43 & 13726 & 10 & 10 \\
J132640.7-472700-4 & 21g & AGN & AB & 0.39 & 1519 & 5 & 4 \\
J132643.2-472816-5 & 11e & AB? & AB & 0.97 & 13726 & 10 & 6 \\
J132648.2-472640-3 & 22a & FG & AB & 0.42 & 13726 & 20 & 7 \\
J132651.2-472110-0 &  &  & AGN & 0.48 & 13726 & 10 & 33 \\
J132651.2-472931-2 &  & AB? & AB & 0.59 & 13726 & 0 & 9 \\
J132652.5-472737-3 & 12b & CV? & AGN & 0.46 & 13727 & 2 & 36 \\
J132659.2-473457-0 & 73c & AGN? & CV & 0.41 & 653 & 3 & 22 \\
J132710.0-473320-0 & 63d &  & AGN & 0.63 & 653 & 32 & 10 \\
J132717.7-472748-0 & 52g &  & AB & 0.47 & 13726 & 50 & 22 \\
\enddata

\tablecomments{Columns: 'Name': CSC name with '2CXO' prefix removed, 'Lit.': Naming scheme used in \cite{henleywillis_deep_2018, cool_hstacs_2013}, 'Manual': Manual classification, 'MUWCLASS': MUWCLASS classification, 'P\_C': MUWCLASS classification probability, 'ObsID': Chandra ObsID of the observation showing significant $>3\sigma$ variability, 'T': Approximate duration of flare or other variability event, 'Counts': Total counts detected in observation showing significant variability. Source 2CXo J132651.2-472931 suspected to be cosmic ray afterglow. }

\end{deluxetable*}

\section{Discussion}
\label{discussion}

In addition to confirming and revising the nature of previously classified X-ray sources, we classified new sources in the $\omega$ Cen field, including 3 candidate CVs, 7 candidate MSPs, 2 confident and 12 candidate giant ABs, as well as a number of background AGN and foreground stars. We discuss our classification results in the context of past observational and theoretical work on the X-ray source distribution of $\omega$ Cen. 

Since AGNs can be relatively easily mistaken for compact-object binaries or isolated pulsars, we compare the expected and classified numbers of AGNs. We follow \cite{henleywillis_deep_2018} and use the density of AGNs in the Chandra Deep Fields \citep{luo_chandra_2017}. As oMEGACat does not cover the entire field of view of the Chandra observations, and as the Chandra source density (and therefore the percentage of sources that are AGNs) varies with distance from the cluster, we first focus on the region $<2 r_c$ of Omega Cen \citep[one core radius $r_c=155 \arcsec$, ][]{henleywillis_deep_2018} to estimate the AGN surface density. In this region, there are 56 CSC sources brighter than \SI{5e-16}{\erg\per\square\cm\per\second} in the 0.5-2.0 keV band, of which 10 are associated with PM-identified galaxies. Based on Figure 5 of \cite{henleywillis_deep_2018}, $\sim 35\%$ of sources ($\approx 20$) are expected to be AGN in this region. Then, based on the number of CDF galaxies with HST counterparts, 70\% ($\approx 14$) of these 20 expected AGN should be detectable with HST. The number of observed AGN in this region is thus slightly lower than the number expected. The 8 candidate AGNs in this region, classified based on X-ray properties and the presence of MeerKAT counterparts, but lacking associations with extragalactic PMs, account for the expected number of AGN not detectable with HST.

At X-ray fluxes lower than $\SI{5e-16}{\erg\per\square\cm\per\s}$, Chandra completeness in this field decreases (see Figures 4, 5 in \cite{henleywillis_deep_2018}), and estimates of AGN surface densities from the Chandra Deep Fields may no longer be appropriate. However, if we assume that a percentage similar to the brighter X-ray sources ($\sim 60\%$) of these faint X-ray sources are AGN, then we expect $\sim 130$ AGN in total. To account for this, all 82 confident and candidate AGNs, plus $\sim 70\%$ of the 64 sources not matched to HST or MeerKAT counterparts must be AGN. Therefore, it is plausible that most CSC sources with MeerKAT counterparts, and most of the other CSC sources without any assigned counterparts, are AGN. 

\cite{henleywillis_deep_2018} estimated a total of $\sim 40$ CVs with $L_X(0.5-2.0 \text{ keV}\geq \qty{1e30}{\erg\per\s}$ to be present in the cluster, extrapolating from their findings of 18 CVs and CV candidates. However, we find that 6 of these CV candidates are likely to be AGNs. Scaling down gives an estimate of $\sim 27$ total CVs, compared to 12 of our CVs and CV candidates. These 12 sources would make $\omega$ Cen one of the GCs with the most known CVs. The only GCs with higher CV counts detected in the CSC are 47 Tuc and NGC 6397, which are very dense, if not completely core collapsed. Still, \cite{henleywillis_deep_2018} notes that the volume density of CVs in $\omega$ Cen is lower than that in the Galactic field. 

Since $\omega$ Cen is not a core-collapsed GC, mass segregation of the relatively heavier WDs toward the cluster center is far from complete \citep{ferraro_pure_2006, henleywillis_deep_2018}. As noted by \cite{henleywillis_deep_2018}, these CVs indeed are not concentrated toward the cluster center. In fact, bright X-ray sources such as 83e, far from the center and outside of oMEGACat coverage, may also be CVs. These sources hint at the existence of fainter unidentified CVs in regions not currently covered by HST observations. 

We searched for optical counterparts to known MSP and BW spider systems, finding a likely WD counterpart (4th association) to MSP Q. Out of the possible counterparts to the five BW systems, that of MSP K (1st association) is the most convincing, as it lies very close to the updated timing position from \cite{bernadich_joint_2026}. The lack of firm detections of the BW MSPs suggests that these companions are indeed very faint.

We also identified 7 candidate MSPs based on X-ray properties and the presence of MeerKAT counterparts, two of which are strong binary WD-MSP candidates. While their radio counterparts were detected from the MeerKAT images, the existing radio timing observations were not able to detect their pulsations. This may be due to their MeerKAT fluxes being lower than the fluxes of most of the known MSPs and spiders in this cluster. In $\omega$ Cen, radio observations were only sensitive enough to discover the first MSPs in 2020 \citep{dai_discovery_2020}. Despite recent MeerKAT discoveries and updates, it seems likely that only the brightest MSPs within the half-light radius have been detected \citep{chen_meerkat_2023, bernadich_joint_2026}; see also Figure \ref{fig:radio_xray_lum}.

The large population of MSPs in $\omega$ Cen also raises questions, especially the large fraction of isolated and BW MSPs \citep{chen_meerkat_2023, bernadich_joint_2026}. Given the low density in $\omega$ Cen, the stellar encounter rate, and the encounter rate per binary are small when compared to a dense GC such as 47 Tuc \citep{chen_meerkat_2023}. The stellar encounter rate, which estimates the total number of MSPs, somewhat under-predicts the observed MSP number, while the isolated MSP fraction is a true outlier considering the low encounter rate per binary (see Figure 7 of \cite{bernadich_joint_2026}). Isolated MSPs are formed through the disruption of accreting MSP systems by three-body interactions. With a low encounter rate per binary, there should not have been enough close encounters to either harden binary MSP systems into BWs, or disrupt them into isolated MSPs, in the numbers observed. More wide binary MSPs with orbital periods over a day, such as MSPs I, N, and Q, are expected, but only these three are currently observed \citep{bernadich_joint_2026}. The MSP population in $\omega$ Cen thus cannot be explained just with present day encounter-rate parameters, and the past history of the cluster as the core of a dwarf galaxy may also be an important factor \citep{chen_meerkat_2023}. Alternatively, isolated MSPs may form from other channels than three-body disruption, e.g. the complete ablation of BW companions \citep{bernadich_joint_2026}.

Our candidate MSPs can help constrain this puzzle. The two binary WD MSP candidates and the huntsman spider candidate would double the number of binary MSPs known that are wider than spiders, but do not change the picture significantly by themselves. The other candidate MSPs lack obvious counterparts, and may be either isolated or binary systems, but continued radio observations should reveal their nature in the near future. 
It is also possible that the apparent excess of bright WDs near reported isolated MSPs such as MSPs D, E, and J are in fact companions in very wide orbits, naturally resolving the issue. The widest known MSP binaries in GCs have orbital periods of hundreds of days \citep{chen_meerkat_2023}. MSPs D and E were timed with the Parkes telescope with a baseline of 3.5 years \citep{dai_timing_2023}, implying that if their orbital periodicity is still undiscovered, then it would be on the order of years. Nevertheless, such unprecedentedly wide orbits may be possible in $\omega$ Cen given its low density.

We also refined and expanded the known population of giant star ABs. From the MUSE metallicity measurements \citep{nitschai_omegacat_2023}, we found no evidence that the previously associated giant ABs all belong to the metal-rich RGB/SGB-a, as suggested by \cite{cool_hstacs_2013}. If our 10 candidates are confirmed (see panel 5 of Figure \ref{fig:csc_omegacat_separation_density}), they will be among the most optically luminous giant ABs in GCs. 

On the other hand, we find a modest population of candidate lower-mass, MS ABs (panel 7, Figure \ref{fig:csc_omegacat_separation_density}). Around half of them appear to lie slightly above the MS in the bluer CMDs, and they are all brighter than $\sim 8$ absolute mag in F336W. This is somewhat puzzling, as MS ABs make up a large fraction of TD sources, and the Chandra sensitivity limits should be deep enough to reveal a larger population, especially fainter ones, if their properties are similar to those of ABs in other GCs. This may be explained by $\omega$ Cen's lower density, as most of the MS ABs in the TD are from 47 Tuc, M4, and NGC 6397, three clusters that are very dense, or core-collapsed. The high stellar encounter rate in these clusters can further ``harden'' MS binaries that are already ``hard'' \citep{de_menezes_origin_2026}, allowing tidal-locking and coronal emissions to occur. The lower density of $\omega$ Cen on the other hand may make this formation channel unlikely. No radio emissions are seen around these sources, as despite the sensitivity of the MeerKAT observations, more than an order of magnitude increase in sensitivity is required to detect radio emissions from coronally active binary stars.

Altogether, we have confidently classified 25 X-ray sources as cluster members, including previously classified ones. Adding our candidate CVs, MSPs, and giant ABs raise this number to 44. All of these sources lie within 2 core radii of the cluster, and they can account for the $\sim 48$ cluster members estimated for this region in Section 4 of \cite{henleywillis_deep_2018} This number is also broadly consistent with the excess PM members seen in Figure \ref{fig:csc_omegacat_separation_density}. Outside this region, \cite{henleywillis_deep_2018} only estimates $\sim 8$ cluster members among detected Chandra sources. Thus, estimates of X-ray source cluster membership based both on X-ray source distributions, and on optical counterpart distributions (see Figure \ref{fig:csc_omegacat_separation_density}) do not leave room for a significant population of MS ABs or other X-ray sources among currently detected X-ray sources. However, due to loss of sensitivity of Chandra at larger off-axis angles, a Chandra mosaic with pointings at these outer regions could uncover significantly more cluster members. Future Chandra and HST observations of these regions will enable a more complete census of X-ray sources in $\omega$ Cen, and can further probe the effect of cluster dynamics on the distribution of X-ray emitting binaries. Improved radio positional accuracy will also enable much more reliable optical counterpart identification, helping to constrain possible source classes.

\section{Summary}

We conducted classification of X-ray sources in $\omega$ Cen using the latest Chandra and HST catalogs as well as JWST, and MeerKAT data, whenever covarage was available. Using astrometric, optical-IR photometric, and X-ray data, we determined the likely optical, infrared, and radio counterparts of X-ray sources. We find or confirm the probable counterparts for 18 AGN, 12 foreground stars, 6 giant ABs, 9 CVs, and the binary MSP Q and the BW spider MSP K. Additionally, we find 7 candidate MSPs, including 2 WD-MSP binaries, 3 candidate CVs, 12 candidate giant ABs, and numerous candidate AGN and foreground stars. The numbers of cluster members and non-cluster members among the X-ray sources are largely consistent with the expected numbers from \cite{henleywillis_deep_2018}.
While the MUWCLASS classification results helped guide our analysis, its performance is limited by the features available to the training dataset. Many features critical for counterpart identification and classification is challenging to incorporate into the pipeline, due to lack of coverage (e.g. MeerKAT data), or incompatible formats in different catalogs (e.g. PM-determined cluster membership). Regardless, the result of our work is the most comprehensive census of X-ray sources in $\omega$ Cen to date, especially within 2 core radii of $\omega$ Cen. The X-ray sources outside this region are expected to be mostly foreground stars and AGN. However, due to incompleteness of Chandra observations at fainter fluxes and large off-axis angles, a significant population of cluster members may still exist within its outskirts. Additional Chandra and HST/JWST observations are necessary to fully constrain X-ray populations, models of compact object formation, and cluster dynamical evolution.

\section{Acknowledgments}

This research is supported by Space Telescope Science Institute award HST-AR-16620.001-A and JWST-GO-04343.001. J.H. acknowledges support from NASA under award number 80GSFC24M0006.

The MeerKAT telescope is operated by the South African Radio Astronomy Observatory, which is a facility of the National Research Foundation, an agency of the Department of Science, Technology and Innovation. OMS's research is supported by the South African Research Chairs Initiative of the NRF and DSTI (grant No. 81737). We acknowledge the financial support of the Breakthrough Listen project. Breakthrough Listen is managed by the Breakthrough Initiatives, sponsored by the Breakthrough Prize Foundation.

This paper made use of the ATNF and SpiderCat catalogs, hosted at \url{https://www.atnf.csiro.au/research/pulsar/psrcat/} and \url{https://astro.phys.ntnu.no/SpiderCAT/} respectively. We thank Maximillian Haberle for helpful discussions about oMEGACat. We thank George Pavlov for constructive feedback on the methodology.

\appendix

\section{Data Availability}

The full electronic table of CSC source classifications in $\omega$ Cen, along with their spectra and light curves, is available online at \url{https://doi.org/10.5281/zenodo.23423086}. Also available are the TD and some supplemental tools that aided us in our analysis, including an interactive plotting tool and an interactive light-curve viewer. Another tool for the visualization of CSC sources and their potential counterpart associations in Chandra and HST images is available at \url{https://schen70.github.io/CSC_Omega_Cen_Viewer/}.

The raw MeerKAT data underlying this article is publicly available via the SARAO archive\footnote{\url{https://archive.sarao.ac.za}} under proposal ID SSV-20181107-FC-02.

\section{Gaussian Process Implementation}
\label{gp_details}

A GP is a set of infinite random variables that form a \textit{distribution over functions}, where any subset of the random variables forms a normal distribution. They are useful for regression tasks, as they are non-parametric in nature, while the choice of the \textit{kernel} that determines the covariance between the points in the GP can be physically motivated by the expected behavior in the data. For example, an Exponential Sine Squared kernel contains a sine squared function, therefore it outputs covariance values close to 1 only if the distance between two points is a multiple of its periodicity hyperparameter, enabling it to model periodic behavior in data. Multiple kernels can be added or multiplied together to model more complex behavior in the data. For more details, refer to a comprehensive overview of GPs by \cite{rasmussen_gaussian_2005}. 

For our purposes of fitting broadband photometry, we used a Radial Basis Function (RBF) kernel, which controls how fast the random variables can vary with a lengthscale hyperparameter $l$ with respect to the wavelength distance $d(x_i, x_j)$ between bands $i$ and $j$, summed with a quadratic kernel (squared dot-product kernel), which tends to model the data as a quadratic function, see Equation \ref{eq:kernel}. The RBF kernel can smoothly model any feature in the data that varies over a wavelength range longer than the lengthscale, but will tend to immediately move the GP towards the mean of the training data used to fit the hyperparameters, when outside the wavelength range of the training data. The quadratic kernel mitigates this issue by modeling the data on a large scale as a quadratic function, but by itself, it lacks the flexibility to fit any local features. The combined kernel is thus capable of fitting smoothly varying blackbody or power-law SEDs well, but is unable to account for absorption or emission features or any temporal variability. We constrain the lengthscale hyperparameter $l$ to be between 100 and 200 nm, and the dot-product kernel inhomogeneity hyperparameter $\sigma_0$ (which biases the quadratic kernel) to be between 3 and 1000.

\begin{align}
    k(x_i, x_j) = \exp{\left(- \frac{d(x_i, x_j)^2}{2l^2} \right)} + (\sigma_0 ^ 2 + x_i \cdot x_j)^2
    \label{eq:kernel}
\end{align}

\begin{figure*}
    \centering
    \includegraphics[width=0.5\textwidth]{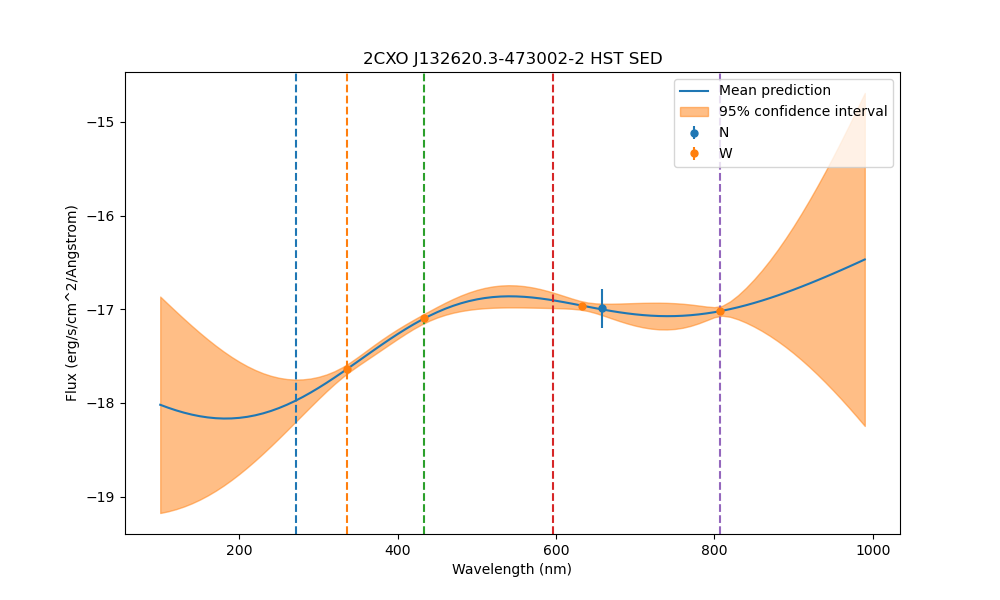}
    \caption{An example GP fit to the SED of a source in $\omega$ Cen. HST fluxes used to fit the GP are shown as markers at their average wavelengths, with colors indicating their band type (wide or narrow). The mean and 95\% confidence interval of the GP fit are shown as the solid blue line and the orange band. For this source, a missing value in the F606W band is fitted using nearby data.}
    \label{fig:gp_sed_fit}
\end{figure*}

After fitting to the available photometry, the GP outputs a mean predicted flux and a corresponding standard deviation at any point within the specified wavelength domain, where the magnitude of the standard deviation depends on the distance to the nearest point used to train the fit. An example GP fit is shown in Figure \ref{fig:gp_sed_fit}, where W indicates wide bands, and N indicates narrow bands, e.g. the F658N band. This feature of GPs naturally integrates into the uncertainty sampling step of MUWCLASS: for every missing flux interpolated by GPs, we can randomly sample the mean interpolated flux using its standard deviation. In this way, any uncertainties in the GP interpolation process will be taken into account by MUWCLASS, similar to a measured flux value with large uncertainties. We emphasize that the purpose of GP interpolation is not to fit source SEDs to a high degree of accuracy, but to fill in reasonable values, with reasonably large uncertainties in bands that would otherwise be missing data.

\newpage

\bibliographystyle{aasjournalv7}
\bibliography{references.bib}

\end{document}